\documentclass[useAMS,usenatbib]{mnras}
\usepackage{graphicx}

\usepackage{journal_names}
\usepackage{subfig}
\usepackage{pdflscape}
\usepackage{amssymb}

\title[Radio spectra of z$>$4.5 sources]{Radio spectra of bright compact sources at z$>$4.5} 
\author[Rocco Coppejans et al.]{Rocco Coppejans$^{1}$\thanks{E-mail: r.coppejans@astro.ru.nl }, Sjoert van Velzen$^{2}$, Huib T. Intema$^{3}$, Cornelia M\"{u}ller$^{1}$, \newauthor S\'{a}ndor Frey$^{4,8}$, Deanne L. Coppejans$^{1}$, D\'{a}vid Cseh$^{1}$, Wendy L. Williams$^{5}$, \newauthor Heino Falcke$^{1,6}$, Elmar G. K\"{o}rding$^{1}$, Emanuela Orr\'{u}$^{6,1}$, Zsolt Paragi$^{7}$, \newauthor and Krisztina \'{E}. Gab\'{a}nyi$^{4,8}$ \\
$^{1}$Department of Astrophysics/IMAPP, Radboud University, P.O. Box 9010, 6500 GL Nijmegen, The Netherlands\\
$^{2}$Department of Physics and Astronomy, The Johns Hopkins University, Baltimore, MD 21218, USA\\
$^{3}$Leiden Observatory, Leiden University, PO Box 9513, 2300 RA, Leiden, The Netherlands\\
$^{4}$F\"{O}MI Satellite Geodetic Observatory, PO Box 585, H-1592 Budapest, Hungary\\
$^{5}$School of Physics, Astronomy and Mathematics, University of Hertfordshire, College Lane, Hatfield AL10 9AB, UK\\
$^{6}$Netherlands Institute for Radio Astronomy (ASTRON), PO Box 2, 7990 AA Dwingeloo, The Netherlands\\
$^{7}$Joint Institute for VLBI ERIC, Postbus 2, 7990 AA Dwingeloo, The Netherlands\\
$^{8}$Konkoly Observatory, MTA Research Centre for Astronomy and Earth Sciences, Konkoly Thege Mikl\'{o}s \'{u}t 15-17, H-1121 Budapest, Hungary\\
}

\begin{document}

\date{}

\pagerange{\pageref{firstpage}--\pageref{lastpage}} \pubyear{2016}

\maketitle

\label{firstpage}

\begin{abstract}
High-redshift quasars are important to study galaxy and active galactic nuclei (AGN) evolution, test cosmological models, and study supermassive black hole growth. Optical searches for high-redshift sources have been very successful, but radio searches are not hampered by dust obscuration and should be more effective at finding sources at even higher redshifts. Identifying high-redshift sources based on radio data is, however, not trivial. Here we report on new multi-frequency Giant Metrewave Radio Telescope (GMRT) observations of eight $z>4.5$ sources previously studied at high angular resolution with very long baseline interferometry (VLBI). Combining these observations with those from the literature, we construct broad-band radio spectra of all 30 $z>4.5$ sources that have been observed with VLBI. In the sample we found flat, steep and peaked spectra in approximately equal proportions. Despite several selection effects, we conclude that the $z>4.5$ VLBI (and likely also non-VLBI) sources have diverse spectra and that only about a quarter of the sources in the sample have flat spectra. Previously, the majority of high-redshift radio sources were identified based on their ultra-steep spectra (USS). Recently a new method has been proposed to identify these objects based on their megahertz-peaked spectra (MPS). Neither method would have identified more than 18\,per\,cent of the high-redshift sources in this sample. More effective methods are necessary to reliably identify complete samples of high-redshift sources based on radio data.
\end{abstract}


%%%%%%%%%%%%%%%%%%%%%%%%%%%%%%%%%%%%%%%%%%%%%%%%%%%%%%%%%%%%%%%%%%%%%%%%%%%%%%%%%%%%%%%%%%%%%%%%%%%%%%%%%%%%%%%%%%%%%%%%%%%%%%%%%%%%%%%%%%%%%%%%%%%%%%%%%%%%%%%%%%%%%%%%%%%%%%%%%%%%%%%%%%%%%%%%%%%%%%%%%%%%%%%%%%%%%%%%%%
%%%%%%%%%%%%%%%%%%%%%%%%%%%%%%%%%%%%%%%%%%%%%%%%%%%%%%%%%%%%%%%%%%%%%%%%%%%%%%%%%%%%%%%%%%%%%%%%%%%%%%%%%%%%%%%%%%%%%%%%%%%%%%%%%%%%%%%%%%%%%%%%%%%%%%%%%%%%%%%%%%%%%%%%%%%%%%%%%%%%%%%%%%%%%%%%%%%%%%%%%%%%%%%%%%%%%%%%%%
\section{Introduction}
\label{sec:introduction}
It is believed that there is a supermassive black hole at the center of nearly every galaxy. These objects power active galactic nuclei (AGN) and were formed in the early Universe. They continue to influence, shape, and grow with their host galaxy via feedback \citep[e.g.][]{2005MNRAS.362...25B,fabian2012,2013Sci...341.1082M}. To understand present-day galaxies, we consequently need to understand AGN evolution \citep[e.g.][]{fabian2012}. A critical aspect of this is identifying AGN at high redshifts. 

In the optical, AGN have been found at distances of up to redshift 7.1 \citep{mortlock2011}. However, due to Ly-alpha absorption, detecting sources beyond $z=6.5$ is very difficult in the optical \citep{mortlock2011,becker2001}. In addition optical searches are hampered by dust obscuration, which does not affect radio observations \citep[e.g.][]{2004cbhg.symp..324O}. With radio observations, we should therefore be able to detect sources at all redshifts more effectively, and detect sources out to higher redshifts. It is worth noting that optical spectroscopy is still essential to determine redshifts of the candidate high-redshift sources detected in the radio. 

One of the main techniques that is used to identify high-redshift sources in radio images, is the ultra-steep-spectrum (USS) method. This method is based on an observed correlation between the spectral index ($\alpha$; defined as $S \propto \nu ^ \alpha$ where $S$ is the flux density at frequency $\nu$) and redshift \cite[e.g.][]{1957MNRAS.117..680W,1979A&A....80...13B,1980MNRAS.190..903L,DeBreuck2000}. According to this correlation, sources that have steeper spectra are at higher redshifts. The USS method has proven successful: most of the high-redshift sources identified through radio observations were selected using this method \citep{DeBreuck2000,2010AstL...36....7V,singh2014}, and it has also succeeded in finding sources out to $z>4$ \citep[e.g.][]{1999ApJ...518L..61V,jarvis2001,2006AstL...32..433K}.

Despite this success, there is no physical explanation for why USS sources should be at higher redshifts than non-USS sources \citep[e.g.][]{miley2008,2006MNRAS.371..852K,2010AstL...36....7V,singh2014}, and several recent studies have failed to find a correlation between the spectral index and redshift \citep{ker2012,singh2014,smolcic2014}. The exact definition of a USS source (based on spectral index) differs between authors, e.g., $\alpha^{\rm 608\,MHz}_{\rm 327\,MHz} < -1.1$ \citep{wieringa1992}, $\alpha^{\rm 4.85\,GHz}_{\rm 151\,MHz} < -0.981$ \citep{blundell1998}, $\alpha^{\rm 1.4\,GHz}_{\rm 843\,MHz}<-1.3$ \citep{deBreuck2004}, $\alpha^{\rm 1.4\,GHz}_{\rm 151\,MHz} < -1.0$ \citep{cruz2006}, $\alpha^{\rm 843\,MHz}_{\rm 408\,MHz} \leq -1.0$ \citep{broderick2007} and $\alpha^{\rm 1.4\,GHz}_{\rm 325\,MHz} \leq-1.0$ \citep{singh2014}. However, \citet{coppejans2015} pointed out that in their sample of sources, in which all of the sources are detected at 153, 325 and 1400\,MHz, when first selecting USS sources between 153 and 325\,MHz and then selecting USS sources between 325 and 1400\,MHz, less than 26\,per\,cent of the sources appear in both selections. \citet{2003NewA....8..805P} has also pointed out that the USS sources may not be representative of the entire high-redshift source population, since USS sources are typically smaller and more powerful than non-USS sources \citep{1999AJ....117..677B}. This argument is supported by the discovery of two non-USS sources at $z=4.4$ and 4.9 with $\alpha^{\rm 8.5\,GHz}_{\rm 1.4\,GHz}=0.94\pm0.06$ and $\alpha^{\rm 1.4\,GHz}_{\rm 325\,MHz}=0.75\pm0.05$, respectively, \citep{1999ApJ...526L..77W,2009MNRAS.398L..83J}. \citet{2003NewA....8..805P} has shown that up to 40\,per\,cent of the high-redshift sources in a survey can be lost by applying a spectral index cut.

\citet{falcke2004} and \citet{coppejans2015} proposed a new method for searching for high-redshift AGN, namely the megahertz peaked-spectrum (MPS) method. Compact steep-spectrum (CSS), MPS, gigahertz peaked-spectrum (GPS) and high-frequency peaked (HFP) sources are all AGN that show spectral turnovers in their synchrotron spectra, that are believed to be produced by synchrotron self-absorption. GPS, MPS and CSS sources together make up between 15 and 30\,per\,cent of the sources in flux density limited catalogues \citep{o'dea1998,2016AN....337....9O}. The observed turnover (or peak) frequencies ($\nu_{\mathrm o}$) of the CSS, MPS, GPS and HFP sources are $\nu_{\mathrm o}<0.5$\,GHz, $\nu_{\mathrm o}<1$\,GHz, $1<\nu_{\mathrm o}<5$\,GHz and $\nu_{\mathrm o}>5$\,GHz \citep{o'dea1998,2000A&A...363..887D,coppejans2015}, respectively. These sources are believed to be young (rather than confined) AGN, some of which will likely evolve into FR\,I and FR\,II radio galaxies \citep{Begelman1996,o'dea1998,snellen2000,Conway2002,devries2002,Murgia2002,2003PASA...20...19M,2009AN....330..120F,2012ApJ...760...77A,2016AN....337....9O}. For the nearby ($z\sim1$) CSS, MPS, GPS and HFP sources, an empirical relation exists between the rest-frame turnover frequencies ($\nu_{\mathrm r}$, where $\nu_{\mathrm r} = \nu_{\mathrm o}(1+z)$) and the linear sizes of the sources \citep{o'dea1998,snellen2000,Orienti2014}. From this relation, sources with lower values of $\nu_{\mathrm r}$ have larger linear sizes. 

The premise of the MPS method is that there are two classes of sources that have peak frequencies below 1\,GHz. The first class, which includes the CSS sources, are nearby sources for which $\nu_{\mathrm o}\simeq\nu_{\mathrm r}$. The second class of sources have $\nu_{\mathrm r}>1$\,GHz, but $\nu_{\mathrm o}<1$\,GHz due to their higher redshifts. There are two differences between these two classes. First, we expect the high-redshift sources to have smaller angular sizes than the CSS sources, as they are at larger redshifts. Second, the high-redshift sources have higher rest-frame turnover frequencies than the nearby sources. From the turnover frequency--linear size relation, we therefore expect the high-redshift sources to have smaller physical sizes than the CSS sources. It should therefore be possible to distinguish between the CSS and the high-redshift sources based on the high-redshift sources having smaller angular sizes than the CSS sources. 
  
To date no new high-redshift sources have been found using the MPS method. However, \citet{coppejans2015} identified 33 MPS sources in the NOAO Bo\"{o}tes field and were able to determine redshifts for 24. Given that the average redshift of the sources is 1.3, that there are five sources at $z>2$ and that four of the sources for which they could not find redshifts are likely also at $z>2$, the authors concluded that there is encouraging evidence in support of the method. Like the USS method, the MPS method likely only selects a subset of the high-redshift sources. However, the MPS method selects a different class of high-redshift sources than the USS method as it is believed that the MPS sources are young AGN \citep{o'dea1998,Murgia2002,Conway2002}. For this reason, the MPS method is important for understanding AGN evolution. The two methods are therefore complementary and will allow for a better understanding of the high-redshift population as a whole.

In \citet[][hereafter CFC2016]{2016MNRAS.tmp.1343C}, we presented very long baseline interferometry (VLBI) observations of ten new $z>4.5$ sources at 1.7 and 5\,GHz with the European VLBI Network (EVN). This increased the number of $z>4.5$ sources that have been observed with VLBI by 50\,per\,cent, from 20 to 30 sources. Using both the VLBI brightness temperatures and 1.4\,GHz luminosities of all 30 $z>4.5$ VLBI sources, we concluded that in one of the sources the radio emission is from star formation, with the emission originating from AGN activity in the other 29 sources\footnote{Typically, brightness temperatures ($T_{\rm b}$) above $10^6$\,K indicate non-thermal emission from AGN \citep[e.g.][]{Kewley2000,Middelberg2011} while thermal emission from star formation has $T_{\rm b}<10^5$\,K \citep{Sramek1986,Condon1991,Kewley2000}. In \citet{Magliocchetti2014} the authors showed that at $z>1.8$ the radio emission in sources with 1.4\,GHz radio luminosities above $4\times10^{24}\mathrm{W\,Hz^{-1}}$ is caused by AGN activity, while the radio emission in sources with 1.4\,GHz radio luminosities lower than $4\times10^{24}\mathrm{W\,Hz^{-1}}$ is caused by star formation.}. This illustrates that even at $z>4.5$, not all sources detected with VLBI are AGN. From the VLBI spectra, brightness temperatures, and 1.4\,GHz variability we also concluded that the $z>4.5$ VLBI sources are a mixture of steep-spectrum sources and flat-spectrum radio quasars (FSRQs), or blazars, i.e. sources in which the jet is aligned within a small angle of our line of sight \citep[e.g.][]{{1999ASPC..159....3U,2013FrPhy...8..609K}}. We finally argued that the steep-spectrum sources are in fact GPS and MPS sources. 

In this paper, we continue our study of all 30 $z>4.5$ VLBI sources by investigating their broad-band radio spectra. The sources were collected from the Optical Characteristics of Astrometric Radio Sources (OCARS) catalogue\footnote{http://www.gao.spb.ru/english/as/ac\_vlbi/ocars.txt}\citep{2008mefu.conf..183M,2016ARep...60..996M} and the literature. To the best of our knowledge, these 30 sources are the only sources with spectroscopic redshifts above 4.5 that have been imaged with VLBI. We restricted ourselves to only studying sources that have been observed with VLBI in this paper for the following reasons: (1) VLBI observations are necessary to get accurate brightness temperatures for the sources. As discussed in CFC2016, this allows us to distinguish between emission from AGN and star formation and is critical to explain the spectra of J1429+5447 and J1205$-$0742 in Sections \ref{subsubsec: J1429+5447} and \ref{subsubsec: J1205$-$0742}. (2) The $z>4.5$ VLBI sources can be seen as forming a flux density limited sample since all $z>4.5$ sources with 1.4\,GHz flux densities above $\sim5$\,mJy in the Very Large Array (VLA) Faint Images of the Radio Sky at Twenty-centimeter (FIRST) survey \citep{FIRST} have been systematically observed with VLBI in published \citep[][and references therein]{2016MNRAS.tmp.1343C} and ongoing VLBI campaigns. We do however note that some authors have specifically targeted fainter sources. In addition not all $z>4.5$ sources with FIRST flux densities above $5$\,mJy are included in our sample of sources, as these sources were only identified as $z>4.5$ sources after the EVN observing proposal for \citet{2016MNRAS.tmp.1343C} had been submitted. These sources are currently being observed in our latest series of EVN observations. (3) This paper is a continuation of the work in CFC2016. The redshifts and VLBI positions of all of the sources are given in Table~\ref{tbl:RA DEC}. The VLBI positions are taken from the highest frequency VLBI observations (listed in Table~\ref{tbl:flux values for each source}) of the sources as these observations will have the highest positional accuracy.

\begin{table}
 \centering
 \hspace{-0.8cm}
 \begin{minipage}{\columnwidth}
  \caption{Source redshifts and positions}
  \begin{tabular}{cccc}
  \hline
  ID & $z$ & RA [J2000] & DEC [J2000] \\
  \hline
  J0011+1446   & 4.96 & 00:11:15.233 & 14:46:01.81\\
  J0131$-$0321 & 5.18 & 01:31:27.347 & $-$03:21:00.08\\
  J0210$-$0018 & 4.65 & 02:10:43.164 & $-$00:18:18.44\\
  J0311+0507$^{\mathrm a}$   & 4.51 & 03:11:47.966 & 05:08:03.87\\
  J0324$-$2918 & 4.63 & 03:24:44.295 & $-$29:18:21.22\\ 
  J0813+3508   & 4.92 & 08:13:33.327 & 35:08:10.77\\ 
  J0836+0054   & 5.77 & 08:36:43.860 & 00:54:53.23\\ 
  J0906+6930   & 5.47 & 09:06:30.750 & 69:30:30.80\\ 
  J0913+5919   & 5.11 & 09:13:16.547 & 59:19:21.67\\ 
  J0940+0526   & 4.50 & 09:40:04.800 & 05:26:30.95\\ 
  J1013+2811   & 4.75 & 10:13:35.440 & 28:11:19.24\\ 
  J1026+2542   & 5.27 & 10:26:23.621 & 25:42:59.43\\
  J1146+4037   & 5.01 & 11:46:57.790 & 40:37:08.63\\ 
  J1205$-$0742 & 4.69 & 12:05:22.977 & $-$07:42:29.75\\ 
  J1235$-$0003 & 4.69 & 12:35:03.046 & $-$00:03:31.76\\ 
  J1242+5422   & 4.73 & 12:42:30.589 & 54:22:57.45\\ 
  J1311+2227   & 4.61 & 13:11:21.321 & 22:27:38.63\\ 
  J1400+3149   & 4.64 & 14:00:25.416 & 31:49:10.68\\ 
  J1427+3312   & 6.12 & 14:27:38.585 & 33:12:41.93\\ 
  J1429+5447   & 6.21 & 14:29:52.176 & 54:47:17.63\\ 
  J1430+4204   & 4.72 & 14:30:23.742 & 42:04:36.49\\ 
  J1454+1109   & 4.93 & 14:54:59.305 & 11:09:27.89\\ 
  J1548+3335   & 4.68 & 15:48:24.014 & 33:35:00.09\\ 
  J1606+3124   & 4.56 & 16:06:08.518 & 31:24:46.46\\ 
  J1611+0844   & 4.54 & 16:11:05.650 & 08:44:35.48\\ 
  J1628+1154   & 4.47 & 16:28:30.465 & 11:54:03.47\\ 
  J1659+2101   & 4.78 & 16:59:13.228 & 21:01:15.81\\ 
  J1720+3104   & 4.62 & 17:20:26.688 & 31:04:31.65\\ 
  J2102+6015   & 4.58 & 21:02:40.219 & 60:15:09.84\\ 
  J2228+0110   & 5.95 & 22:28:43.526 & 01:10:31.91\\ 
  \hline
  \multicolumn{4}{p{8cm}}{\footnotesize{\textbf{Notes:} $^{\mathrm a}$ \citet{2014MNRAS.439.2314P} found that J0311+0507 is composed of eight components and conclude that the third component is the core. The RA and DEC values are therefore for the third component.}}\\
  \end{tabular}
  \label{tbl:RA DEC}
 \end{minipage}
\end{table}

For a source at $z=4.5$, its entire rest-frame spectrum below 5.5\,GHz will be redshifted into observed frequencies below 1\,GHz. Consequently, to accurately characterize the spectrum, multi-frequency observations of the source below 1\,GHz are required. In Section \ref{sec:GMRT}, we present multi-frequency Giant Metrewave Radio Telescope (GMRT) observations below 1\,GHz of eight $z>4.5$ sources that have been observed at two frequencies with the EVN. Section \ref{sec:litrature} contains a description of how we matched all 30 $z>4.5$ VLBI sources to previous radio observations. The spectra and classifications are presented for each source individually in Section \ref{sec:spectra}. In Section \ref{sec:discus} we discuss the spectral classification of the $z>4.5$ VLBI sources, before presenting a summary and conclusion in Section \ref{sec:summary}. Throughout this paper we assume the following cosmological model parameters: $\Omega_{\rm m}=0.3$, $\Omega_{\lambda}=0.7$, $H_0=72$\,km\,s$^{-1}$\,Mpc$^{-1}$.

%%%%%%%%%%%%%%%%%%%%%%%%%%%%%%%%%%%%%%%%%%%%%%%%%%%%%%%%%%%%%%%%%%%%%%%%%%%%%%%%%%%%%%%%%%%%%%%%%%%%%%%%%%%%%%%%%%%%%%%%%%%%%%%%%%%%%%%%%%%%%%%%%%%%%%%%%%%%%%%%%%%%%%%%%%%%%%%%%%%%%%%%%%%%%%%%%%%%%%%%%%%%%%%%%%%%%%%%%%
%%%%%%%%%%%%%%%%%%%%%%%%%%%%%%%%%%%%%%%%%%%%%%%%%%%%%%%%%%%%%%%%%%%%%%%%%%%%%%%%%%%%%%%%%%%%%%%%%%%%%%%%%%%%%%%%%%%%%%%%%%%%%%%%%%%%%%%%%%%%%%%%%%%%%%%%%%%%%%%%%%%%%%%%%%%%%%%%%%%%%%%%%%%%%%%%%%%%%%%%%%%%%%%%%%%%%%%%%%
\section{Observations with the GMRT}
\label{sec:GMRT}
The sources presented in Table~\ref{tbl:GMRT values} were observed with the GMRT during two projects: 21\_013 and 29\_007. During project 21\_013 the following three sources were observed: J1146+4037, J1242+5422 and J1659+2101. The remaining five sources were observed during project 29\_007. The sources for project 21\_013 were selected from \citet{2010AandA...524A..83F}, while the sources for project 29\_007 were selected from CFC2016. In these two publications, the observations of 15 $z>4.5$ sources with the EVN at 1.6 and 5\,GHz, or 1.7\,GHz and 5\,GHz are described. In project 21\_013, sources were only considered for observation if they had steep radio spectra ($\alpha<-0.5$) based on their VLBI flux densities. To ensure that the sources were sufficiently bright to be detected with the GMRT, in project 29\_007, we selected sources based on their 1.4\,GHz flux densities in FIRST, and based on whether they were detected at 325 or 148\,MHz with the Westerbork Northern Sky Survey \citep[WENSS;][]{wenss} and the Tata Institute of Fundamental Research GMRT Sky Survey alternative data release 1 \citep[TGSS;][]{2016arXiv160304368I}, respectively.

During project 21\_013, the observations of J1146+4037, J1242+5422 and J1659+2101 were carried out using 32\,MHz of bandwidth in the 325\,MHz band and 16\,MHz of bandwidth in the 610, 235 and 150\,MHz bands. The central frequencies in each of these bands were 612, 322, 235 and 148\,MHz. In project 29\_007, J0210$-$0018, J0940+0526, J1400+3149, J1548+3335 and J1628+1154 were observed using 32\,MHz of bandwidth in the 610, 325 and 150\,MHz bands, which had central frequencies of 608, 323 and 148\,MHz. In both projects the observations of the target sources were flanked (where possible), or preceded or followed (where not possible), by 5--10\,minute observations of one or two of the following calibrator sources: 3C48, 3C147, 3C286, J1146+399, J1219+484, J1427+3312, J1506+375 and J1719+177. In total, 24.5\,hours of observations were taken for project 21\_013 and 13.5\,hours for project 29\_007. 

The data were reduced using the \textsc{SPAM} pipeline as described by \citet{2016arXiv160304368I}. The flux density scale was set by 3C48, 3C147 or 3C286 and was tied to the \citet{Scaife2012} standard with an accuracy of $\sim10$\,per\,cent \citep[e.g.][]{2004ApJ...612..974C}. The initial phase calibration of the target fields was done using a source model derived from the TGSS survey \citep{2016arXiv160304368I}. The source parameters in Table~\ref{tbl:GMRT values} were extracted from the images using the \textsc{pybdsm} source detection package \citep{2015ascl.soft02007M}. As the VLBI positions of all of the sources are known \citep[][and references therein]{2016MNRAS.tmp.1343C}, we set the source detection threshold, defined as the source's peak brightness divided by the local root mean square (rms) noise ($\sigma_{\rm{local}}$), to $3\sigma_{\rm{local}}$. All of the sources, except J0210$-$0018, were detected in all the observations as single components. J0210$-$0018 had two components in the GMRT610 image and one component in the GMRT325 and GMRT150 images. This is discussed in detail in Section \ref{subsubsec: J0210$-$0018}. Following \citet{2016arXiv160304368I}, the uncertainties on the flux densities in Table~\ref{tbl:GMRT values} were increased by adding 10\,per\,cent of the flux densities to the uncertainties in quadrature to account for systematic uncertainties.

\begin{table*}
 \hspace{-10cm}
 \centering
 \begin{minipage}{\columnwidth}
  \caption{GMRT image parameters}
  \begin{tabular}{ccccccccc}
  \hline
  ID & Observation &  Flux density & Local noise    & \multicolumn{2}{c}{Deconvolved source size}   & \multicolumn{2}{c}{Restoring beam}\\
  \cline{5-6}
  \cline{7-8}
               &   name  &  [mJy]                     & [mJy\,beam$^{-1}$] & [arcsec]$^{\mathrm a}$ & PA [$^{\circ}$]$^{\mathrm b}$ & [arcsec] & PA [$^{\circ}$]\\
  (1)          & (2)     & (3)                        & (4)                & (5)            & (6)             & (7)            & (8) \\   
  \hline  
  J0210$-$0018 & GMRT610S & $10.5\pm1.1$   & 0.04 & $(1.1\pm0.1)\times(0.0\pm0.1)$ & $25\pm1$ & $7.1\times4.0$ & 100 \\
               & GMRT610N & $4.4\pm0.5$    & 0.04 & $(1.7\pm0.2)\times(0.0\pm0.1)$ & $55\pm2$ & $7.1\times4.0$ & 100 \\
               & GMRT325  & $19.0\pm2.1$   & 0.32 & $(0.0\pm0.1)\times(0.0\pm0.1)$ & $0\pm3$ & $9.3\times6.7$ & 63\\
               & GMRT150  & $23.0\pm8.0$   & 4.50 & $(0.0\pm8.9)\times(0.0\pm2.5)$ & $0\pm16$ & $32.2\times16.7$ & 64\\
  \hline
  J0940+0526   & GMRT610  & $102.8\pm10.3$ & 0.09 & $(1.2\pm0.1)\times(0.9\pm0.1)$ & $115\pm1$ & $4.8\times4.0$ & 87\\
               & GMRT325  & $135.1\pm13.6$ & 0.62 & $(2.2\pm0.1)\times(2.1\pm0.1)$ & $115\pm2$ & $10.1\times8.9$ & 0\\
  \hline
  J1146+4037   & GMRT610  & $6.8\pm0.7$    & 0.08 & $(0.8\pm0.1)\times(0.0\pm0.1)$ & $55\pm2$ & $5.8\times4.1$ & 112\\
               & GMRT325  & $4.6\pm0.5$    & 0.05 & $(2.1\pm0.1)\times(0.9\pm0.1)$ & $18\pm2$ & $9.7\times7.4$ & 61\\
               & GMRT235  & $4.9\pm1.1$    & 0.61 & $(0.0\pm2.9)\times(0.0\pm1.3)$ & $0\pm19$ & $14.7\times10.8$ & 113\\
               & GMRT150  & $4.6\pm1.4$    & 0.73 & $(0.0\pm7.1)\times(0.0\pm1.9)$ & $0\pm13$ & $25.0\times16.8$ & 13\\
  \hline
  J1242+5422   & GMRT610  & $29.7\pm3.0$   & 0.10 & $(1.5\pm0.1)\times(0.8\pm0.1)$ & $129\pm1$ & $5.8\times4.1$ & 138 \\
               & GMRT325  & $30.0\pm3.0$   & 0.10 & $(1.2\pm0.1)\times(0.9\pm0.1)$ & $52\pm1$ & $10.8\times7.6$ & 45\\
               & GMRT235  & $27.6\pm2.9$   & 0.56 & $(2.1\pm0.1)\times(0.0\pm0.1)$ & $47\pm3$ & $14.8\times10.7$ & 146\\
               & GMRT150  & $26.1\pm2.9$   & 0.69 & $(0.0\pm0.8)\times(0.0\pm0.1)$ & $0\pm3$ & $27.2\times17.3$ & 1\\
  \hline
  J1400+3149   & GMRT610  & $24.6\pm2.5$   & 0.08 & $(0.9\pm0.1)\times(0.7\pm0.1)$ & $177\pm1$ & $4.6\times3.6$ & 51\\
               & GMRT150  & $56.2\pm6.5$   & 2.11 & $(22.3\pm2.6)\times(9.9\pm0.9)$ & $63\pm5$ & $24.9\times15.6$ &70 \\
  \hline
  J1548+3335   & GMRT610  & $77.6\pm7.8$   & 0.19 & $(1.9\pm0.1)\times(1.3\pm0.1)$ & $66\pm1$ & $9.4\times4.0$ & 83\\
  \hline
  J1628+1154   & GMRT610  & $107.7\pm10.8$ & 0.13 & $(1.9\pm0.1)\times(0.3\pm0.1)$ & $25\pm1$ & $6.0\times3.5$ & 82\\
               & GMRT325  & $152.4\pm15.3$ & 0.63 & $(1.8\pm0.1)\times(0.5\pm0.1)$ & $171\pm1$ & $11.6\times7.1$ & 83\\
  \hline
  J1659+2101   & GMRT610 & $48.1\pm4.8$    & 0.13 & $(1.2\pm0.1)\times(0.5\pm0.1)$ & $73\pm1$ & $4.6\times3.6$ & 24\\
               & GMRT325 & $53.0\pm5.3$    & 0.13 & $(3.0\pm0.1)\times(1.2\pm0.1)$ & $44\pm1$ & $10.2\times6.7$ & 65\\
               & GMRT235 & $54.7\pm5.7$    & 0.84 & $(0.0\pm0.1)\times(0.0\pm0.1)$ & $0\pm3$ & $12.0\times9.5$ & 22\\
               & GMRT150 & $48.2\pm5.4$    & 1.45 & $(8.4\pm0.9)\times(2.0\pm0.4)$ & $47\pm4$ & $21.6\times15.1$ & 17\\
  \hline
  \multicolumn{8}{p{18cm}}{\footnotesize{\textbf{Columns:} Col.~1 -- source name (J2000); Col.~2 -- observation name; Col.~3 -- integrated flux densities and uncertainties; Col.~4 -- rms noise at the source position; Col.~5 -- deconvolved source size (FWHM); Col.~6 -- deconvolved major axis position angle (measured from north through east); Col.~7 -- Gaussian restoring beam size (FWHM); Col.~8 -- Gaussian restoring beam major axis position angle (measured from north through east). }}\\ 
  \multicolumn{8}{p{18cm}}{\footnotesize{\textbf{Notes:} $^{\mathrm a}$ Uncertainties that would round down to zero are reported as 0.1\,arcsec. $^{\mathrm b}$ Uncertainties that would round down to zero are reported as 1\,$^{\circ}$. }}\\
  \end{tabular}
  \label{tbl:GMRT values}
 \end{minipage}
\end{table*}

%%%%%%%%%%%%%%%%%%%%%%%%%%%%%%%%%%%%%%%%%%%%%%%%%%%%%%%%%%%%%%%%%%%%%%%%%%%%%%%%%%%%%%%%%%%%%%%%%%%%%%%%%%%%%%%%%%%%%%%%%%%%%%%%%%%%%%%%%%%%%%%%%%%%%%%%%%%%%%%%%%%%%%%%%%%%%%%%%%%%%%%%%%%%%%%%%%%%%%%%%%%%%%%%%%%%%%%%%%
%%%%%%%%%%%%%%%%%%%%%%%%%%%%%%%%%%%%%%%%%%%%%%%%%%%%%%%%%%%%%%%%%%%%%%%%%%%%%%%%%%%%%%%%%%%%%%%%%%%%%%%%%%%%%%%%%%%%%%%%%%%%%%%%%%%%%%%%%%%%%%%%%%%%%%%%%%%%%%%%%%%%%%%%%%%%%%%%%%%%%%%%%%%%%%%%%%%%%%%%%%%%%%%%%%%%%%%%%%
\section{Flux densities from the literature}
\label{sec:litrature}
In this section we describe the procedure we followed to obtain previously recorded radio observations (10\,MHz$<\nu<$250\,GHz) for all 30 $z>4.5$ VLBI sources from the literature. These literature values are included with our observations (Section \ref{sec:GMRT}) to produce the final spectra in Section \ref{sec:spectra}. 

For each source, we obtained the detected radio flux densities from the NASA/IPAC Extragalactic Database (NED)\footnote{http://ned.ipac.caltech.edu/}. Additionally, we recorded all unique matches to the source in the catalogues in the VizieR database \citep{2000A&AS..143...23O} and in articles in the SAO/NASA Astrophysics Data System (ADS)\footnote{http://adsabs.harvard.edu/}. In each case, a matching radius of 20\,arcsec from the VLBI position was used.

A number of our targets were observed, but not detected, in the following large surveys: The VLA Low-Frequency Sky Survey Redux \citep[VLSSr, 74\,MHz;][]{2014MNRAS.440..327L}, TGSS, WENSS, the Green Bank 4.85\,GHz survey \citep[GB6, 4850\,MHz;][]{1996ApJS..103..427G}, the 62\,MHz Low-Frequency Array (LOFAR) image of the Bo\"{o}tes field made by \citet{vanweeren2014} and the 3\,GHz Caltech--NRAO Stripe 82 Survey \citep[CNSS;][]{2016ApJ...818..105M}. To determine consistent upper limits for these non-detections, we downloaded the survey images and measured $\sigma_{\rm{local}}$ within the 10$\times$10\,arcmin area surrounding the VLBI position. The flux density upper limit was then recorded as $3\sigma_{\rm{local}}$. As there were no images available for the GB6 survey, we used the detection threshold of 18\,mJy \citep{1996ApJS..103..427G} as an upper limit.

As we have known VLBI coordinates for our targets, we used a lower detection threshold ($3\sigma_{\rm{local}}$) than the VLSSr, WENSS ($5\sigma_{\rm{local}}$) and TGSS surveys ($7\sigma_{\rm{local}}$). To include the $3\sigma_{\rm{local}}$ detections from these surveys, we ran source extraction on the survey images using \textsc{pybdsm} as described in Section \ref{sec:GMRT}. The flux densities of sources that were detected at a significance (defined as the sources peak brightness divided by $\sigma_{\rm{local}}$) greater than $3\sigma_{\rm{local}}$, and for which the source position differed by less than half the FWHM of the restoring beam of the image were recorded as detections. These detections are listed in Table~\ref{tbl:lit extraced fluxes}. For these sources, the uncertainties on the 148\,MHz TGSS and 74\,MHz VLSSr flux densities were increased by 10 and 12\,per\,cent, respectively, to account for systematic uncertainties, as was done in \citet{2016arXiv160304368I} and \citet{2014MNRAS.440..327L}.

\begin{table}
 \centering
 \hspace{-1.0cm}
 \begin{minipage}{\columnwidth}
  \caption{Flux densities of sources that are not in the survey catalogues but that were detected}
  \begin{tabular}{ccccc}
  \hline
  ID & Observation & $\nu$  & Flux density & Detection \\
     &  name       & [MHz]  & [mJy]        & significance \\
     &             &        &              & [$\sigma_{\rm{local}}$] $^{\mathrm a}$\\
  \hline
  J0131$-$0321 & TGSS  & 148 & $24.6\pm4.5$ & $\sim7.5$\\
  J0210$-$0018 & TGSS  & 148 & $30.3\pm6.0$ & $\sim6.6$\\
  J1026+2542   & VLSSr & 74  & $631\pm237$  & $\sim4.1$\\
  J1628+1154   & VLSSr & 74  & $611\pm239$  & $\sim4.3$\\
  \hline
  \multicolumn{5}{p{\columnwidth}}{\footnotesize{\textbf{Notes:} $^{\mathrm a}$ The detection significance was calculated by dividing the source peak brightness by the local rms noise.}}\\
  \end{tabular}
  \label{tbl:lit extraced fluxes}
 \end{minipage}
\end{table}

The observations and surveys have different angular resolutions, so we checked for possible blended sources. Using the 1.4\,GHz FIRST survey, we recorded the separation between each of our targets and their nearest neighbouring source. If the target was not in the 1.4\,GHz FIRST survey, we used TGSS (148\,MHz) or the 1.4\,GHz Sloan Digital Sky Survey (SDSS) STRIPE82 \citep[][]{2011AJ....142....3H} catalogue (which have resolutions of 25 and 1.8\,arcsec, respectively) instead. For each of the detections we then checked whether the nearest neighbour could be distinguished from the target. All blended sources were discarded. These cases are discussed individually for each source in Section \ref{sec:spectra}.

As a final step, we plotted each of the spectra (Section \ref{sec:spectra}) and discarded the upper limits that were too high to valuably constrain the spectra. All upper limits that were used are given in Table~\ref{tbl:flux values for each source}.

%%%%%%%%%%%%%%%%%%%%%%%%%%%%%%%%%%%%%%%%%%%%%%%%%%%%%%%%%%%%%%%%%%%%%%%%%%%%%%%%%%%%%%%%%%%%%%%%%%%%%%%%%%%%%%%%%%%%%%%%%%%%%%%%%%%%%%%%%%%%%%%%%%%%%%%%%%%%%%%%%%%%%%%%%%%%%%%%%%%%%%%%%%%%%%%%%%%%%%%%%%%%%%%%%%%%%%%%%%
%%%%%%%%%%%%%%%%%%%%%%%%%%%%%%%%%%%%%%%%%%%%%%%%%%%%%%%%%%%%%%%%%%%%%%%%%%%%%%%%%%%%%%%%%%%%%%%%%%%%%%%%%%%%%%%%%%%%%%%%%%%%%%%%%%%%%%%%%%%%%%%%%%%%%%%%%%%%%%%%%%%%%%%%%%%%%%%%%%%%%%%%%%%%%%%%%%%%%%%%%%%%%%%%%%%%%%%%%%
\section{Radio spectra}
\label{sec:spectra}
In this section we will discuss each of the sources individually, and classify their spectra into one of the following classes: flat-spectrum sources, steep-spectrum sources, peaked-spectrum sources, and sources with unusual spectra (or spectra that could be classified into more than one class). A summary of the classification of each source is given in Table \ref{tbl:source clasification}.

\begin{table}
 \centering
 \hspace{-0.8cm}
 \begin{minipage}{\columnwidth}
  \caption{Summary of the spectral classification of each source}
  \begin{tabular}{cc}
  \hline
  ID & Classification$^{\mathrm a}$ \\
  \hline
  J0011+1446   & Flat\\
  J0131$-$0321 & Flat\\
  J0210$-$0018 & Flat (steep)\\
  J0311+0507   & Steep (USS)\\
  J0324$-$2918 & Peaked\\
  J0813+3508   & Steep\\
  J0836+0054   & Steep (USS)\\
  J0906+6930   & Peaked\\
  J0913+5919   & Peaked\\
  J0940+0526   & Steep\\
  J1013+2811   & Flat or peaked\\
  J1026+2542   & Flat\\
  J1146+4037   & Peaked (inverted)\\
  J1205$-$0742 & Concave\\
  J1235$-$0003 & Peaked\\
  J1242+5422   & Peaked\\
  J1311+2227   & Inverted or flat or peaked\\
  J1400+3149   & Flat\\
  J1427+3312   & Steep (flat)\\
  J1429+5447   & Steep\\
  J1430+4204   & Flat\\
  J1454+1109   & Unknown\\
  J1548+3335   & Steep\\
  J1606+3124   & Peaked\\
  J1611+0844   & Inverted or flat or peaked\\
  J1628+1154   & Steep\\
  J1659+2101   & Peaked\\
  J1720+3104   & Flat or peaked \\
  J2102+6015   & Peaked\\
  J2228+0110   & Peaked\\
  \hline
  \multicolumn{2}{p{9cm}}{\footnotesize{\textbf{Notes:} $^{\mathrm a}$Wording such as `Flat (steep)' indicates that the source has a flat spectral index, but that it could be steep within the uncertainties. Wording such as `Flat or peaked' is used when there is insufficient information to classify the spectrum of the source, but (often using upper limits) it is possible to exclude certain spectral types.}}\\
  \end{tabular}
  \label{tbl:source clasification}
 \end{minipage}
\end{table}

Each flux density point in the spectra is labelled with the name of the survey, or else according to the following convention: the first characters are the initial letters of the surnames for the lead authors of the article in which the flux density was published. These characters are followed by the year of publication. If the flux density is from a VLBI observation, the year is followed by `(V)'. In the spectra (Figures~\ref{fig:J0011+1446}, \ref{fig:J0131$-$0321} and \ref{fig:J0210$-$0018}--\ref{fig:J2228+0110} ) VLBI flux densities are also shown as filled grey symbols to distinguish them from non-VLBI flux densities. Upper limits are indicated by an unfilled downward arrow originating at the symbol. We note that for some publications and catalogues, no flux density errors are available. This is the case for the PBW1992, B2.2 and B3 catalogues, however, following \citet{2005A&A...431.1177V}, we assumed errors of 10\,per\,cent for PBW1992 and 20\,per\,cent for B2.2 and B3. A table containing all of the flux density labels, the observing frequency at which the measurement was taken, and the literature reference is given in Appendix~\ref{appendix: Flux density references}. A table containing the flux density values in the spectra of each source is given as online-only material. A sample of the table is shown in Table \ref{tbl:flux values for each source}.

\begin{table*}
 \centering
 \begin{minipage}{12cm}
 \centering
  \caption{Example entries in the online-only table containing the flux density values for each source}
  \begin{tabular}{cccccc}
  \hline
   Source name & Observation name & $\nu$ [MHz] & Upper limit $^{\mathrm a}$ & Flux density & Flux density \\
               &                  &             &                            & [mJy]        & error [mJy]  \\
  \hline
   J0011+1446 & FIRST & 1400 & N & 24.3 & 1.2\\
   J0011+1446 & CFC2016(V) & 1658 & N & 18.6 & 1.0\\
   J0011+1446 & CFC2016(V) & 4990 & N & 10.3 & 0.6\\
   J0011+1446 & CLASS & 8460 & N & 15.6 & 3.1\\
   J0131$-$0321 & TGSS & 148 & N & 24.6 & 4.5\\
   J0131$-$0321 & FIRST & 1400 & N & 33.7 & 1.7\\
   J0131$-$0321 & NVSS & 1400 & N & 31.4 & 1.0\\
   J0131$-$0321 & GCF2015(V) & 1658 & N & 64.4 & 0.3\\
  \hline
  \multicolumn{6}{p{12cm}}{\footnotesize{\textbf{Notes:} $^{\mathrm a}$ ``Y'' indicates that the value is an upper limit, ``N'' indicates that the value is not an upper limit. }}\\
  \end{tabular}
  \label{tbl:flux values for each source}
 \end{minipage}
\end{table*}

Throughout this section, when fitting the spectra we used a linear least-squares fitting routine. Because of their much higher angular resolution, VLBI measurements are insensitive to the large-scale radio emission. VLBI flux densities are therefore usually underestimates of the total flux densities, unless the source is very compact. Consequently, unless specifically noted, the spectral fits do not include VLBI flux densities, flux densities without uncertainties and flux density upper limits. Note that the values in the spectra are integrated flux densities unless only the peak brightness was available. We finally point out that in most cases the flux density measurements used here are taken at different epochs. In the case of source variability, this may affect the estimated spectral index.

All of the sources have single components in their non-VLBI images unless noted otherwise in the discussion of the source. The VLBI morphological classifications of all of the sources are given in CFC2016.

\subsection{Flat-spectrum sources}
\label{subsec:flat spectrum}
The following six sources all have flat spectra (they can be fitted by a single power law with $-0.5<\alpha<0.5$).
%%%%%%%%%%%%%%%%%%%%%%%%%%%%%%
\subsubsection{J0011+1446}
\label{subsubsec: J0011+1446}
\begin{figure}
  \includegraphics[width=\columnwidth]{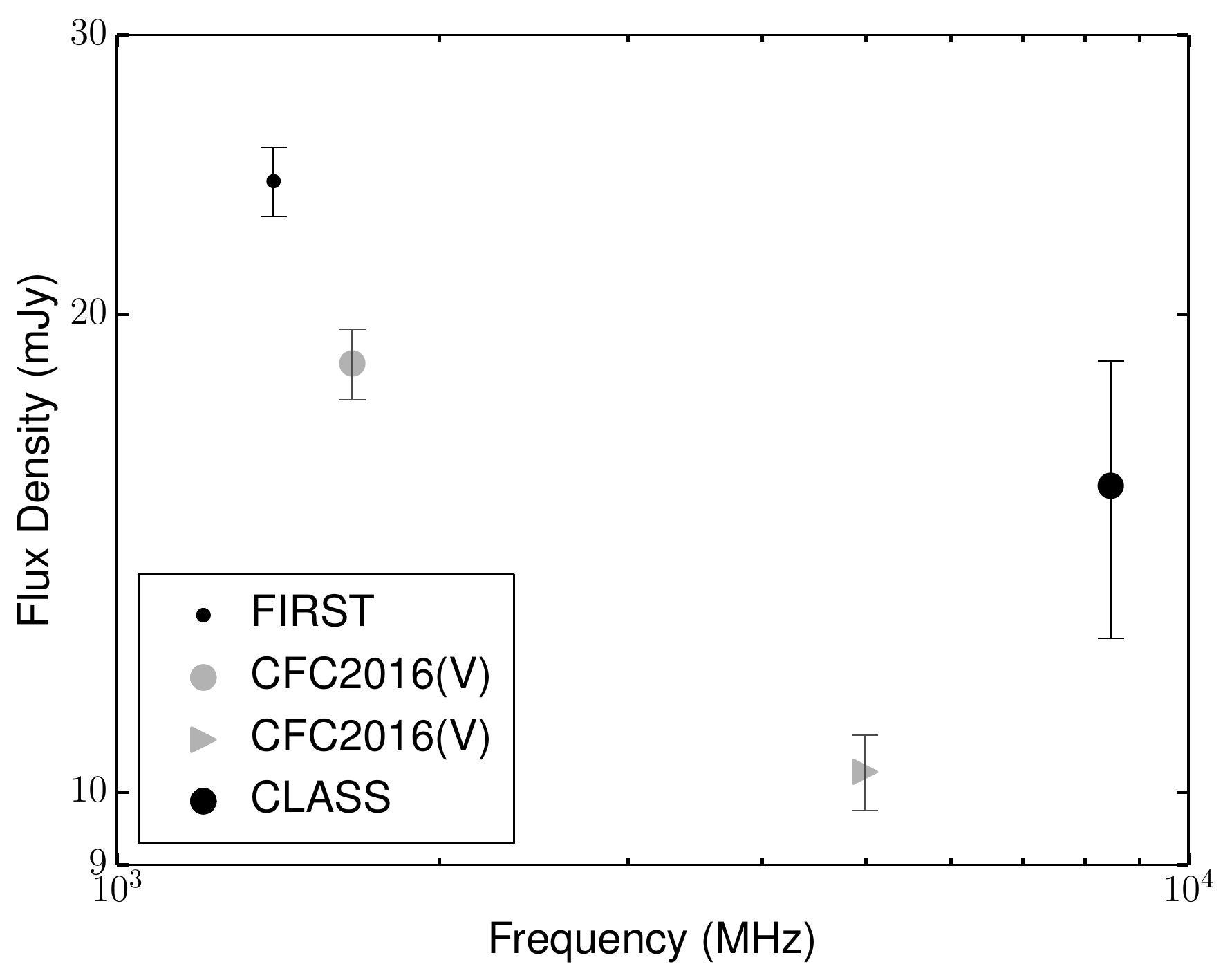}
  \caption{The radio spectrum of J0011+1446.}
  \label{fig:J0011+1446}
\end{figure}

We matched J0011+1446 to sources in the 148\,MHz TGSS, National Radio Astronomy Observatory (NRAO) VLA Sky Survey \cite[NVSS;][]{nvss} and 4.9\,GHz GB6 catalogues. However, in the 1.4\,GHz FIRST catalogue there are two sources that are 16.4 and 29.3\,arcsec away from the J0011+1446 VLBI position. Since the flux density of these sources will blend with that of J0011+1446 in the lower resolution TGSS, 1.4\,GHz NVSS and GB6 catalogues, we discarded these matches. The spectrum is shown in Fig.~\ref{fig:J0011+1446}. Fitting a power law between the FIRST and 8.5\,GHz CLASS flux densities gives a spectral index of $\alpha=-0.25\pm0.11$. J0011+1446 is therefore a flat-spectrum source, although, because non-VLBI flux densities are only available at two frequencies, it is possible that it could also have a peaked or concave spectrum. From the spectrum it is clear that some of the source's flux density was resolved out in the VLBI observations, or the source is variable.

%%%%%%%%%%%%%%%%%%%%%%%%%%%%%%
\subsubsection{J0131$-$0321}
\label{subsubsec: J0131$-$0321}
\begin{figure}
  \includegraphics[width=\columnwidth]{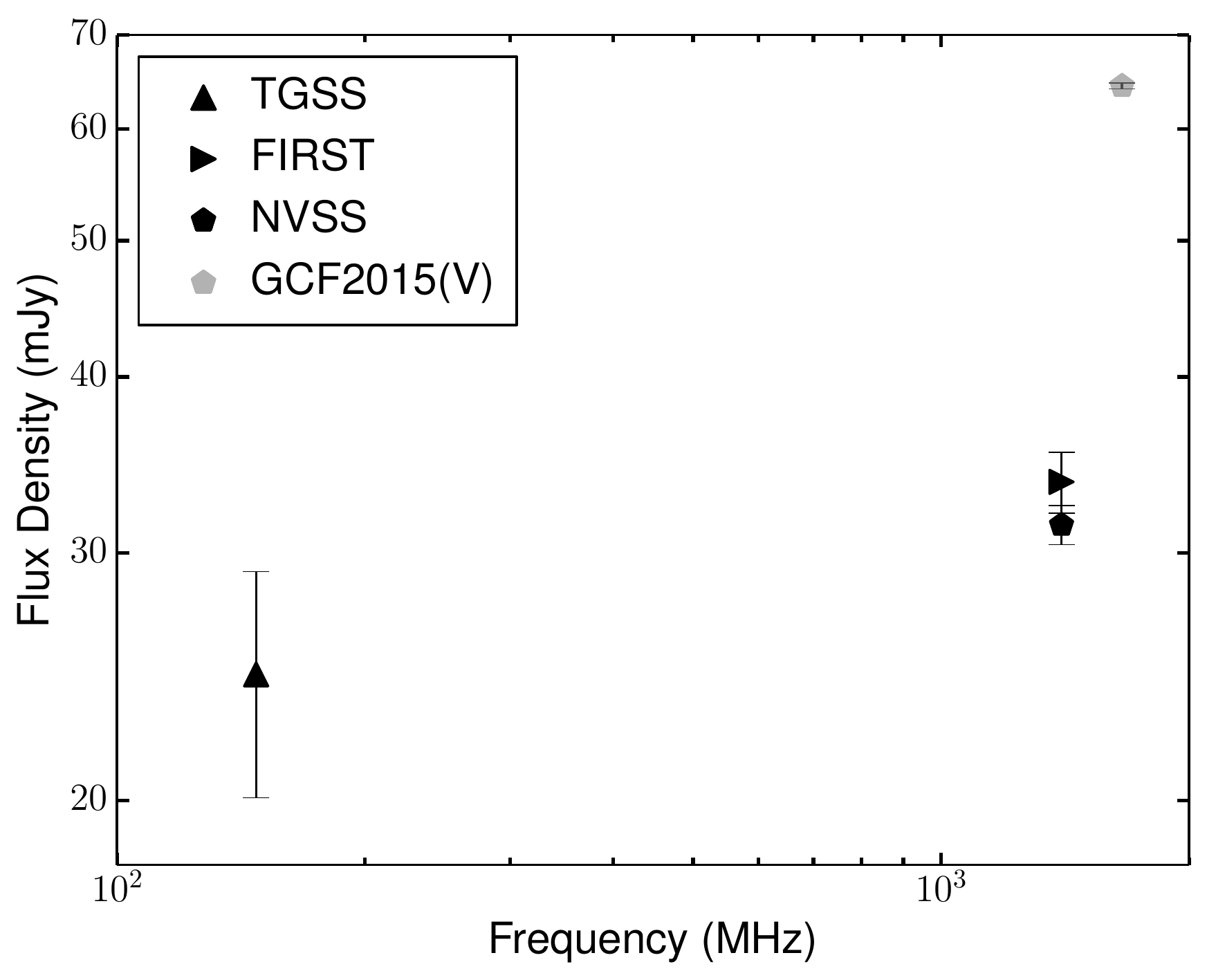}
  \caption{The radio spectrum of J0131$-$0321.}
  \label{fig:J0131$-$0321}
\end{figure}

A power-law fit for the spectrum of J0131$-$0321 (Fig.~\ref{fig:J0131$-$0321}) gives $\alpha=0.12\pm0.10$. J0131$-$0321 is therefore a flat-spectrum source, although, because non-VLBI flux densities are only available at two frequencies, it is possible that it could also have a peaked or concave spectrum. GCF2015(V) observed this source with the EVN at 1.7\,GHz and found it to be unresolved, with a flux density of $64.4\pm0.3$\,mJy. Comparing this to the 1.4\,GHz FIRST and NVSS flux densities of $33.7\pm1.7$ and $31.4\pm1.0$\,mJy, respectively, GCF2015(V) concluded that J0131$-$0321 is likely variable. However, since the epochs when FIRST and NVSS observed J0131$-$0321 differ by about 15.25\,years \citep{2011ApJ...737...45O,2015ApJ...801...26H}, if J0131$-$0321 is variable it means that the FIRST and NVSS observations were serendipitously done on two epochs when J0131$-$0321 happened to have the same flux density. The argument that J0131$-$0321 is variable is, however, supported by our finding that J0131$-$0321 has a flat spectrum, and GCF2015(V)'s conclusion that the VLBI emission is Doppler-boosted. 

%%%%%%%%%%%%%%%%%%%%%%%%%%%%%%
\subsubsection{J0210$-$0018}
\label{subsubsec: J0210$-$0018}
\begin{figure}
  \includegraphics[width=\columnwidth]{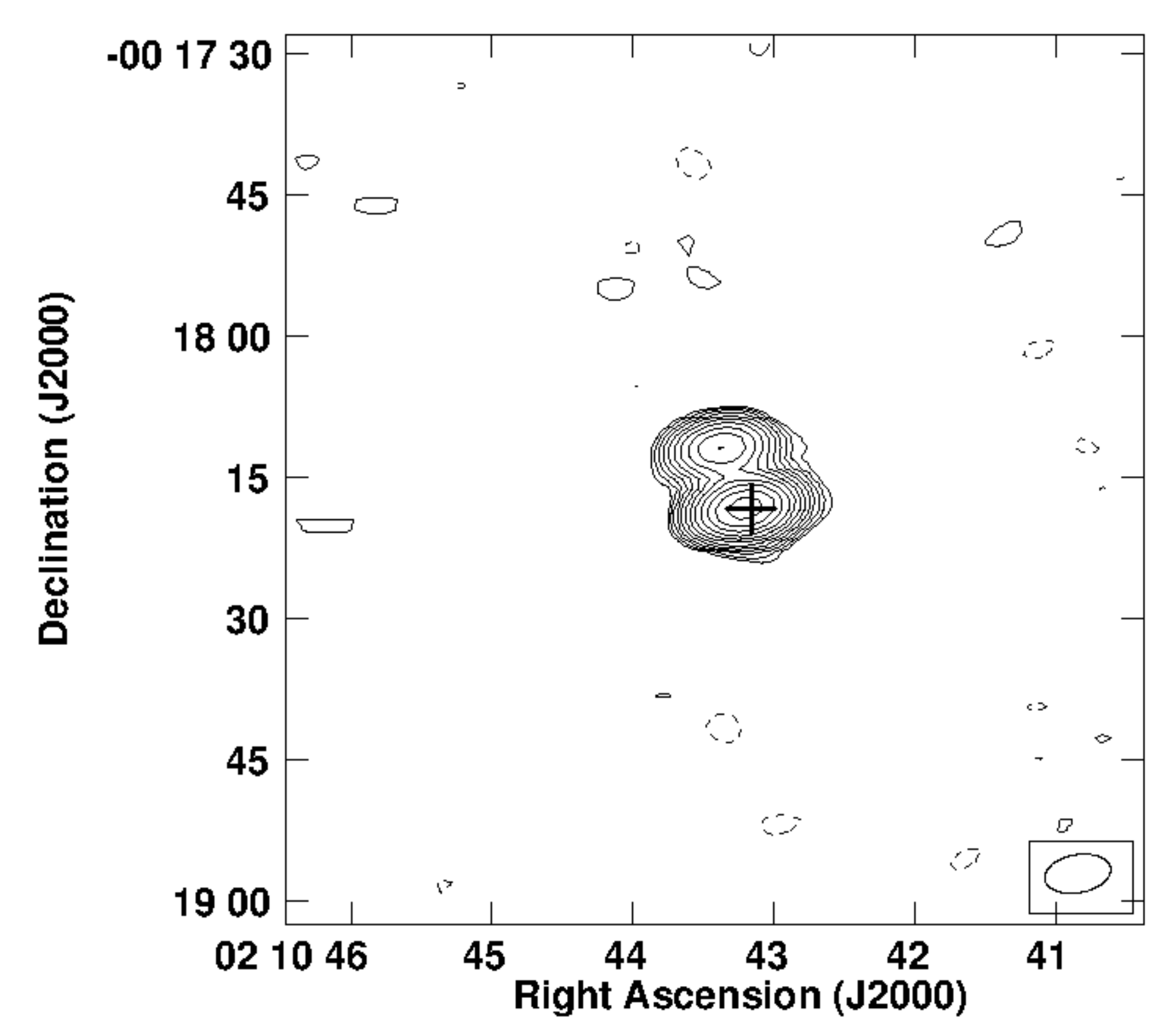}
  \caption{608\,MHz GMRT610 image of J0210$-$0018. The lowest contours are drawn at $-0.18$ and 0.18\,mJy\,beam$^{-1}$, the positive contours increase in factors of $\sqrt 2$ thereafter. The restoring beam (FWHM) is shown in the bottom right corner and the position of the optical AGN is indicated by a cross.}
  \label{fig:J0210$-$0018 gmrt610 image}
\end{figure}

\begin{figure}
  \includegraphics[width=\columnwidth]{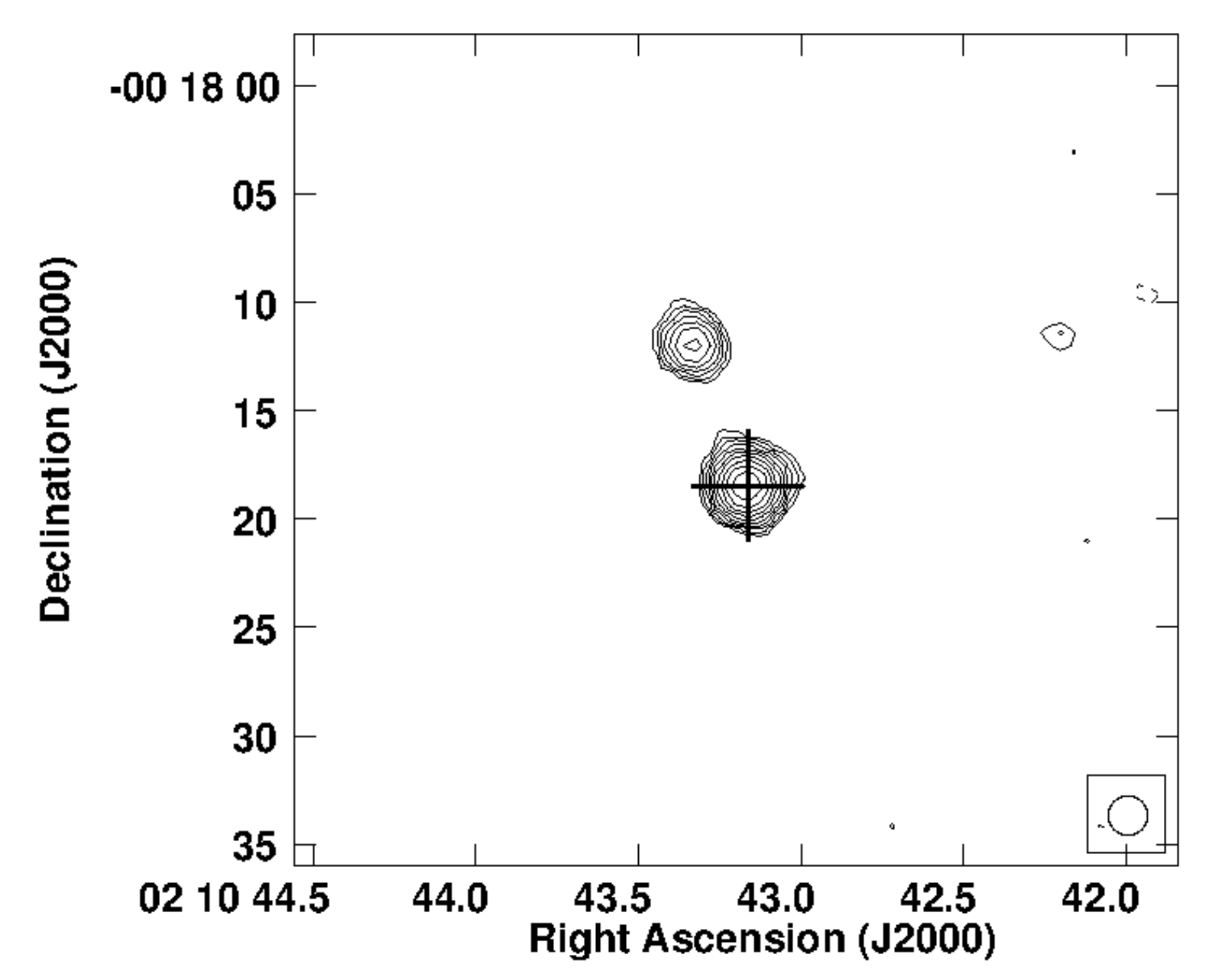}
  \caption{1.4\,GHz STRIPE82 image of J0210$-$0018. The lowest contours are drawn at $-0.21$ and 0.21\,mJy\,beam$^{-1}$. The positive contours increase in factors of $\sqrt 2$ thereafter. The restoring beam (FWHM) is shown in the bottom right corner and the position of the optical AGN is indicated by a cross.}
  \label{fig:J0210$-$0018 stripe82 image}
\end{figure}

Fig.~\ref{fig:J0210$-$0018 gmrt610 image} and \ref{fig:J0210$-$0018 stripe82 image} show the 608\,MHz GMRT610 and 1.4\,GHz VLA STRIPE82 images of J0210$-$0018. In both of these images the source has two components. Table~\ref{tbl:J0210$-$0018 componet flux} gives the flux densities of the individual components. Using the GMRT610 and STRIPE82 flux densities we calculate spectral indices of $-0.79\pm0.21$ and $-0.36\pm0.13$ for the northern and southern components, respectively.

\begin{table}
 \centering
 \begin{minipage}{\columnwidth}
  \caption{J0210$-$0018 component flux densities}
  \begin{tabular}{ccccc}
  \hline
  Image & Component & Flux density \\
        &          &  [mJy]       \\ 
  \hline
  GMRT610  & north & $4.36\pm0.45$  \\
           & south & $10.46\pm1.05$ \\
  STRIPE82 & north & $2.22\pm0.33$  \\
           & south & $7.72\pm0.34$  \\
  \hline
  \end{tabular}
  \label{tbl:J0210$-$0018 componet flux}
 \end{minipage}
\end{table}

In all the other observations (except for the 1.4\,GHz FIRST observations), J0210$-$0018 only has a single component due to a lack of resolution. Although FIRST has sufficient resolution to resolve J0210$-$0018, the source is fit by a single component with deconvolved major and minor axes of 4.3 and 1.3\,arcsec, respectively. The FIRST image does show an indication of a second component at the position of the northern component. It is not detected however, because the separation between the two components is small, and the northern component is significantly fainter than the southern component. At 1.4\,GHz the two components are therefore only detected in the STRIPE82 catalogue, which has both higher resolution and sensitivity than FIRST. Using the STRIPE82 positions of the two components, the angular separation between the components is 7.0\,arcsec, which translates to a linear separation of $\sim45.6$\,kpc.

The southern component coincides positionally with the optical AGN (Figures~\ref{fig:J0210$-$0018 gmrt610 image} and \ref{fig:J0210$-$0018 stripe82 image}). In principle there are four possibilities for what J0210$-$0018 could be: (1) the two components are unrelated sources at different redshifts; (2) the northern and southern components are gravitationally lensed images of the same source; (3) J0210$-$0018 is a one-sided source where one of the components is a hotspot or a lobe of the other; (4) the two components are separate, unrelated AGN at the same redshift.

The possibility that the two components of J0210$-$0018 are formed by gravitational lensing is unlikely given that the southern component positionally coincides with the optical AGN. In addition, if they are formed by gravitational lensing, the two components will have the same radio spectral index, which is not the case. We therefore conclude that the components are not gravitationally lensed images of the same source. One way to confirm that the two components are related is to search for a jet between them. Using our previous 1.7 and 5\,GHz EVN observations of J0210$-$0018 \citep{2016MNRAS.tmp.1343C}, in which the southern component was detected at both frequencies, we searched for a jet and did not find anything. We do, however, note that the 1.7\,GHz EVN flux density is only 22\,per\,cent of the 1.4\,GHz STRIPE82 flux density of the southern component. This indicates that the VLBI observations resolved out a significant fraction of the source's flux density. Consequently it is possible that this flux density is contained in a jet between the components that was resolved out. This possibility is further supported by the fact that the southern and northern components have flat and steep spectra, respectively. This likely indicates that the southern component is the AGN core (which will have a flat spectrum), and the northern component is a lobe or a hotspot (which typically have steep spectra) in the southern component's jet. This interpretation is also supported by there being no optical counterpart to the northern component in the co-add of SDSS Stripe 82 imaging data \citep{2009ApJS..182..543A}, which reach a typical depth of $m_r\approx 24.5$ \citep{2014ApJS..213...12J}.

\begin{figure}
  \includegraphics[width=\columnwidth]{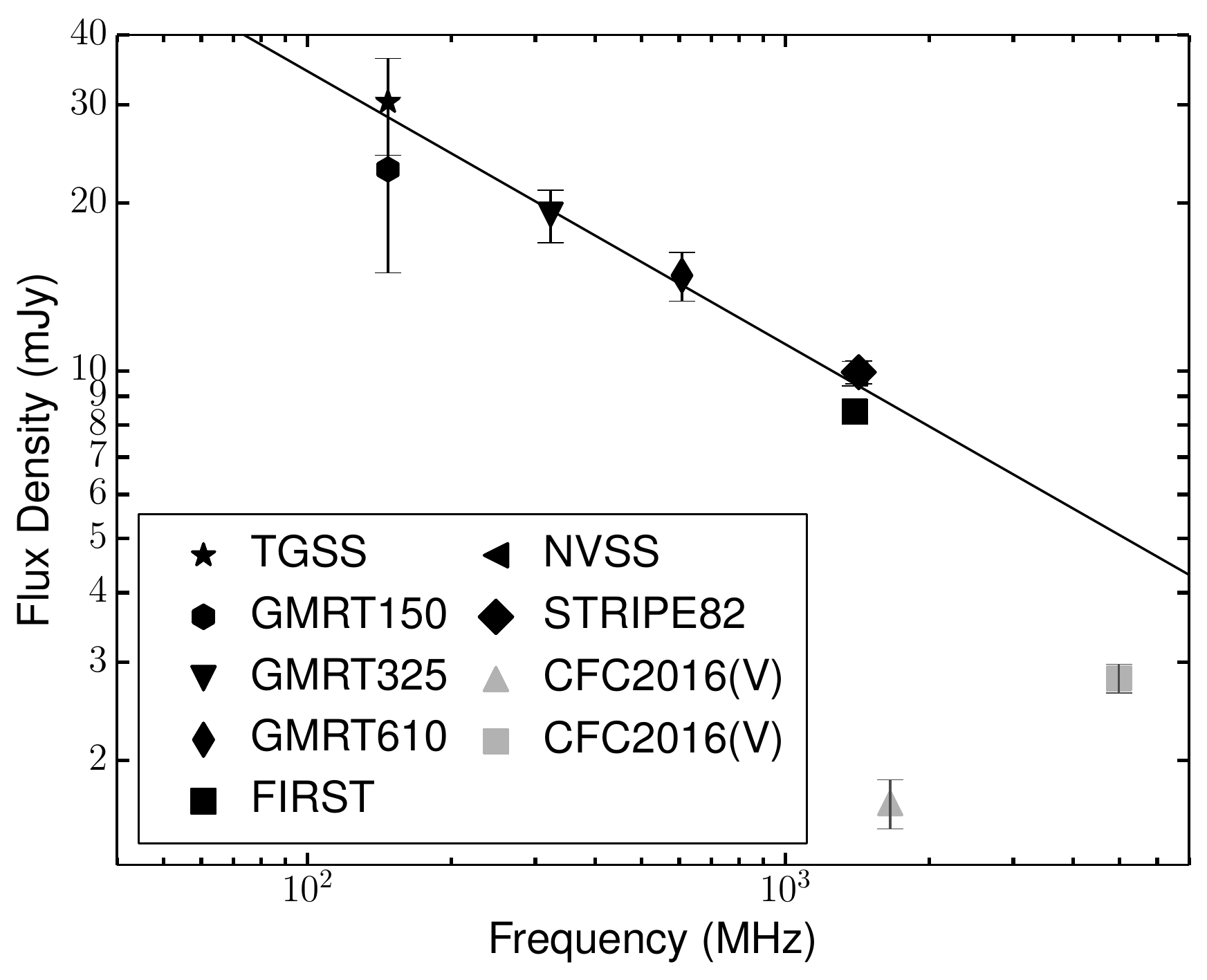}
  \caption{The radio spectrum of J0210$-$0018. The fit to the spectrum is shown as a solid line.}
  \label{fig:J0210$-$0018}
\end{figure}

In Fig.~\ref{fig:J0210$-$0018} we show the spectrum of J0210$-$0018. In the spectrum, the GMRT610 and STRIPE82 flux densities are the sums of the flux densities of the two components. Fig.~\ref{fig:J0210$-$0018} is therefore the sum of the spectra of both components. A power law fit to the spectrum gives $\alpha=-0.49\pm0.07$. We therefore classify J0210$-$0018 as having an overall flat spectrum. We do, however, note that J0210$-$0018 can be a steep-spectrum source (defined in Section \ref{subsec:negative spectrum}) within the uncertainties.

%%%%%%%%%%%%%%%%%%%%%%%%%%%%%%
\subsubsection{J1026+2542}
\label{subsubsec: J1026+2542}
\begin{figure*}
  \includegraphics[width=12cm]{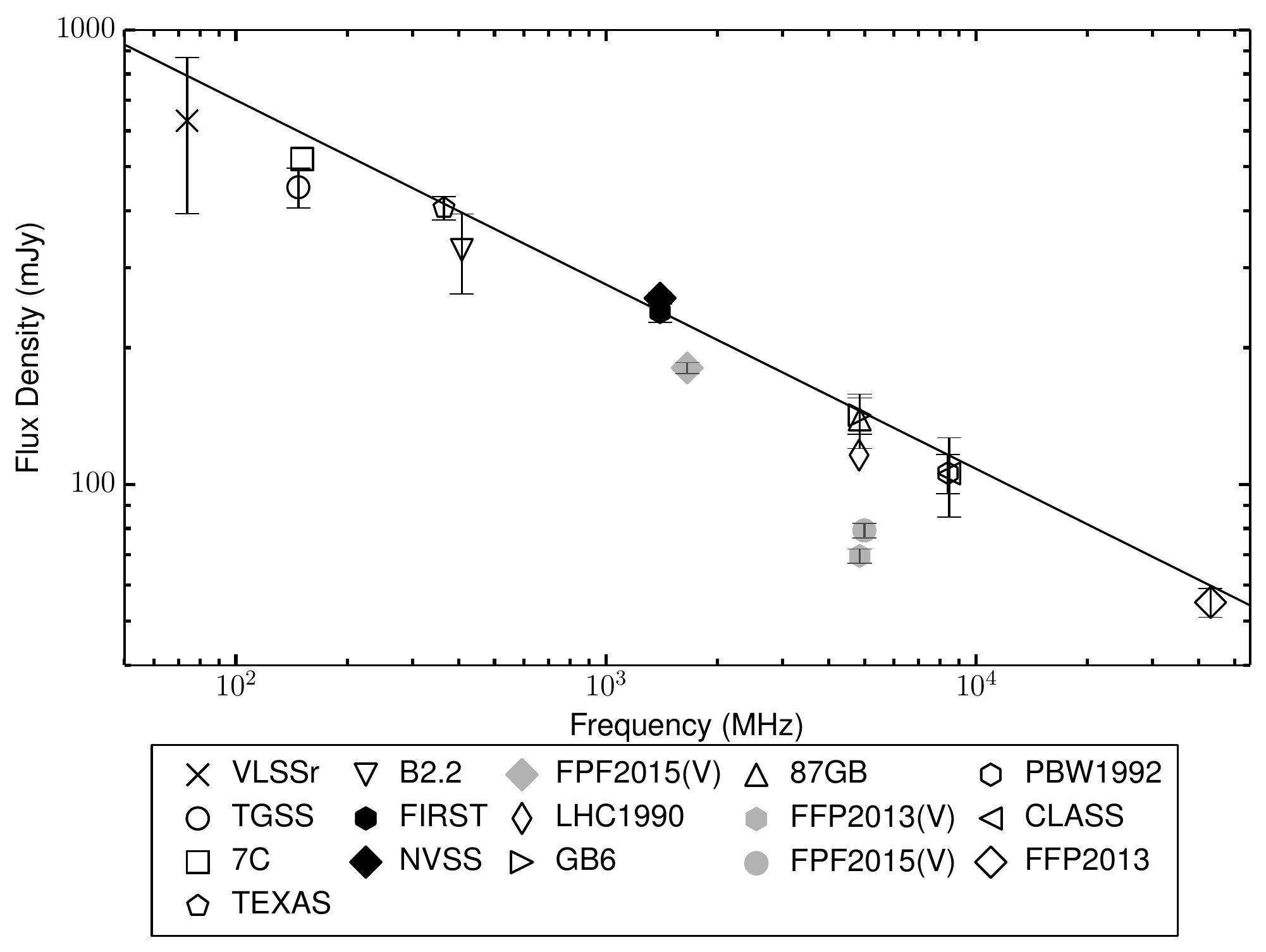}
  \caption{The radio spectrum of J1026+2542. The fit to the spectrum is shown as a solid line.}
  \label{fig:J1026+2542}
\end{figure*}

We fitted the spectrum of J1026+2542 (Fig.~\ref{fig:J1026+2542}) with a single power law with a spectral index of $\alpha=-0.41\pm0.02$. This is consistent with the value of $\alpha=-0.4$ found by FFP2013(V), and the fact that the source is Doppler-boosted \citep{2016MNRAS.tmp.1343C}. 

%%%%%%%%%%%%%%%%%%%%%%%%%%%%%%
\subsubsection{J1400+3149}
\label{subsubsec: J1400+3149}
\begin{figure}
  \includegraphics[width=\columnwidth]{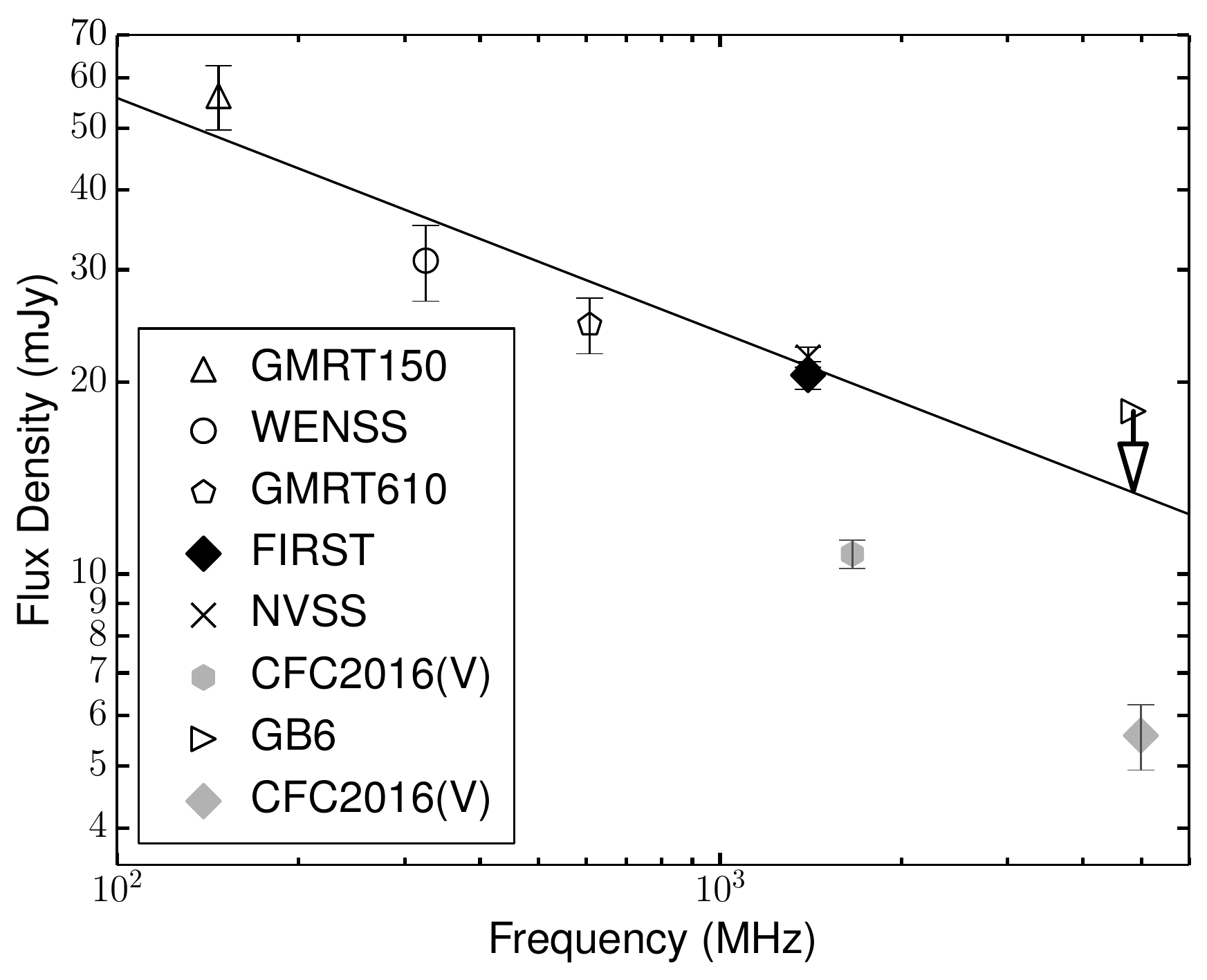}
  \caption{The radio spectrum of J1400+3149. The fit to the spectrum is shown as a solid line.}
  \label{fig:J1400+3149}
\end{figure}

We fitted the spectrum of J1400+3149 (Fig.~\ref{fig:J1400+3149}) with a power law with a spectral index of $-0.36\pm0.07$. 

%%%%%%%%%%%%%%%%%%%%%%%%%%%%%%
\subsubsection{J1430+4204}
\label{subsubsec: J1430+4204}
\begin{figure*}
  \includegraphics[width=13cm]{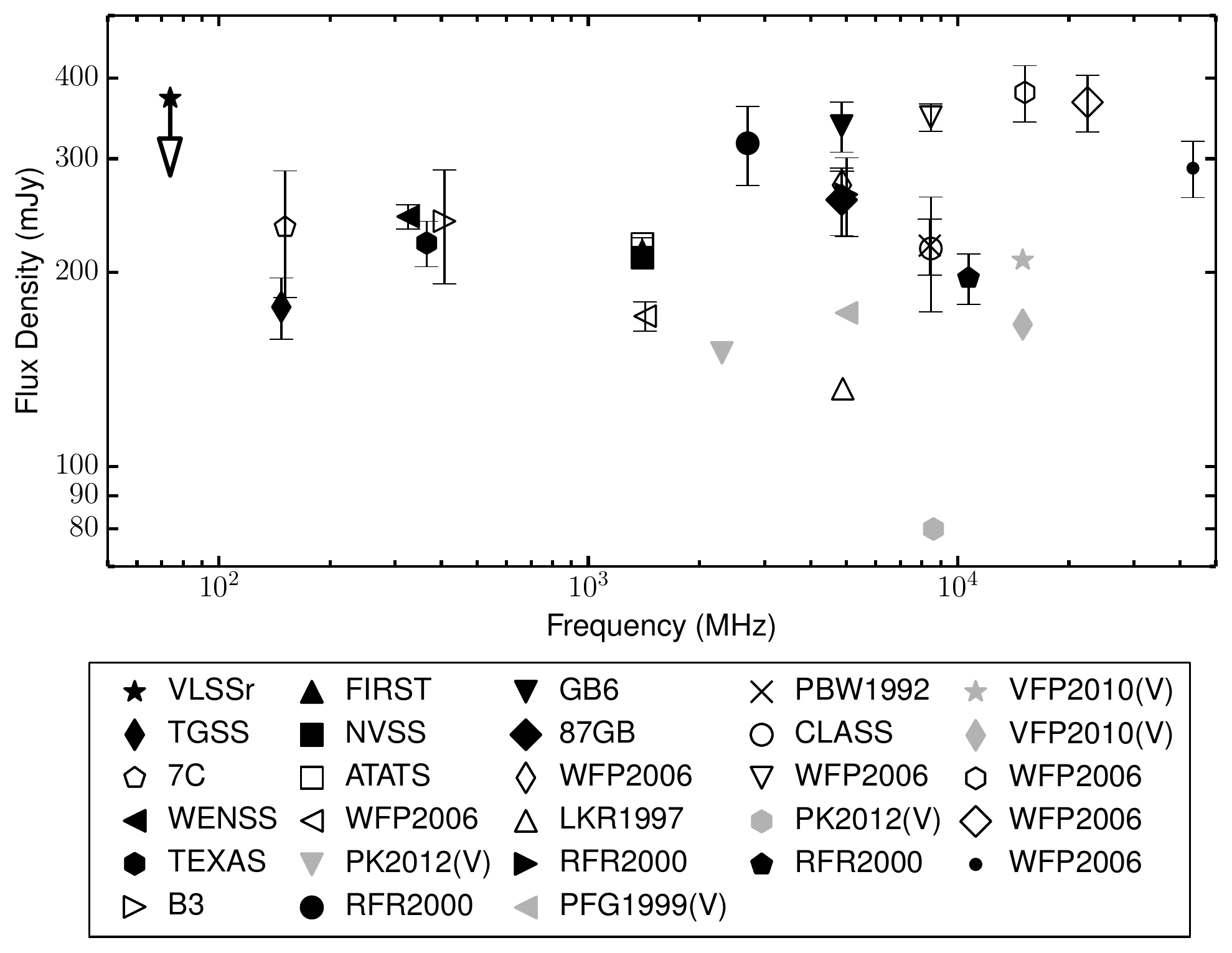}
  \caption{The radio spectrum of J1430+4204.}
  \label{fig:J1430+4204}
\end{figure*}

WFP2006 observed J1430+4204 at 15.2\,GHz over a period of $\sim7.5$\,years, during which time they found the flux density to vary between $\sim70$ and $\sim430$\,mJy. Based on these findings and the spectrum of J1430+4204 (Fig.~\ref{fig:J1430+4204}), we conclude that J1430+4204 is extremely variable. Fig.~\ref{fig:J1430+4204} gives the average 15.2\,GHz WFP2006 flux density. Fitting a power law to the spectrum, we find a spectral index of $0.10\pm0.03$. While this spectral index is likely not a good indication of the spectral index of the source at any given time, it can be considered as an average spectral index. Combining this with the finding that J1430+4204 is Doppler-boosted \citep{2016MNRAS.tmp.1343C}, we conclude that J1430+4204 is a flat-spectrum radio quasar.  

%%%%%%%%%%%%%%%%%%%%%%%%%%%%%%%%%%%%%%%%%%%%%%%%%%%%%%%%%%%%%%%%%%%%%%%%%%%%%
%%%%%%%%%%%%%%%%%%%%%%%%%%%%%%%%%%%%%%%%%%%%%%%%%%%%%%%%%%%%%%%%%%%%%%%%%%%%%
\subsection{Steep-spectrum and ultra-steep-spectrum sources}
\label{subsec:negative spectrum}
The eight sources discussed in this section are all fitted with a single power-law spectrum with $\alpha<-0.5$. Included in this class of sources are the USS sources, which we will define as objects with $\alpha<-1.0$ across their entire spectral range. 
%%%%%%%%%%%%%%%%%%%%%%%%%%%%%%
\subsubsection{J0311+0507}
\label{subsubsec: J0311+0507}
Matching the VLBI position for J0311+0507 to FIRST (1.4\,GHz), we find that there are 15 sources within two arcminutes of the source and that the nearest neighbour is 5.2\,arcsec away. In the survey catalogue these sources are indicated to have side lobe probabilities between 0.063 and 0.528 \citep{2015ApJ...801...26H}. Looking at the image of J0311+0507 in FIRST, the VLA beam pattern is clearly visible around the source, with the neighbouring sources all lying on the beam pattern\footnote{http://third.ucllnl.org/cgi-bin/firstcutout}. Comparing the 1.4\,GHz FIRST and NVSS images and based on the probabilities of the sources being side lobes, we conclude that the nearest real source to J0311+0507 is 330 arcsec away, and that the 15 neighbouring sources in FIRST are all artefacts. We matched J0311+0507 to the source 4C+04.11 in the 178\,MHz 4C survey \citep{1967MmRAS..71...49G}. However, because the 4C survey has a resolution of 11.5\,arcmin, the flux density of the nearby sources will blend with that of J0311+0507, we discarded the match. We, for the same reason, discarded the matches to \citet[][]{1996BSAO...40..128B} (at 0.96, 2.3, 3.94 and 7.69\,GHz), \citet[][]{2010ARep...54..675P} (at 0.5, 1.4 and 3.94\,GHz), \citet[][]{1996BSAO...40....5P} (at 1.425\,GHz), \citet[][]{1992AandAS...96..583P} (at 3.945\,GHz) and \citet[][]{1979Ap&SS..64...73B} (at 16.7\,MHz).

\begin{figure*}
  \includegraphics[width=13cm]{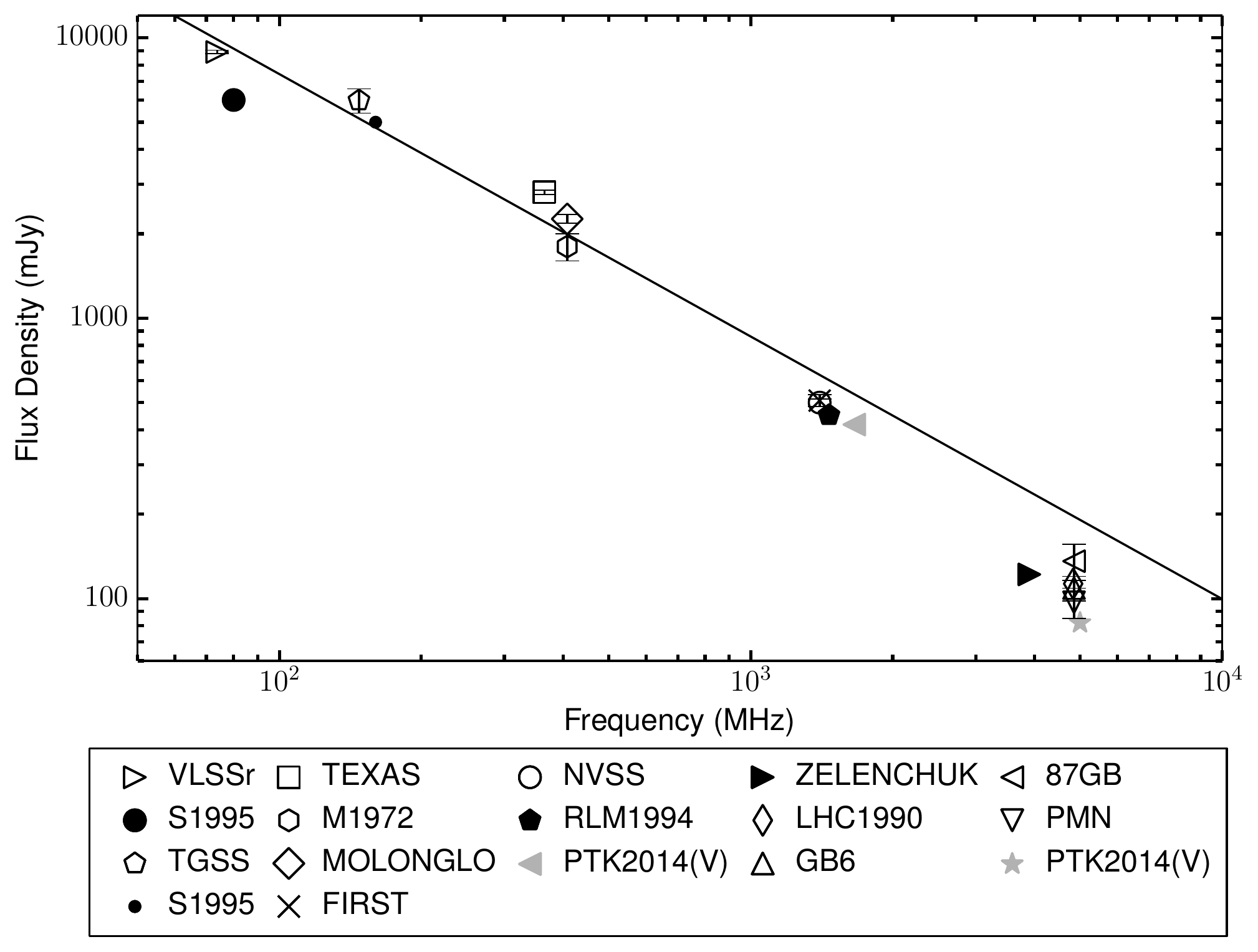}
  \caption{The radio spectrum of J0311+0507. The fit to the spectrum is shown as a solid line.}
  \label{fig:J0311+0507}
\end{figure*}

J0311+0507 was classified as a USS source by \citet{1994AandAS..108...79R} who found it to have a spectral index of $-1.17\pm0.03$ between 150\,MHz and 4.85\,GHz. We fitted the spectrum (shown in Fig.~\ref{fig:J0311+0507}) with a single power law with a spectral index of $\alpha=-0.94\pm0.06$, and therefore classify J0311+0507 as a steep-spectrum source that could also be a USS source. We do note that our spectral index is higher than the spectral index of $-1.31$ between 365 and 4850\,MHz found by \citet{1992AZh....69..673G} and \citet[][and references therein]{2014MNRAS.439.2314P}. As a final point, we note that the 1.7 and 5\,GHz PTK2014(V) VLBI observations of J0311+0507 showed that it has a FR\,II structure, and an angular and linear size of 2.8\,arcsec and 18.7\,kpc, respectively. 

%%%%%%%%%%%%%%%%%%%%%%%%%%%%%%
\subsubsection{J0813+3508}
\label{subsubsec: J0813+3508}
\begin{figure}
  \includegraphics[width=\columnwidth]{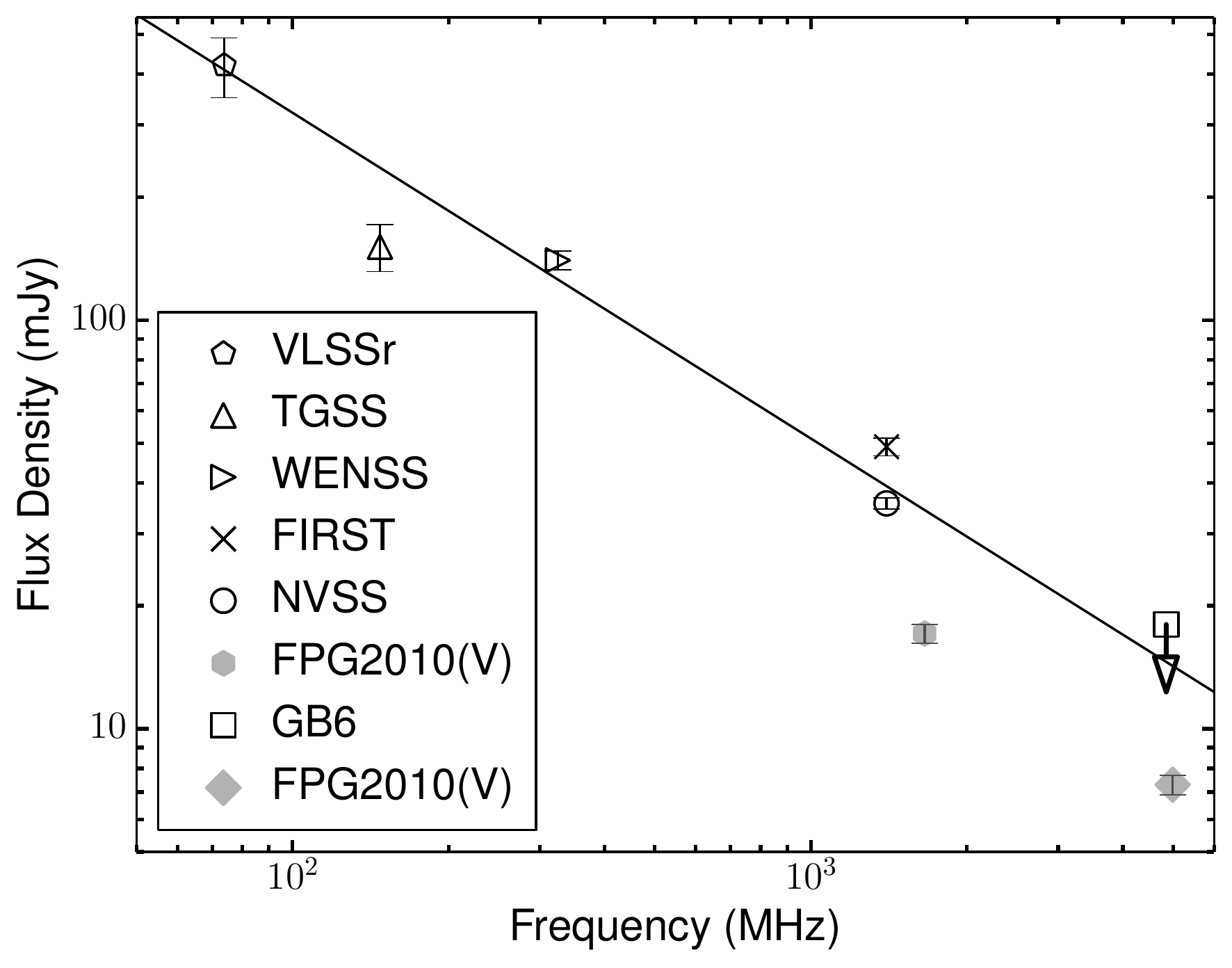}
  \caption{The radio spectrum of J0813+3508. The fit to the spectrum is shown as a solid line.}
  \label{fig:J0813+3508}
\end{figure}

In FIRST (1.4\,GHz) there is a second source due northwest of the source matched to J0813+3508 that is 6.9\,arcsec distant from the J0813+3508 VLBI position, which translates to a linear size of $\sim43.7$\,kpc. FPG2010(V) observed both sources with the EVN at 1.7 and 5\,GHz. While the second source was not detected, the authors did find a jet pointing from J0813+3508 towards the second source in the 1.7\,GHz image. From this, FPG2010(V) concluded that the second source is a lobe of J0813+3508 that is resolved out by the VLBI observations. The only non-VLBI observation that has high enough resolution to resolve the two components is FIRST, in which the main and second components have flux densities of $37.5\pm1.9$ and $11.5\pm0.6$\,mJy, respectively. In the source spectrum (shown in Fig.~\ref{fig:J0813+3508}), the FIRST flux density is therefore the sum of the flux densities of the two components. Fitting a power law to the spectrum we find $\alpha=-0.80\pm0.12$. We note that 148\,MHz TGSS has a resolution of $25\times25$\,arcsec and that J0813+3508 has a fitted source size of $(28.8\pm1.4)\times(18.8\pm0.6)$\,arcsec in the survey \citep{2016arXiv160304368I}. The TGSS flux density being lower than the predicted value can therefore be explained by J0813+3508 being partially resolved or by variability.

%%%%%%%%%%%%%%%%%%%%%%%%%%%%%%
\subsubsection{J0836+0054}
\label{subsubsec: J0836+0054}
\begin{figure}
  \includegraphics[width=\columnwidth]{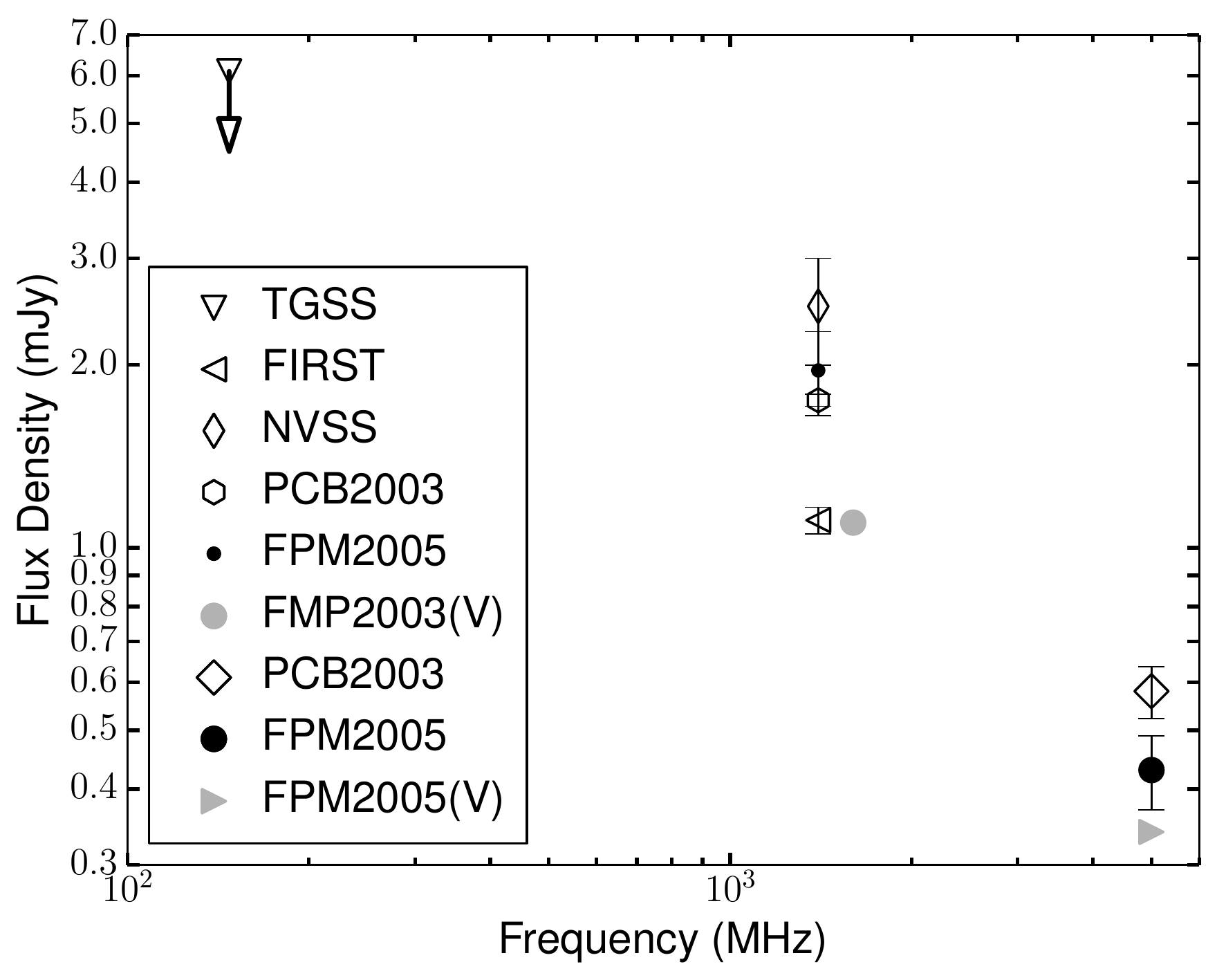}
  \caption{The radio spectrum of J0836+0054.}
  \label{fig:J0836+0054}
\end{figure}

Fitting the spectrum of J0836+0054 (Fig.~\ref{fig:J0836+0054}) with a power law gives a spectral index of $\alpha=-0.89\pm0.29$. This indicates that the source can be a USS source within the uncertainties. J0836+0054 has 1.4\,GHz FIRST and NVSS flux densities of $1.11\pm0.06$\,mJy and $2.5\pm0.5$\,mJy, respectively. In addition, PCB2003 found a 1.4\,GHz flux density of $1.75\pm0.04$\,mJy during their observations with the VLA at a resolution of 1.5\,arcsec. Since the PCB2003 observations have a higher resolution than FIRST, and a $\sim60$\,per\,cent higher flux density, this, along with the flux density difference between FIRST and NVSS, could indicate that J0836+0054 is variable. However, the NVSS source is positionally offset from the FIRST source by about 15\,arcsec to the northeast. Since NVSS has a resolution of 45\,arcsec compared to the 5\,arcsec of FIRST, the flux density and positional difference could also be because of resolution effects. This interpretation is supported by the PCB2003 flux density being consistent with the NVSS value and the PCB2003 observations having a $1\sigma$ noise level of 0.0216\,mJy\,beam$^{-1}$ compared to the 0.15\,mJy\,beam$^{-1}$ of FIRST. Additionally the 1.4\,GHz FPM2005 flux density is consistent with both the NVSS and PCB2003 values but not with the FIRST value. While the FPM2005 observations have a resolution of $6.3\times4.4$\,arcsec, which is similar to FIRST, they have a lower noise level of 0.083\,mJy\,beam$^{-1}$. We therefore conclude that J0836+0054 is likely not variable, but cannot rule out the possibility. 

We finally note that the fitted spectrum predicts a 148\,MHz flux density of $\sim12.0$\,mJy, while the 148\,MHz TGSS upper limit indicates that the flux density is below 6.1\,mJy. This could be due to the uncertainty introduced in the fitted spectral index by the resolution effects mentioned above, variability, or a potential spectral turnover.

%%%%%%%%%%%%%%%%%%%%%%%%%%%%%%
\subsubsection{J0940+0526}
\label{subsubsec: J0940+0526}
\begin{figure}
  \includegraphics[width=\columnwidth]{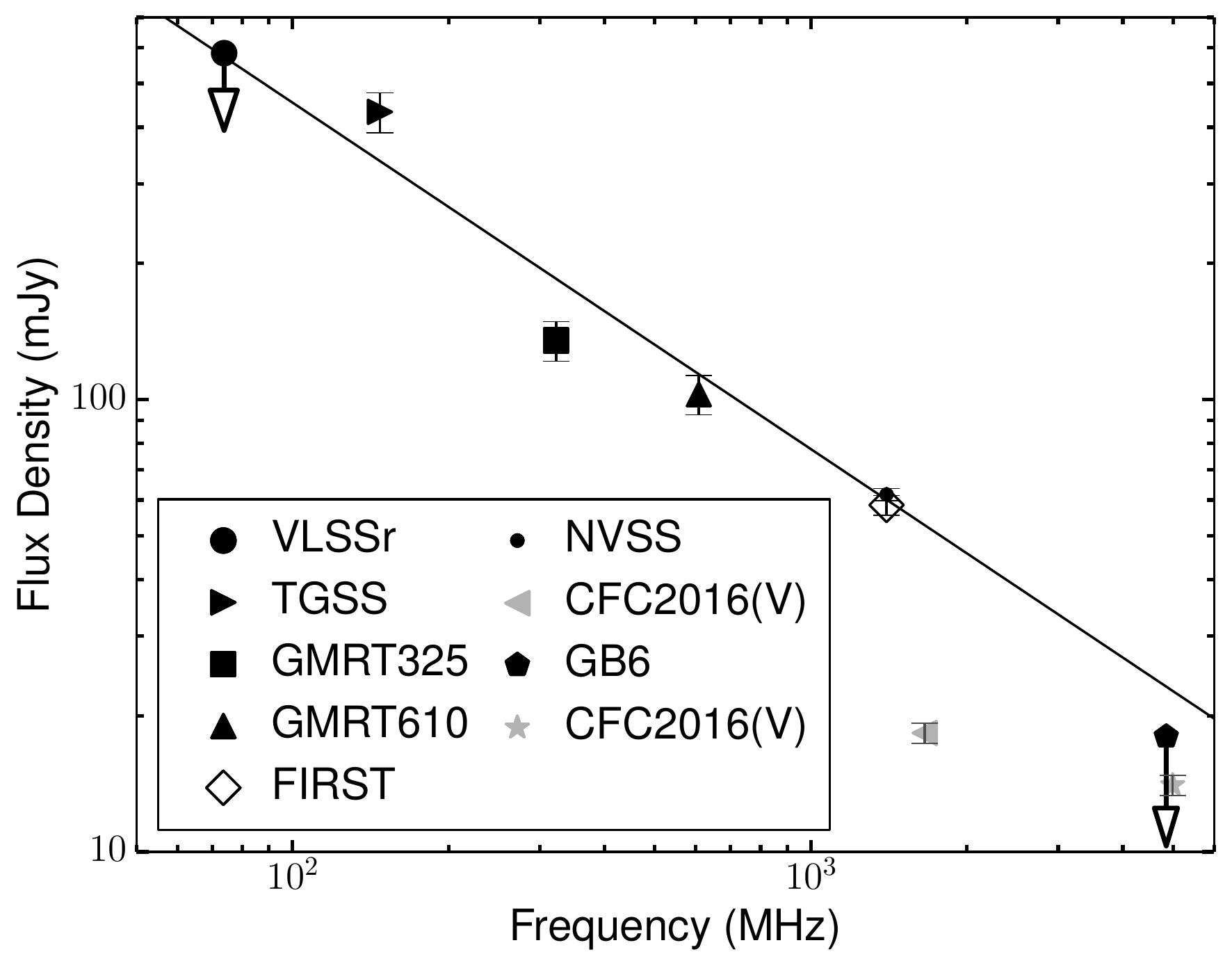}
  \caption{The radio spectrum of J0940+0526. The fit to the spectrum is shown as a solid line.}
  \label{fig:J0940+0526}
\end{figure}

We fitted the spectrum of J0940+0526 (Fig.~\ref{fig:J0940+0526}) with a single power law with a spectral index of $\alpha=-0.77\pm0.10$. 

%%%%%%%%%%%%%%%%%%%%%%%%%%%%%%
\subsubsection{J1427+3312}
\label{subsubsec: J1427+3312}
\begin{figure}
  \includegraphics[width=\columnwidth]{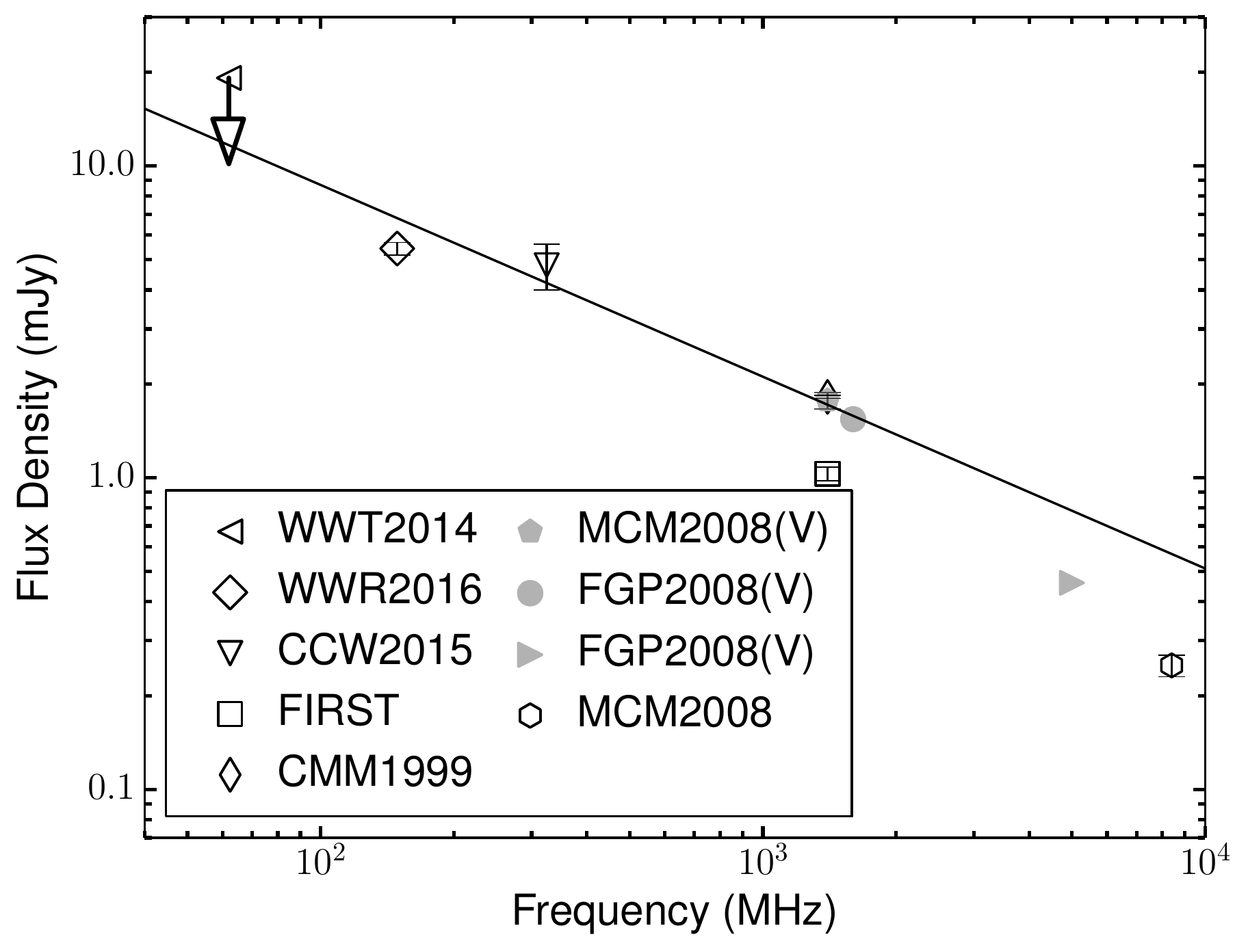}
  \caption{The radio spectrum of J1427+3312. The fit to the spectrum is shown as a solid line.}
  \label{fig:J1427+3312}
\end{figure}

We fitted the spectrum of J1427+3312 (Fig.~\ref{fig:J1427+3312}) with a single power law with $\alpha=-0.62\pm0.17$. Although we classify the source as having a steep spectrum, it is also possible that it has a flat spectrum within the errors. Take note that the reason why the fitted line does not fit the 8.4\,GHz MCM2008 point very well is because the smaller errors on the 149\,MHz WWR2016 and 1.4\,GHz CMM1999 flux densities give these points larger weights during the fitting. Finally, we also note that the 1.4\,GHz FIRST and CMM1999 flux densities differ ($1.03\pm0.05$ and $1.82\pm0.02$\,mJy, respectively), and the 1.6\,GHz FGP2008(V) and 1.4\,GHz MCM2008(V) flux densities are higher than the FIRST flux density. The difference between the FIRST and CMM1999 flux densities could be caused by the CMM1999 observations having a resolution of $\sim15$\,arcsec, which is about three times lower than that of FIRST. The difference, specifically between FIRST and the VLBI flux densities, could also indicate that J1427+3312 is variable. 

%%%%%%%%%%%%%%%%%%%%%%%%%%%%%%
\subsubsection{J1429+5447}
\label{subsubsec: J1429+5447}
\begin{figure}
  \includegraphics[width=\columnwidth]{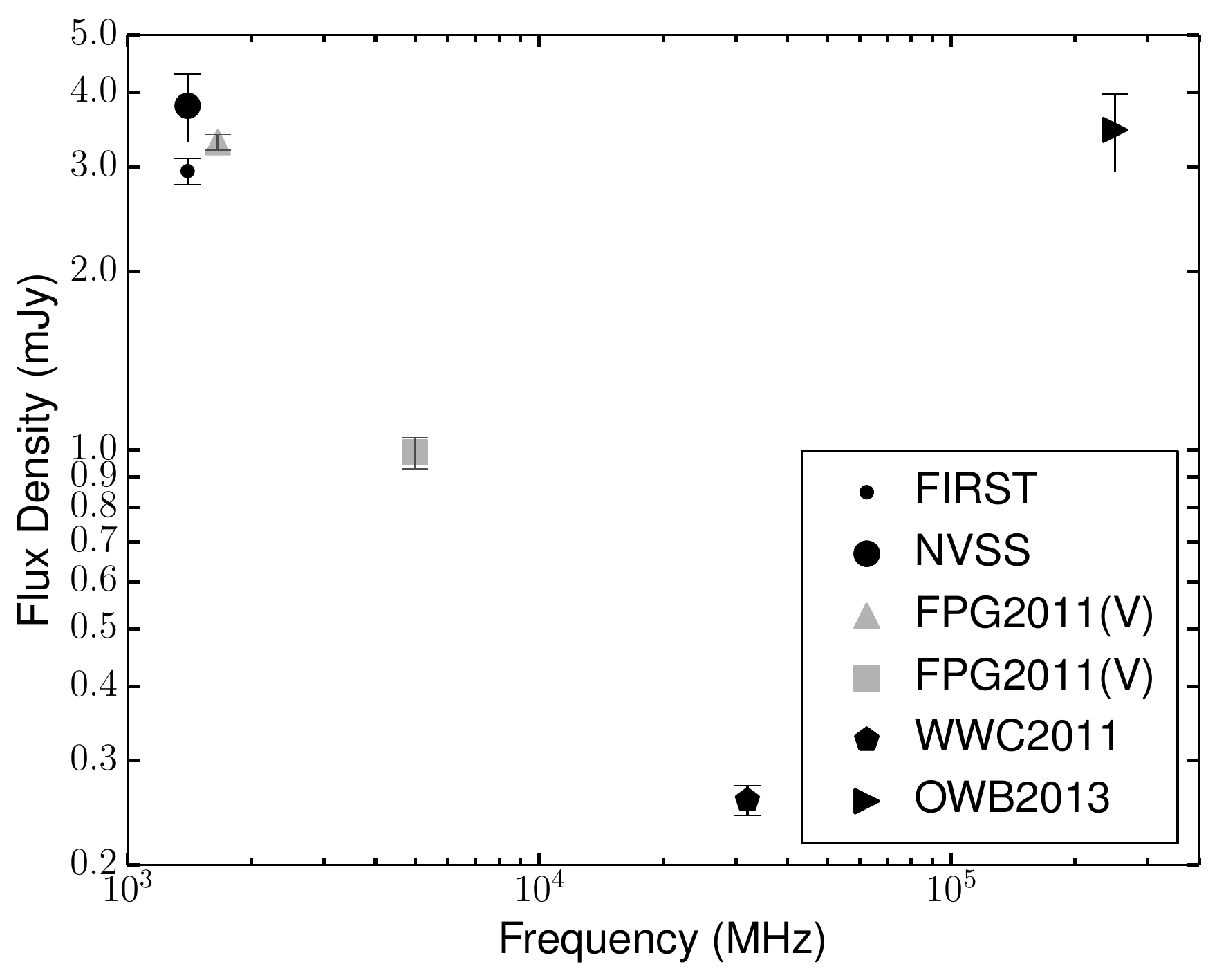}
  \caption{The radio spectrum of J1429+5447.}
  \label{fig:J1429+5447}
\end{figure}

OWB2013 and CFC2016 concluded that in the spectrum of J1429+5447 (Fig.~\ref{fig:J1429+5447}) the emission below 100\,GHz is from AGN activity. WWC2011 found that the CO line emission of the source is resolved into two components that are separated by 1.2\,arcsec ($\sim6.9$\,kpc), with the optical and continuum source positions being consistent with the western peak. The authors also note that the eastern component is possibly extended with a size of $(1.1\pm0.2)\times(0.7\pm0.2)$\,arcsec, which could explain why it is not detected in the continuum observations. OWB2013 also observed J1429+5447 at 250\,GHz and concluded that the majority of the 250\,GHz emission is thermal emission from hot dust. The authors do however note that it is possible that a significant fraction of the 250\,GHz emission could be from AGN driven synchrotron emission. Excluding the 250\,GHz OWB2013 value and fitting the spectrum with a power law gives $\alpha=-0.79\pm0.04$. We therefore classify J1429+5447 as a steep-spectrum source.

%%%%%%%%%%%%%%%%%%%%%%%%%%%%%%
\subsubsection{J1548+3335}
\label{subsubsec: J1548+3335}
\begin{figure}
  \includegraphics[width=\columnwidth]{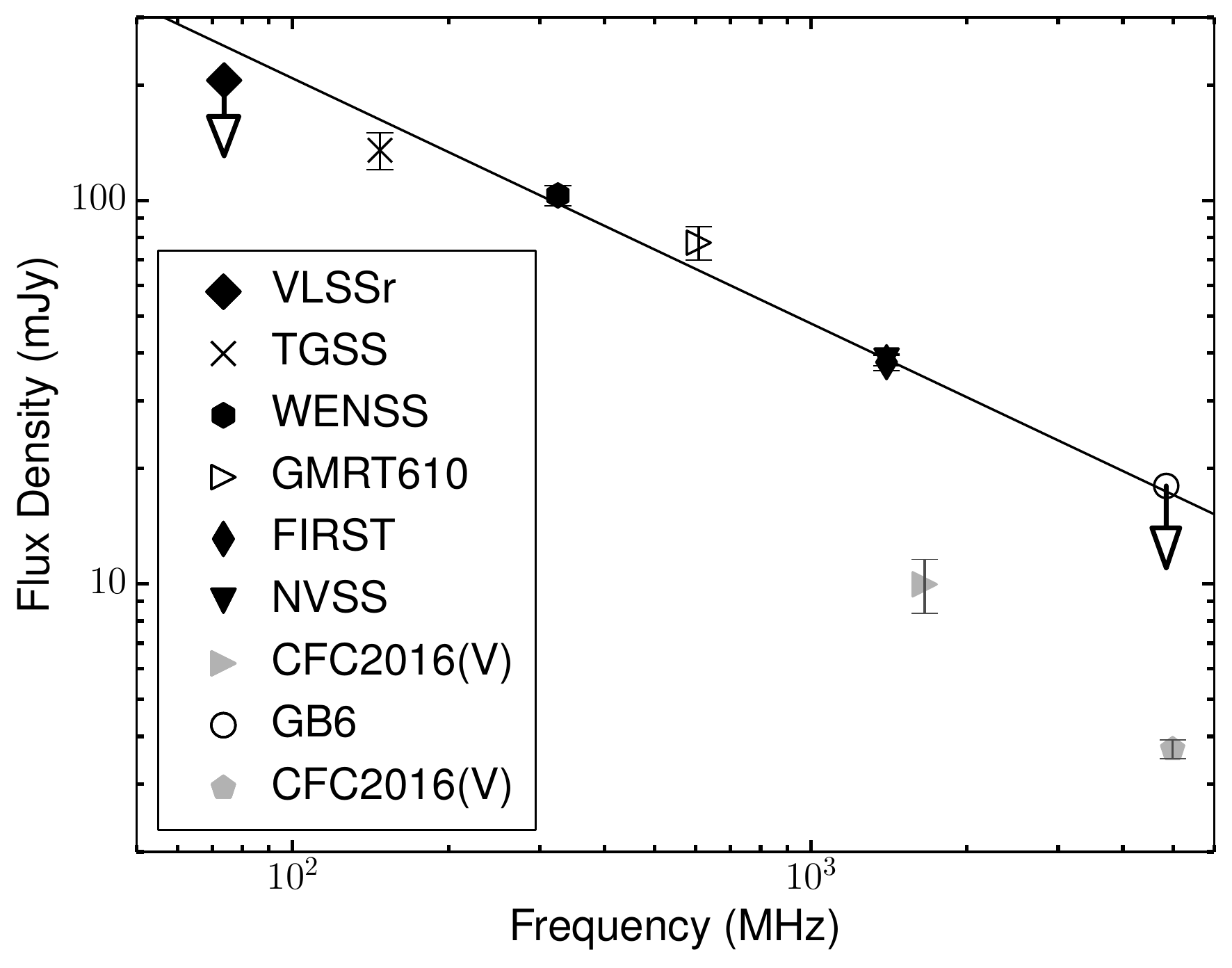}
  \caption{The radio spectrum of J1548+3335. The fit to the spectrum is shown as a solid line.}
  \label{fig:J1548+3335}
\end{figure}

We fitted a power law to the spectrum of J1548+3335 (Fig.~\ref{fig:J1548+3335}) with a spectral index of $\alpha=-0.64\pm0.05$. We note that the 74\,MHz VLSSr and 4.9\,GHz GB6 upper limits could indicate that the spectrum is peaked. However, because there is an equal probability that the flux density of the source is at any value below (including only slightly below) the upper limits, additional observations are required to confirm or refute this.

In the 1.7\,GHz EVN observations, J1548+3335 was found to have two components that are separated by $812\pm3$\,mas, which translates to a projected linear size of $5267\pm17$\,pc \citep{2016MNRAS.tmp.1343C}. The second (fainter) component is not detected in the 5\,GHz EVN observations \citep{2016MNRAS.tmp.1343C}. The primary component coincides with the SDSS position and no jet was detected between the two components. It is, therefore, possible that the second component is a lobe or hotspot of the first component, an unrelated AGN at the same redshift, a foreground or background source that is unrelated to J1548+3335, or that the two components are gravitationally lensed images of the same source \citep{2016MNRAS.tmp.1343C}. From the spectrum it is clear that some of the source's flux density was resolved out in the 1.7\,GHz CFC2016(V) observations, or the source is variable.

%%%%%%%%%%%%%%%%%%%%%%%%%%%%%%
\subsubsection{J1628+1154}
\label{subsubsec: J1628+1154}
\begin{figure}
  \includegraphics[width=\columnwidth]{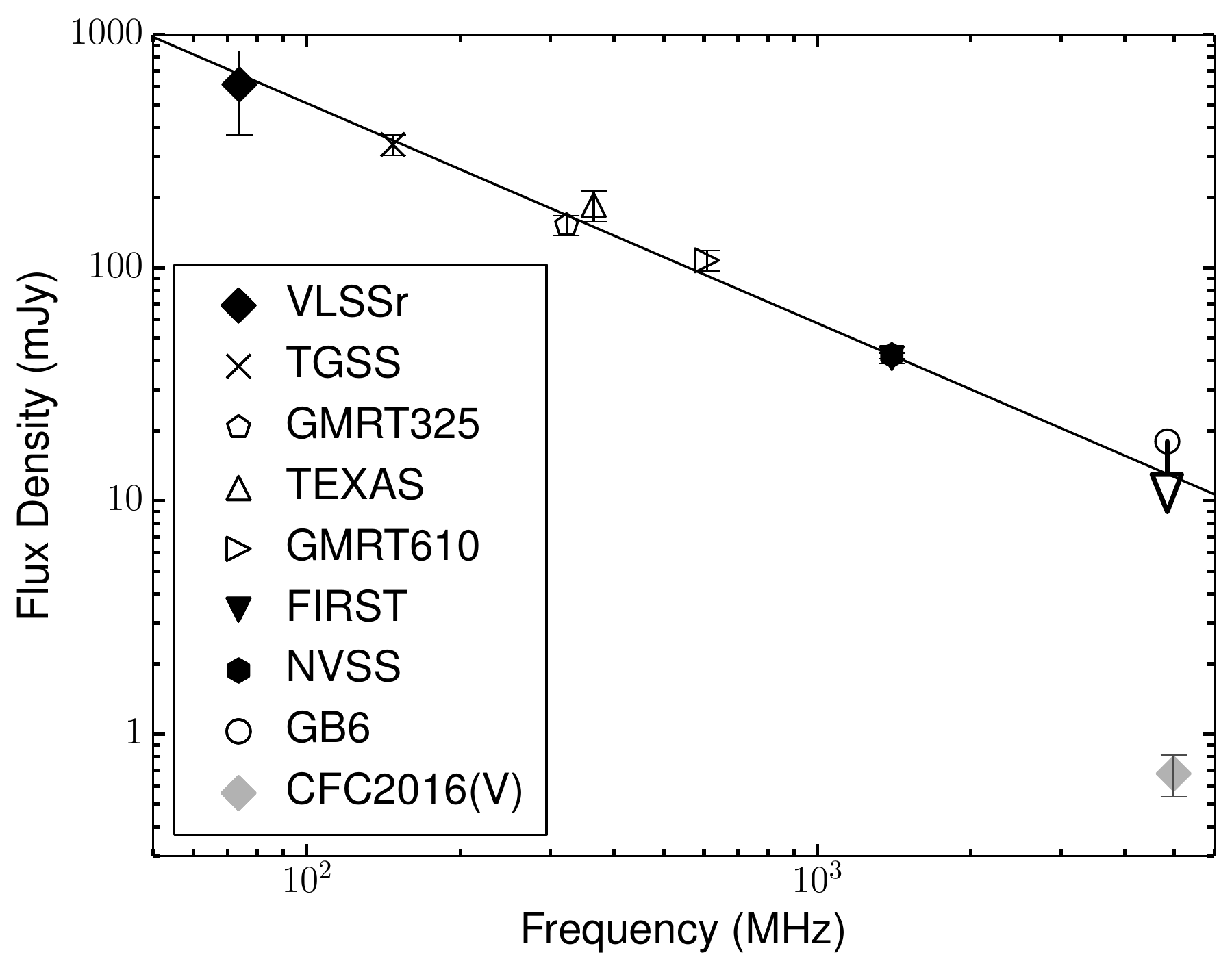}
  \caption{The radio spectrum of J1628+1154. The fit to the spectrum is shown as a solid line.}
  \label{fig:J1628+1154}
\end{figure}

We fitted the spectrum of J1628+1154 (Fig.~\ref{fig:J1628+1154}) with a power law with $\alpha=-0.94\pm0.04$. 

%%%%%%%%%%%%%%%%%%%%%%%%%%%%%%%%%%%%%%%%%%%%%%%%%%%%%%%%%%%%%%%%%%%%%%%%%%%%%
%%%%%%%%%%%%%%%%%%%%%%%%%%%%%%%%%%%%%%%%%%%%%%%%%%%%%%%%%%%%%%%%%%%%%%%%%%%%%
\subsection{Peaked-spectrum sources}
\label{subsec:peaked sources}
The following ten sources all have peaked spectra. Where appropriate, and following \citet{Orienti2007}, \citet{Scaife2012} and \citet{Orienti2014}, we fitted the spectra with log parabolas of the form $\log_{10}(S)= a[\log_{10}(\nu)-\log_{10}(\nu_{\mathrm o})]^2 + b$, where $a$ and $b$ are constants and $S$ is flux density.

%%%%%%%%%%%%%%%%%%%%%%%%%%%%%%
\subsubsection{J0324$-$2918}
\label{subsubsec: J0324$-$2918}
\begin{figure}
  \includegraphics[width=\columnwidth]{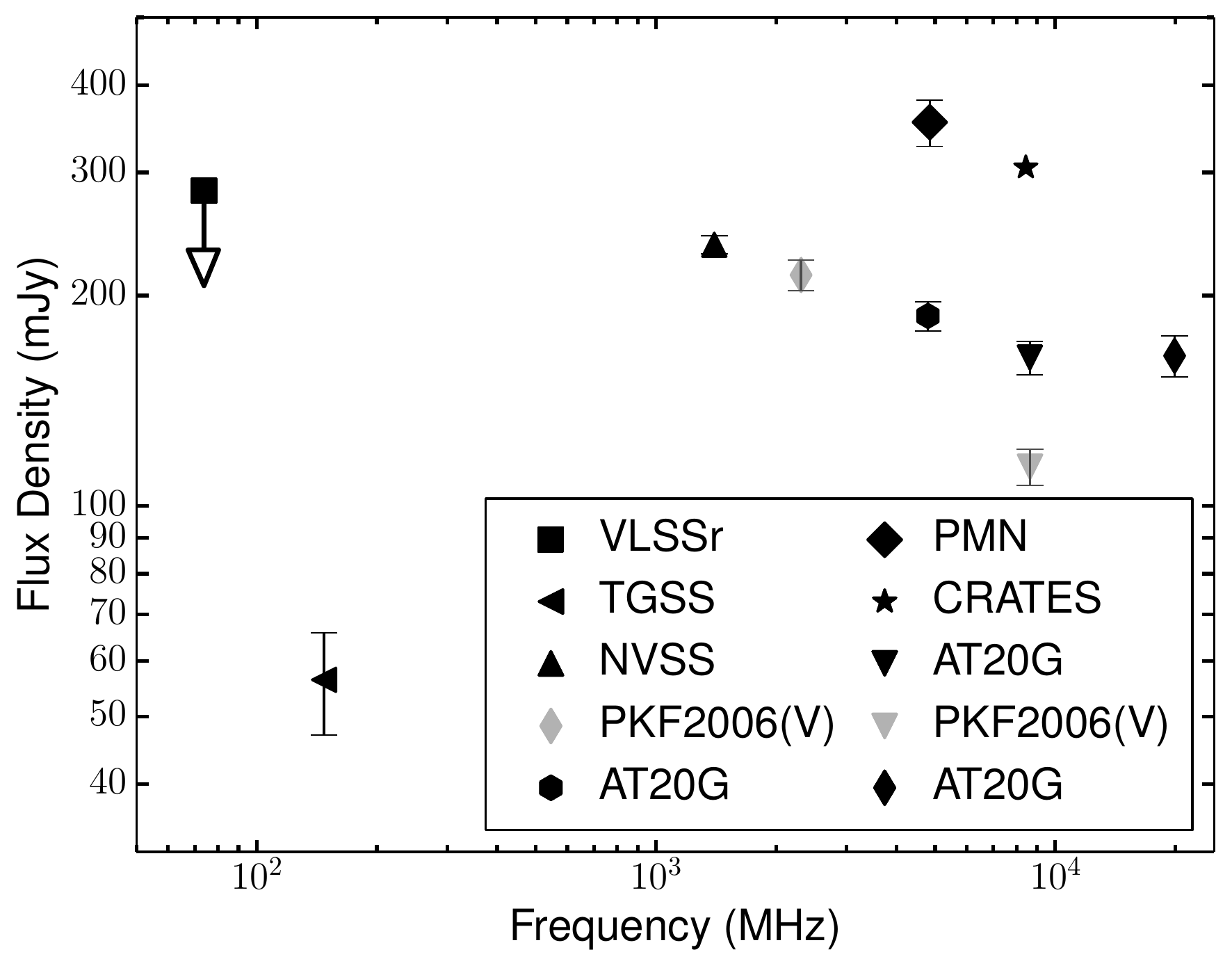}
  \caption{The radio spectrum of J0324$-$2918.}
  \label{fig:J0324$-$2918}
\end{figure}

There is a discrepancy between the 4.8 and 8.6\,GHz AT20G flux densities, and the 8.4\,GHz CRATES and 4.9\,GHz PMN flux densities in the spectrum of J0324$-$2918 (Fig.~\ref{fig:J0324$-$2918}). Regardless of which set of points are considered, it is clear from the 148\,MHz TGSS flux density that J0324$-$2918 is a peaked-spectrum source. The spectral turnover would be at $\sim1.4$\,GHz or 7\,GHz (depending on which observations are considered).

There are two possible explanations for the discrepancy in flux densities between these observations. First, the AT20G values are peak brightnesses, rather than integrated flux densities. Second, the AT20G observations have a resolution between $\sim30$ and $\sim2$\,arcsec \citep{2010MNRAS.402.2403M}, the 4.9\,GHz PMN observations have a resolution of 4.2\,arcmin and we could not determine the resolution of the 8.4\,GHz CRATES observations. Resolution effects could consequently have produced the difference in flux densities. The second possibility is that the difference is due to variability. J0324$-$2918 is a VLBI calibrator \citep{2006AJ....131.1872P} and in CFC2016 we concluded that its VLBI emission is Doppler-boosted, which strengthens the argument that it is variable.

%%%%%%%%%%%%%%%%%%%%%%%%%%%%%%
\subsubsection{J0906+6930}
\label{subsubsec: J0906+6930}
\begin{figure}
  \includegraphics[width=\columnwidth]{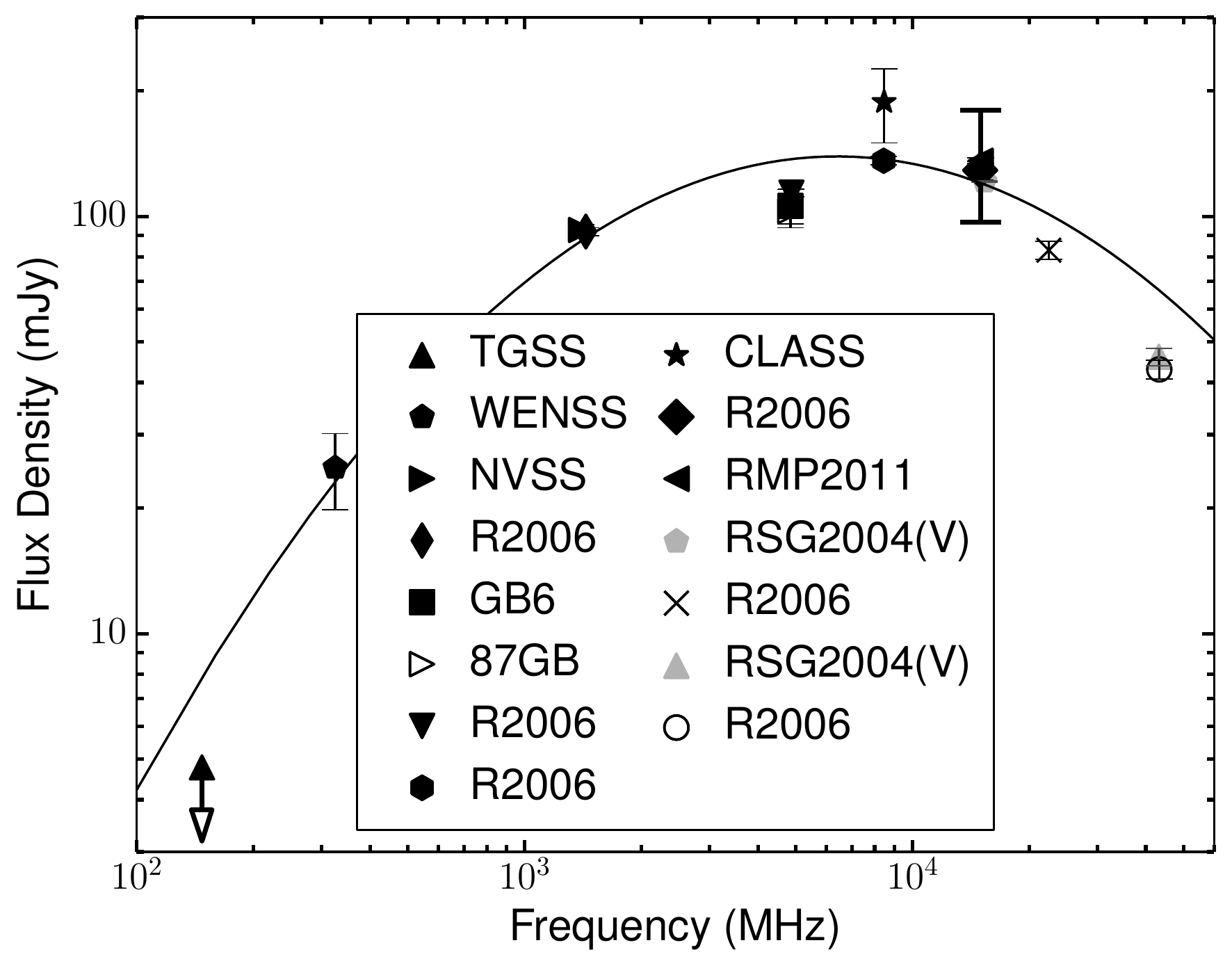}
  \caption{The radio spectrum of J0906+6930. The solid line shows the fitted log parabola. The range of flux density values between which RMP2011 observed 15\,GHz variability is indicated by the thick uncertainty bar.}
  \label{fig:J0906+6930}
\end{figure}

The spectrum of J0906+6930 (Fig.~\ref{fig:J0906+6930}) shows a clear spectral turnover. RMP2011 observed J0906+6930 55 times at 15\,GHz between 2009 March 19 and 2009 December 29. During this time they observed the flux density to vary between 97 and 180\,mJy. As the source is variable, the value in Fig.~\ref{fig:J0906+6930} is the intrinsic mean 15\,GHz flux density ($136\pm2$\,mJy) calculated by RMP2011. Fitting the spectrum, we find a turnover frequency of $6.4\pm0.8$\,GHz. Since J0906+6930 is at $z=5.47$, this translates to a rest-frame turnover frequency of $41.4\pm5.2$\,GHz. Considering that J0906+6930 is variable and that the fitted function does not fit the 148\,MHz TGSS upper limit and the flux densities above 20\,GHz very well, the uncertainty on the turnover frequency is likely underestimated.

%%%%%%%%%%%%%%%%%%%%%%%%%%%%%%
\subsubsection{J0913+5919}
\label{subsubsec: J0913+5919}
\begin{figure}
  \includegraphics[width=\columnwidth]{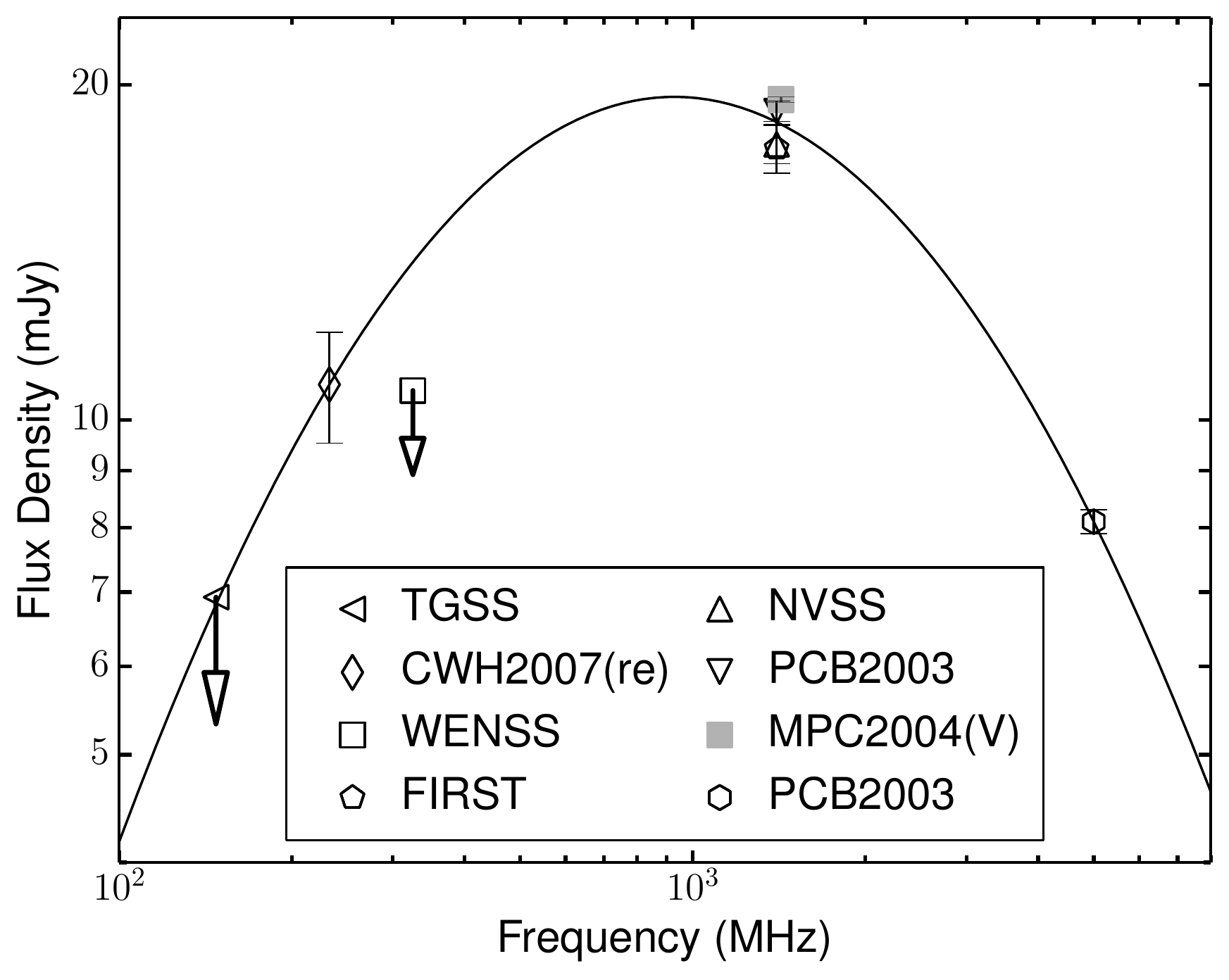}
  \caption{The radio spectrum of J0913+5919. The solid line shows the fitted log parabola.}
  \label{fig:J0913+5919}
\end{figure}

CWH2007 found a 233\,MHz flux density of $30\pm3$\,mJy for J0913+5919 which is incompatible with the 148 and 325\,MHz upper limits of 6.9 and 10.6\,mJy from TGSS and WENSS, respectively, in the spectrum of J0913+5919 (Fig.~\ref{fig:J0913+5919}). To check this apparent discrepancy, we re-processed the same data used by CWH2007. The raw visibility data, available from the GMRT archive under project code 04CCA01, consist of three observing sessions (2003 September 15 to 17) with a total of 11.4\,hours on source. It was recorded over 4\,MHz of bandwidth centered on 232.5\,MHz and used the calibrator 3C48. We extracted the flux densities in the same way as described in Section \ref{sec:GMRT}. This yielded an image with a local rms noise level of 0.36\,mJy\,beam$^{-1}$ at a resolution of $16.4\times10.5$\,arcsec, with a beam position angle of $3^{\circ}$.

The integrated flux density of J0913+5919 in the reprocessed image is $10.7\pm1.2$\,mJy, which is a factor $\sim3$ lower than what was found by CWH2007. The new value is compatible with the TGSS and WENSS upper limits. In the initial (preliminary) image created by our pipeline, there were strong image-plane ripples in the central region near the source. This was a rather common feature in older (hardware-correlator-based) GMRT data, and is likely the result of baseline-based errors. It is not straightforward to suppress, and might have affected the flux density measurement in CWH2007. The \textsc{SPAM} pipeline has dedicated image-based flagging routines to excise the visibility data causing these artefacts, yielding ripple-free images. We will therefore continue using the new flux density which is labeled as CWH2007(re) in Fig.~\ref{fig:J0913+5919}.

Fitting a log parabola to the spectrum gives $\nu_{\mathrm o}=928\pm89$\,MHz, which translates to a rest-frame turnover frequency of $5670\pm544$\,MHz. We note that due to the lack of spectral coverage, the uncertainty on the turnover frequency is likely underestimated.

%%%%%%%%%%%%%%%%%%%%%%%%%%%%%%
\subsubsection{J1146+4037}
\label{subsubsec: J1146+4037}
\begin{figure}
  \includegraphics[width=\columnwidth]{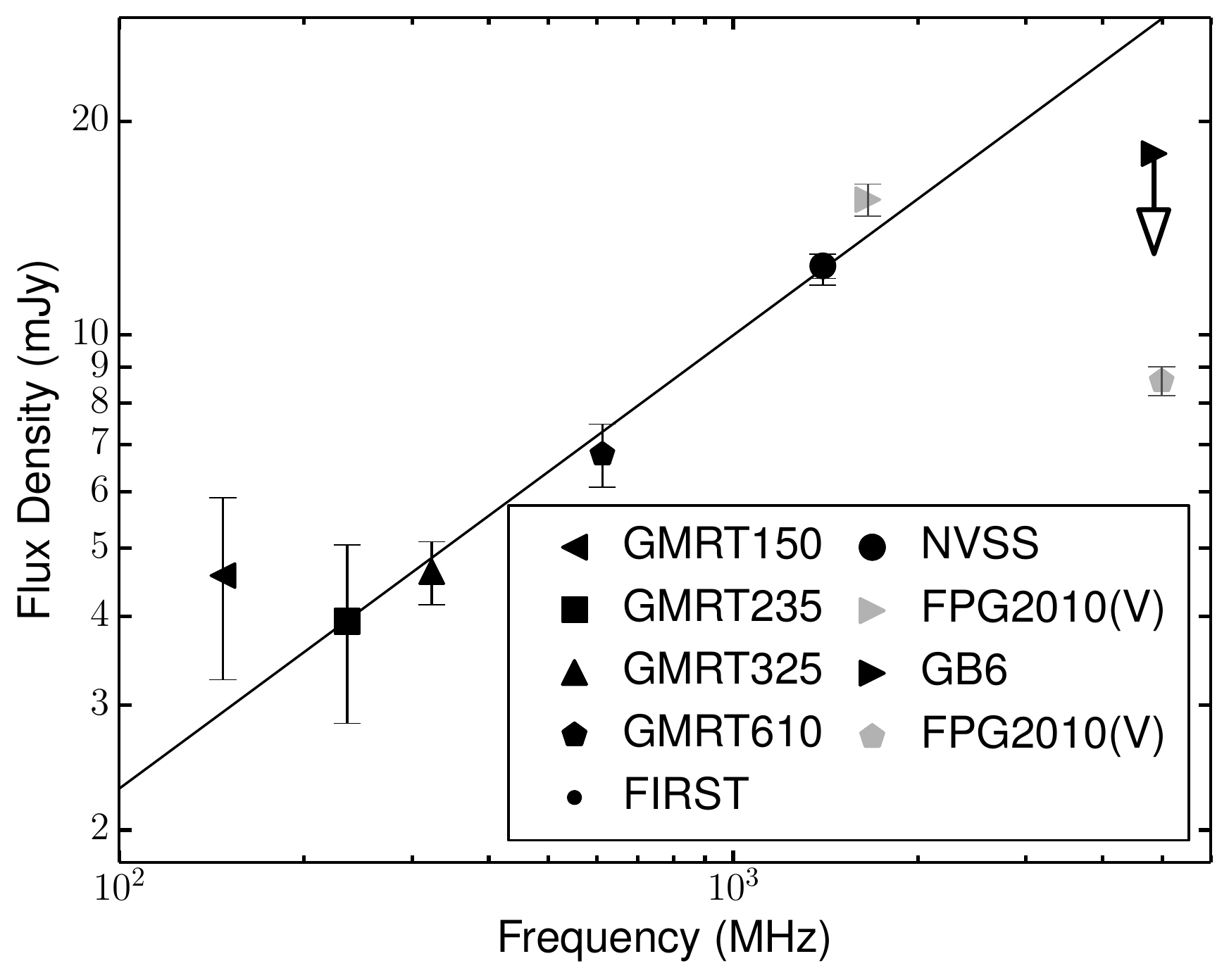}
  \caption{The radio spectrum of J1146+4037. A power law fit to the spectrum is shown as a solid line.}
  \label{fig:J1146+4037}
\end{figure}

If we were to fit a power law to the spectrum of J1146+4037 (excluding the upper limits and VLBI observations), it would give a spectral index of $\alpha=0.64\pm0.05$ (see Fig.~\ref{fig:J1146+4037}). However, the predicted flux density at 4850\,MHz would then be $\sim27$\,mJy, which is well above the 4.9\,GHz GB6 upper limit of 18\,mJy. It is therefore most likely that the spectrum flattens towards higher frequencies, and considering that the spectral index between the 1.7 and 5\,GHz of the FPG2010(V) VLBI points is $-0.53\pm0.06$ \citep{2016MNRAS.tmp.1343C}, it appears to turn over. While care should be taken when comparing non-VLBI and VLBI spectral indices, we believe it is justified in this case, as the 1.4\,GHz FIRST and 1.6\,GHz FPG2010(V) flux densities are comparable ($12.4\pm0.6$ and $15.5\pm0.8$\,mJy, respectively). Crucially, the GB6 upper limit also indicates a turnover. We therefore conclude that J1146+4037 likely has a spectral turnover around 1.4\,GHz and we classify it as a peaked-spectrum source.

%%%%%%%%%%%%%%%%%%%%%%%%%%%%%%
\subsubsection{J1235$-$0003}
\label{subsubsec: J1235$-$0003}
\begin{figure}
  \includegraphics[width=\columnwidth]{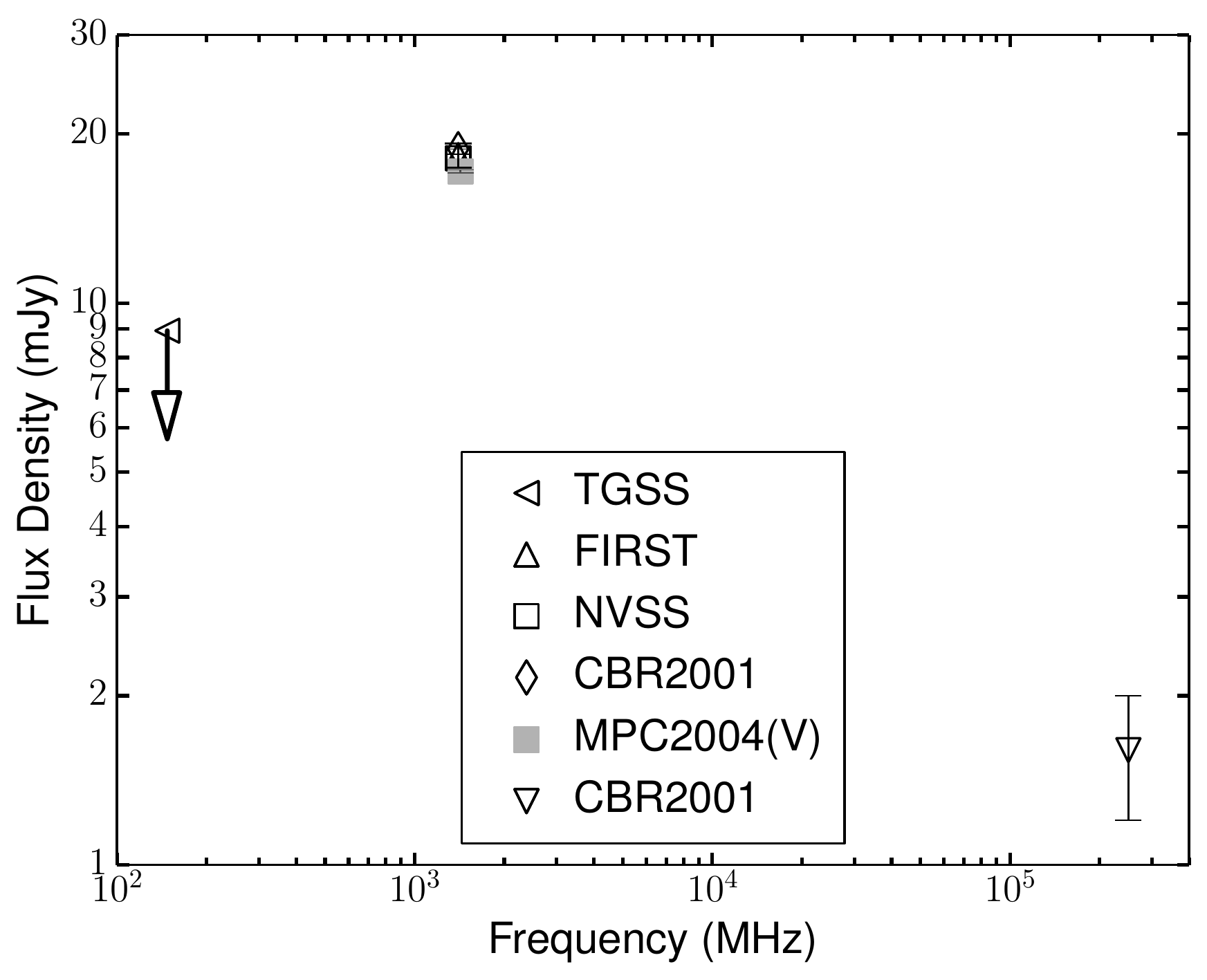}
  \caption{The radio spectrum of J1235$-$0003.}
  \label{fig:J1235$-$0003}
\end{figure}

It is clear that J1235$-$0003 has a peaked spectrum (Fig.~\ref{fig:J1235$-$0003}). However, due to a lack of spectral coverage, we can not constrain the location of the spectral peak.

%%%%%%%%%%%%%%%%%%%%%%%%%%%%%%
\subsubsection{J1242+5422}
\label{subsubsec: J1242+5422}
\begin{figure}
  \includegraphics[width=\columnwidth]{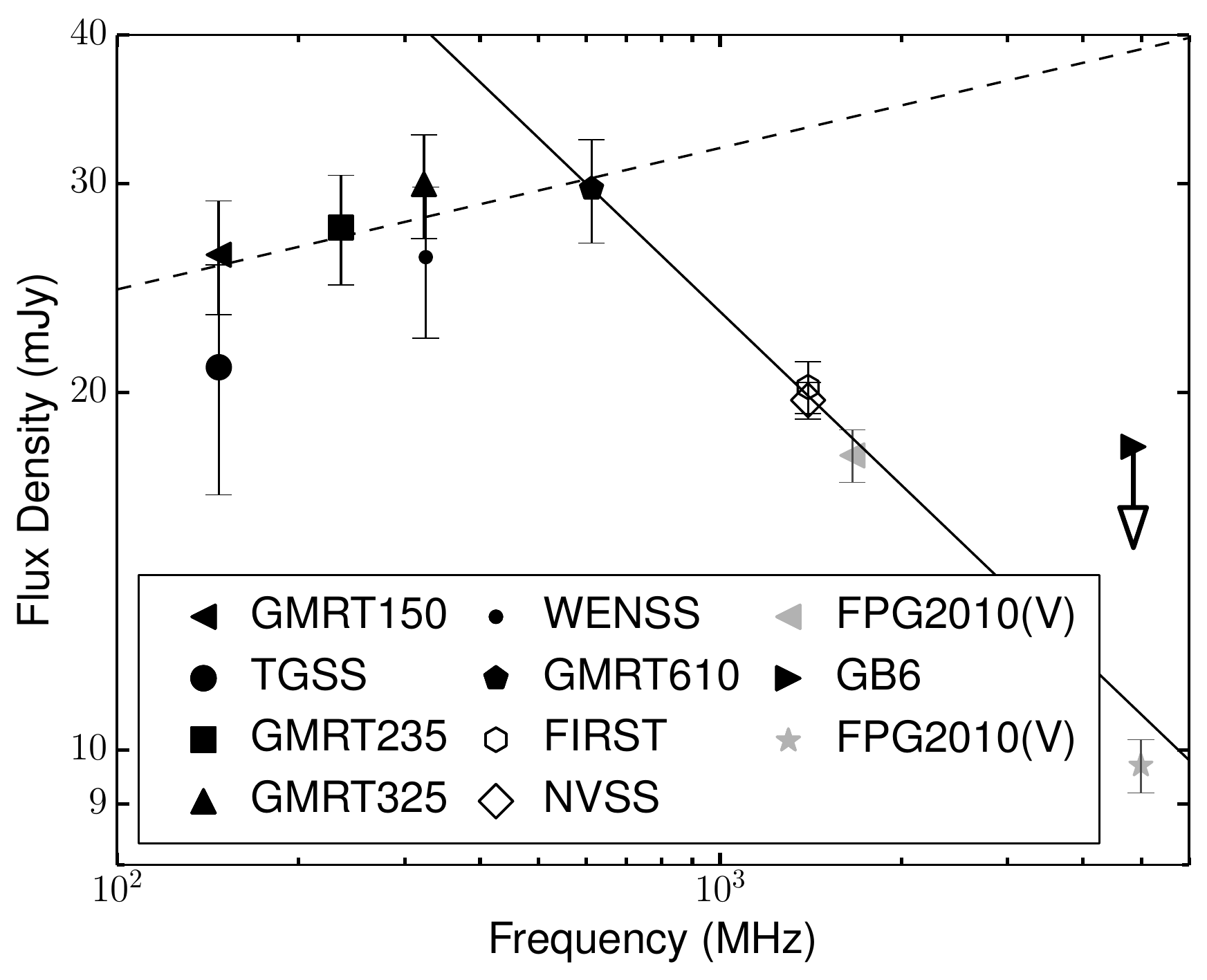}
  \caption{The radio spectrum of J1242+5422. The solid line is fitted between the 612\,MHz GMRT610, FIRST (1.4\,GHz) and NVSS (1.4\,GHz) flux densities, while the dashed line is fitted between all of the non-VLBI flux densities excluding FIRST and NVSS.}
  \label{fig:J1242+5422}
\end{figure}

Fitting a power law between the 1.4\, GHz FIRST, 1.4\,GHz NVSS and 612\,MHz GMRT610 flux densities in the spectrum of J1242+5422 (Fig.~\ref{fig:J1242+5422}) gives $\alpha=-0.49\pm0.05$. Fitting a power law (the dashed line in Fig.~\ref{fig:J1242+5422}) between all of the non-VLBI flux densities excluding FIRST and NVSS, gives $\alpha=0.12\pm0.06$. J1242+5422 therefore has a positive spectral index below $\sim610$\,MHz and a negative spectral index above $\sim610$\,MHz, and is therefore a peaked-spectrum source. This conclusion is supported by the 4.9\,GHz GB6 upper limit.

%%%%%%%%%%%%%%%%%%%%%%%%%%%%%%
\subsubsection{J1606+3124}
\label{subsubsec: J1606+3124}
Matching the VLBI position of J1606+3124 to FIRST, we find that there are five sources within three arcminutes, with the nearest neighbour at a distance of 70\,arcsec. In the survey catalogue these sources are indicated to have side lobe probabilities between 0.272 and 0.439 \citep{2015ApJ...801...26H}. The VLA beam pattern is also clearly visible in the image, and all five neighbouring sources lie on this beam pattern\footnote{http://third.ucllnl.org/cgi-bin/firstcutout}. As the 1.4\,GHz NVSS and 325\,MHz WENSS images show that the nearest neighbour is at a distance of 232\,arcsec from J1606+3124 and based on the probabilities of the sources being side lobes, we conclude that the five neighbouring sources in the 1.4\,GHz FIRST image are all image artefacts. We matched J1606+3124 to sources in the 0.96, 2.3, 3.9, 7.7, 11.2 and 21.65\,GHz catalogues of \citet{1999AandAS..139..545K} and the 1.1, 2.3, 4.8, 7.7, 11.2 and 21.7\,GHz catalogues of \citet{2012AandA...544A..25M}. However, since these observations were taken with the RATAN-600 telescope, the resolution of all of the observations is lower than the distance to the nearest neighboring source. The flux density of the nearby sources will therefore blend with that of J1606+3124 and we discarded the matches.

\begin{figure*}
  \includegraphics[width=13cm]{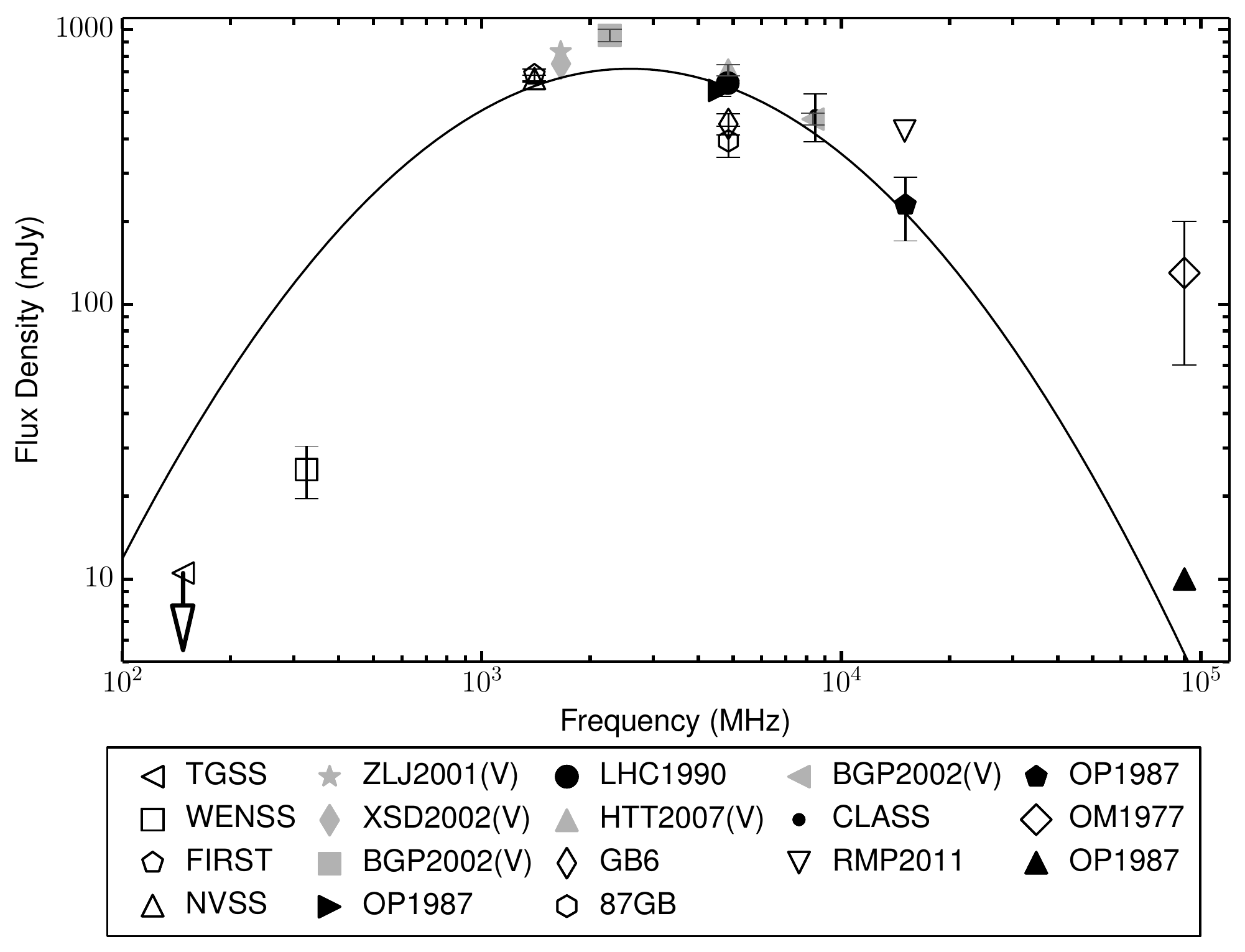}
  \caption{The radio spectrum of J1606+3124. The solid line shows the fitted log parabola.}
  \label{fig:J1606+3124}
\end{figure*}

The spectrum of J1606+3124 is shown in Fig.~\ref{fig:J1606+3124}. RMP2011 observed J1606+3124 98 times at 15\,GHz between 2008 January 1 and 2009 December 28 with the 40\,m telescope at the Owens Valley Radio Observatory. From this they concluded that J1606+3124 is not variable. While we discarded the matches to \citet{2012AandA...544A..25M}, we note that the authors did observe J1606+3124 six times with the RATAN-600 telescope between 2006 July and 2010 May at 21.7, 11.2, 7.7, 4.8 and 2.3\,GHz, and five times at 1\,GHz over the same period. These observations also indicate that J1606+3124 is not variable at these frequencies. The average 15\,GHz flux density of RMP2011 at each frequency are plotted in Fig.~\ref{fig:J1606+3124}. In OP1987 the authors give the 90\,GHz flux density as $10\pm150$\,mJy. Since the uncertainty is nonphysically large we omitted it in Fig.~\ref{fig:J1606+3124}. We do however note that it is possible that the uncertainty is correct and the value itself is wrong.

It has been known for some time that J1606+3124 has a peaked spectrum \citep[e.g.][]{1985A&A...152...38S}, with \citet{1997A&A...321..105D} and \citet{2013AstBu..68..262M} reporting peak frequencies of 1.5 and 3.5\,GHz, respectively. Fitting a log parabola to the spectrum, we found $\nu_{\mathrm o}=2581\pm536$\,MHz. Taking into account the redshift of J1606+3124, our observed turnover frequency translates to a rest-frame turnover frequency of $14.4\pm3.0$\,GHz. We finally note that in the 4.8\,GHz HTT2007(V) and in the 2.2 and 8.3\,GHz BGP2002(V) VLBI observations, J1606+3124 has a Compact Symmetric Object (CSO) structure. CSOs are characterised by unbeamed emission from their steep-spectrum radio lobes on either side of a central position, and have sizes smaller than their host galaxy \citep{Fanti1995, 2009AN....330..120F}.

%%%%%%%%%%%%%%%%%%%%%%%%%%%%%%
\subsubsection{J1659+2101}
\label{subsubsec: J1659+2101}
\begin{figure}
  \includegraphics[width=\columnwidth]{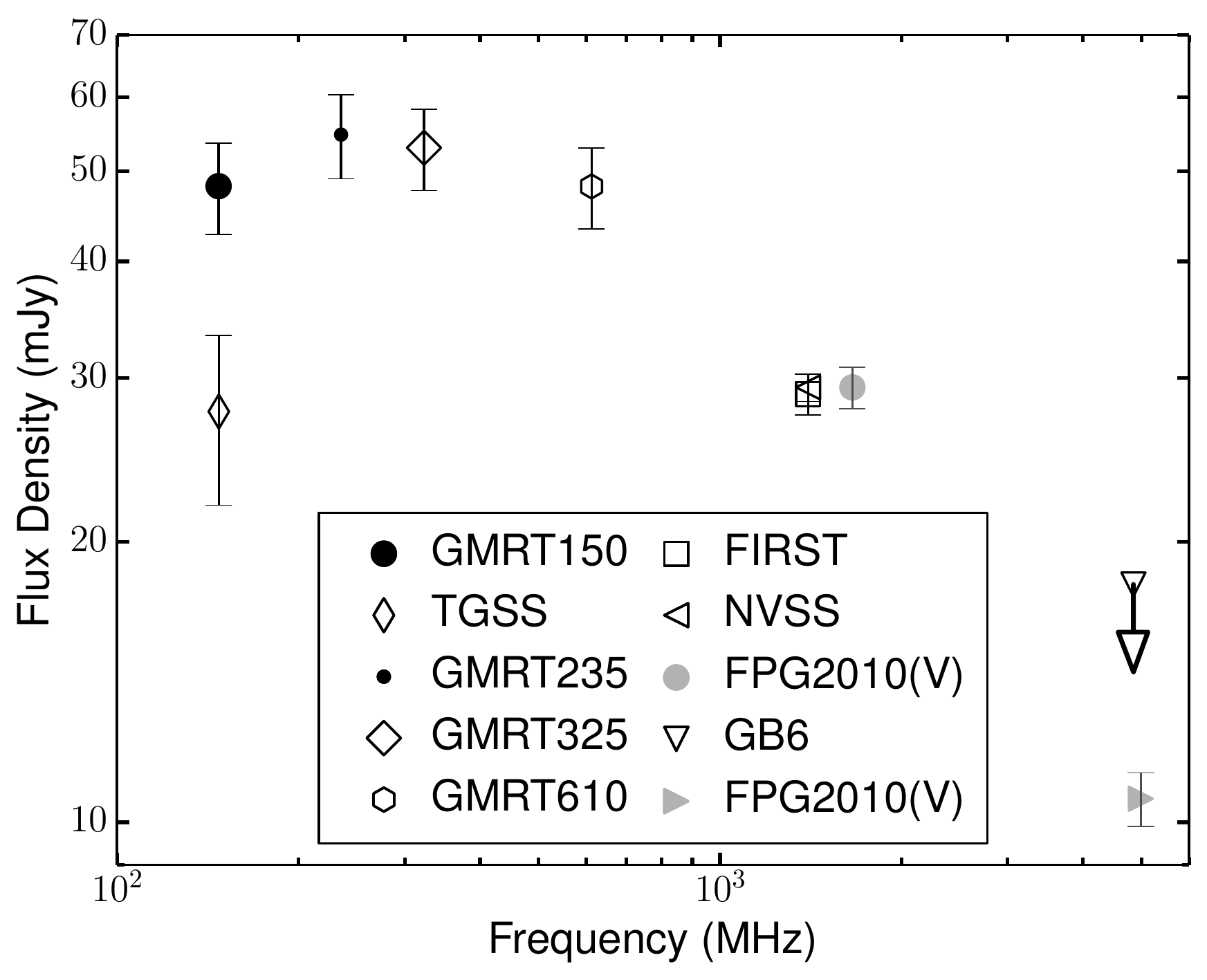}
  \caption{The radio spectrum of J1659+2101.}
  \label{fig:J1659+2101}
\end{figure}

The 148\,MHz TGSS and 147\,MHz GMRT150 flux densities in the spectrum of J1659+2101 (Fig.~\ref{fig:J1659+2101}) are $27.6\pm5.7$ and $48.2\pm5.4$\,mJy, respectively. This translates to a difference of $1.9\sigma$ or 75\,per\,cent in flux density. Visual inspection of the images did not reveal an explanation for the offset. To try find an explanation, we matched the sources in the 147\,MHz GMRT150 image to those in TGSS using a 10\,arcsec search radius. We found 22 matches within a square of $1\times1$\,deg centred on J1659+2101. For each of these sources, we calculated the ratio between the 147\,MHz GMRT150 and the 148\,MHz TGSS flux densities: The median of all of the ratios was 0.95, and the average was 1.02. The discrepancy can consequently not be attributed to a systematic flux density offset between the catalogues. Another possible explanation for the difference could be that J1659+2101 is variable. This is contradicted, but not ruled out, by the 1.4\,GHz FIRST and NVSS flux densities that are within 2\,per\,cent of each other despite the epochs when FIRST and NVSS observed J1659+2101 differing by about 3.4\,years \citep{2011ApJ...737...45O,2015ApJ...801...26H}. Resolution effects also cannot explain the difference, as the resolutions of the surveys are similar ($25\times25$\,arcsec and $23\times16$\,arcsec, respectively). We can therefore not explain the difference between the TGSS and GMRT150 flux densities.

Fitting a power law to the spectrum, and excluding the TGSS and GMRT150 flux densities, gives $\alpha=-0.40\pm0.05$. Repeating the fit using only the GMRT150 and 235\,MHz GMRT235 values give $\alpha=0.27\pm0.33$, while fitting only the TGSS and GMRT235 values gives $\alpha=1.47\pm0.49$. It is therefore clear that irrespective of whether the TGSS or the GMRT150 flux densities are correct, at the very least the spectrum flattens, and it likely turns over around 235\,MHz. We therefore classify J1659+2101 as having a peaked spectrum.

%%%%%%%%%%%%%%%%%%%%%%%%%%%%%%
\subsubsection{J2102+6015}
\label{subsubsec: J2102+6015}
\begin{figure}
  \includegraphics[width=\columnwidth]{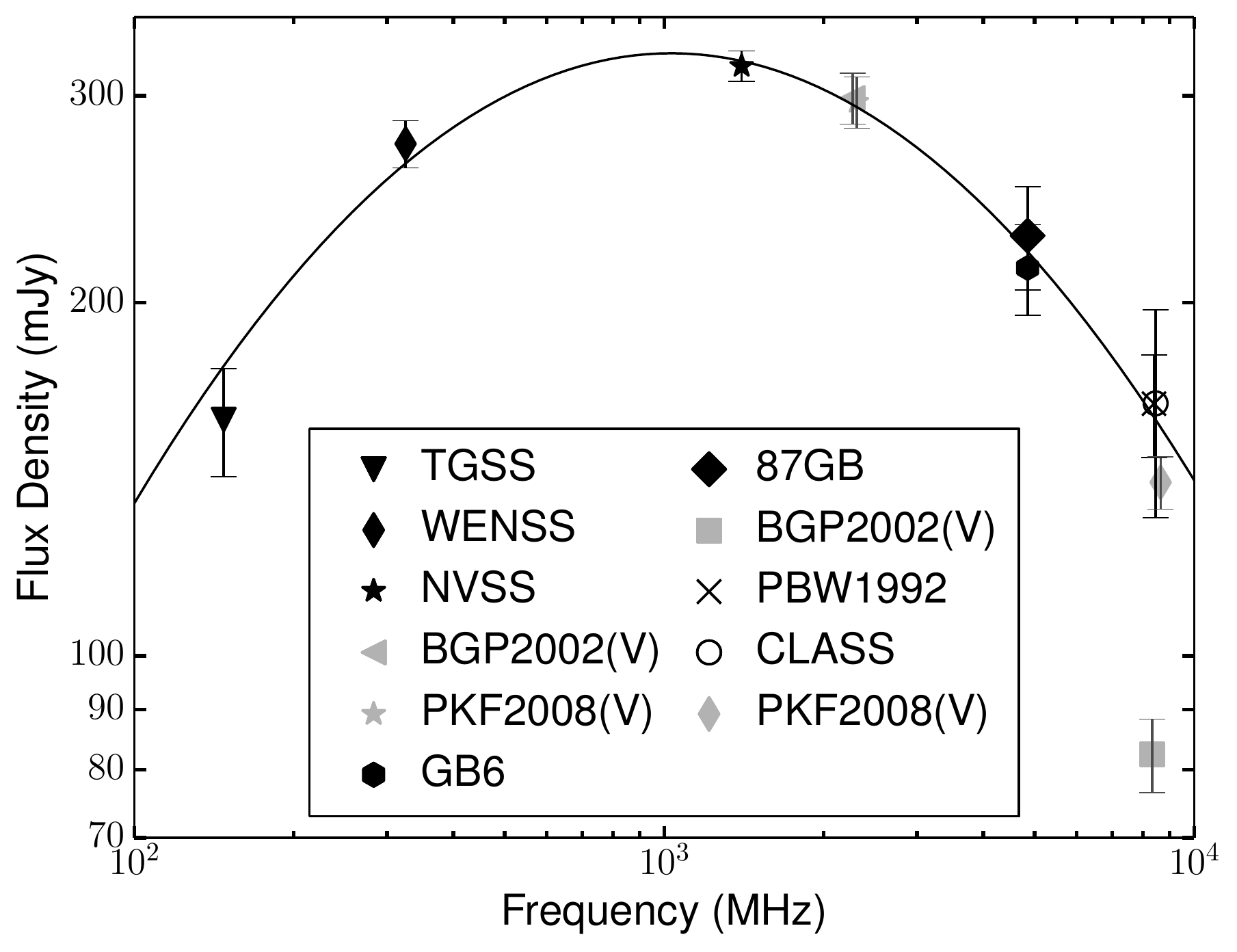}
  \caption{The radio spectrum of J2102+6015. The solid line shows the fitted log parabola.}
  \label{fig:J2102+6015}
\end{figure}

The spectrum of J2102+6015 (Fig.~\ref{fig:J2102+6015}) shows a clear turnover. Fitting the spectrum with a log parabola gives $\nu_{\mathrm o}=1031\pm51$\,MHz. This corresponds to a rest-frame turnover frequency of $5753\pm283$\,MHz.

%%%%%%%%%%%%%%%%%%%%%%%%%%%%%%
\subsubsection{J2228+0110}
\label{subsubsec: J2228+0110}
\begin{figure}
  \includegraphics[width=\columnwidth]{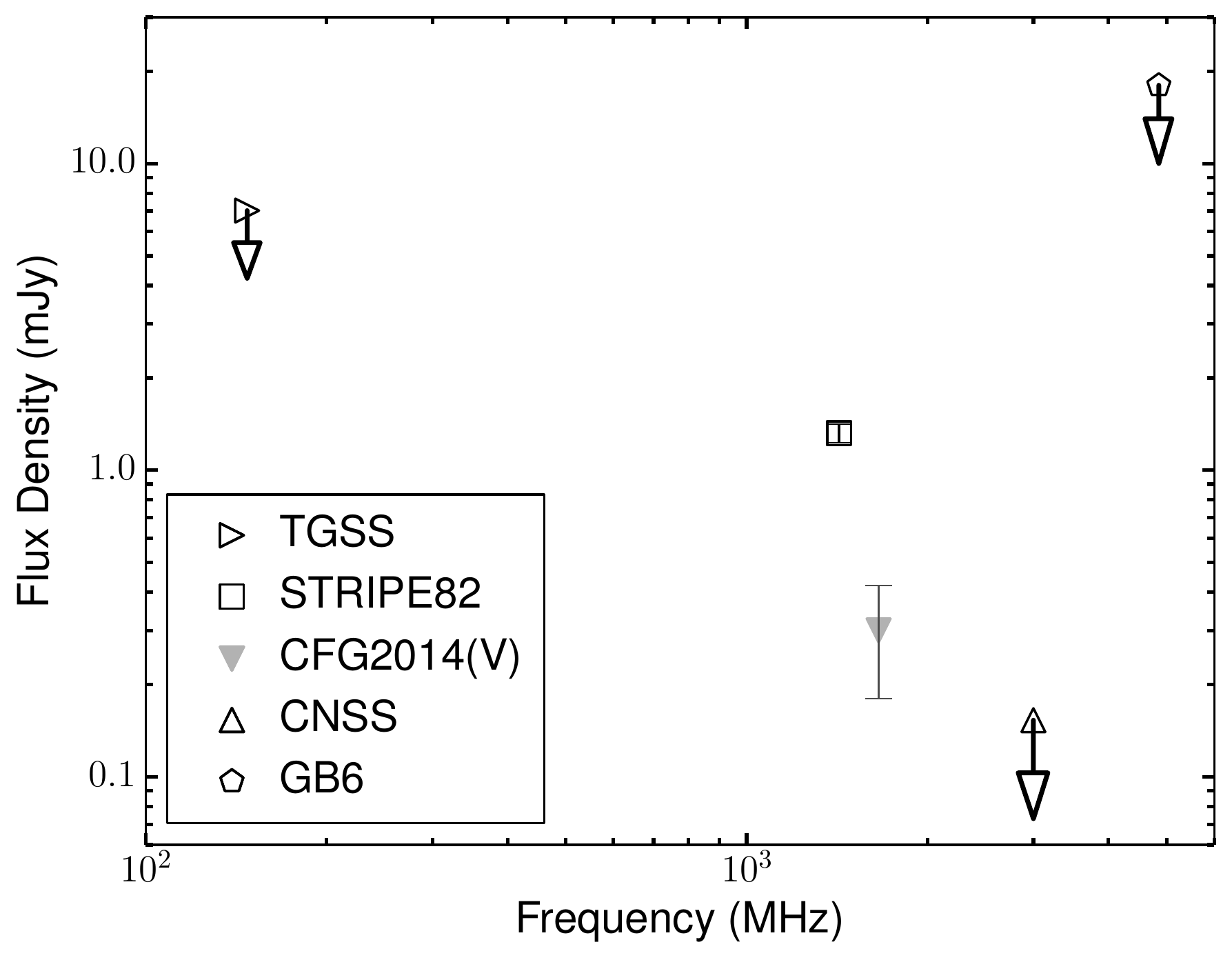}
  \caption{The radio spectrum of J2228+0110.}
  \label{fig:J2228+0110}
\end{figure}

Despite J2228+0110 only being detected in the 1.4\,GHz STRIPE82 survey, the 3\,GHz CNSS and 148\,MHz TGSS upper limits show that its spectrum (Fig.~\ref{fig:J2228+0110}) peaks below 1.4\,GHz.

%%%%%%%%%%%%%%%%%%%%%%%%%%%%%%%%%%%%%%%%%%%%%%%%%%%%%%%%%%%%%%%%%%%%%%%%%%%%%
%%%%%%%%%%%%%%%%%%%%%%%%%%%%%%%%%%%%%%%%%%%%%%%%%%%%%%%%%%%%%%%%%%%%%%%%%%%%%
\subsection{Unusual and unclassified spectra}
\label{subsec:unusual+undefined spectrum}
The last class contains the six sources that cannot be classified into one of the three previous classes, and those that (due to a lack of spectral coverage) could have spectra that fall into more than one of the classes. 
%%%%%%%%%%%%%%%%%%%%%%%%%%%%%%
\subsubsection{J1013+2811}
\label{subsubsec: J1013+2811}
\begin{figure}
  \includegraphics[width=\columnwidth]{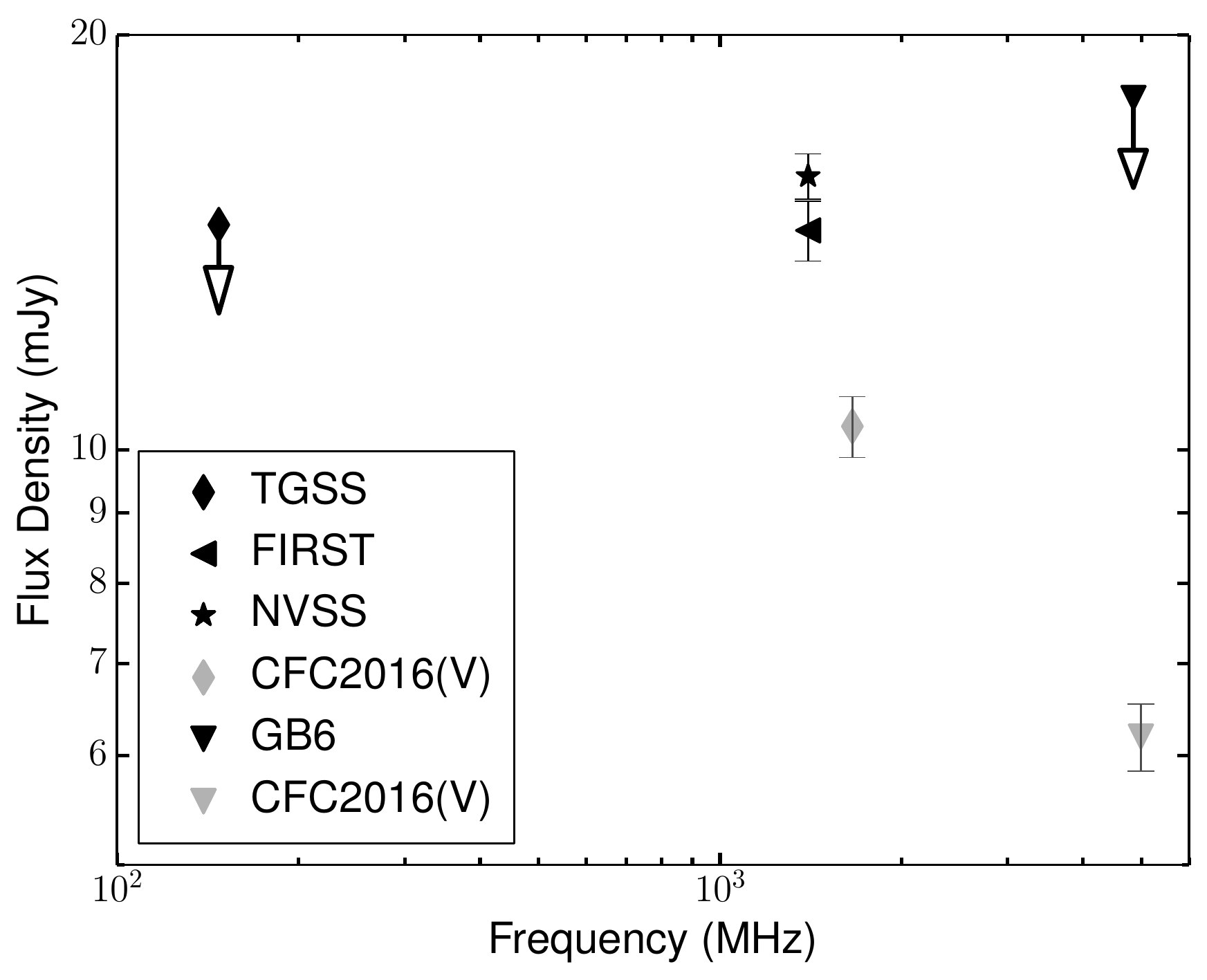}
  \caption{The radio spectrum of J1013+2811.}
  \label{fig:J1013+2811}
\end{figure}

Assuming that the spectrum of J1013+2811 (Fig.~\ref{fig:J1013+2811}) can be fitted with a single power law, and using only the 1.4\,GHz FIRST flux density and the 4.9\,GHz GB6 upper limit, produces a spectral index $\alpha<0.18$. Similarly, a fit using only the FIRST flux density and the 148\,MHz TGSS upper limit, produces a spectral index greater than zero. Based on these limits, J1013+2811 can either have a flat or a peaked spectrum. 

%%%%%%%%%%%%%%%%%%%%%%%%%%%%%%
\subsubsection{J1205$-$0742}
\label{subsubsec: J1205$-$0742}
\begin{figure}
  \includegraphics[width=\columnwidth]{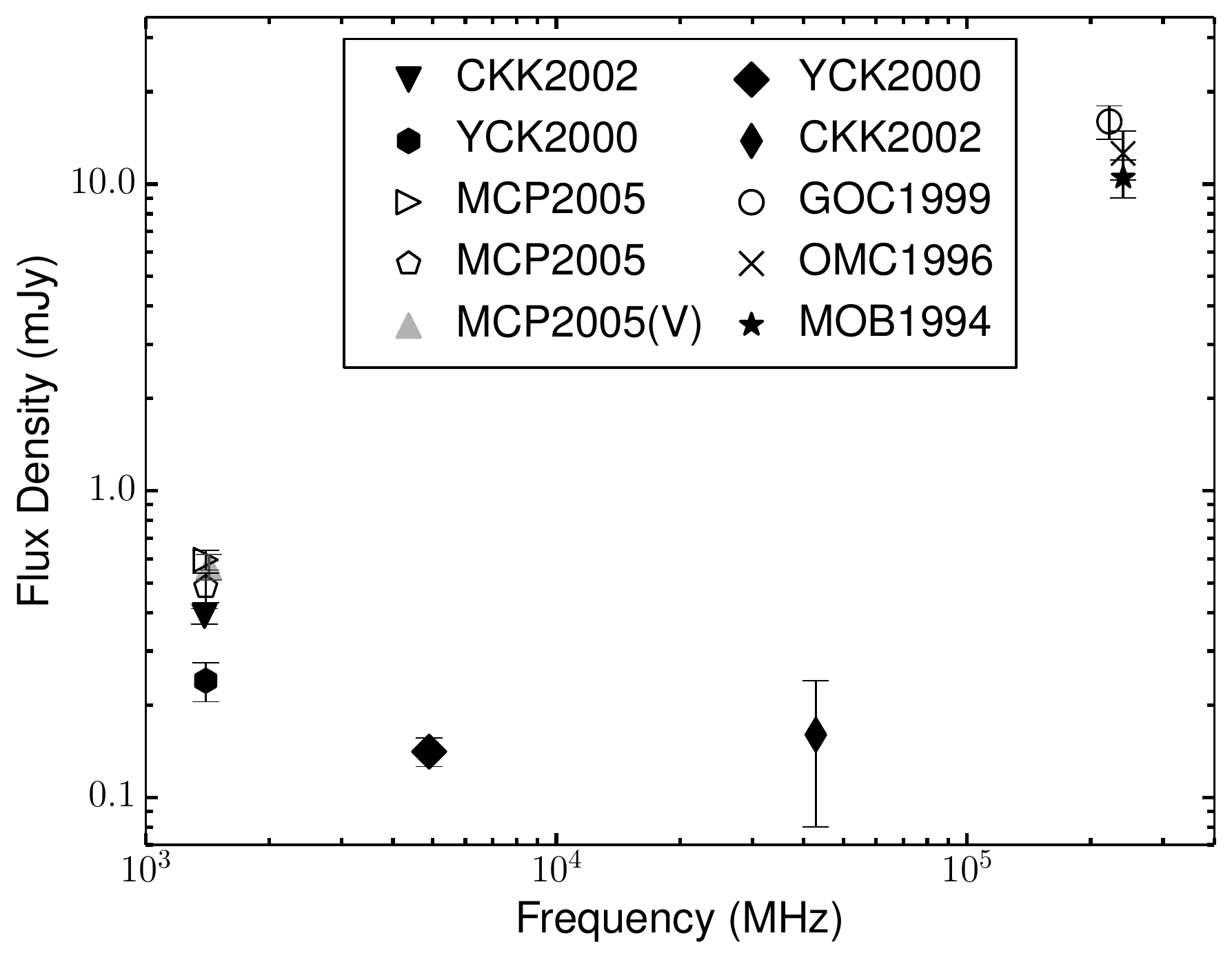}
  \caption{The radio spectrum of J1205$-$0742.}
  \label{fig:J1205$-$0742}
\end{figure}

The spectrum of J1205$-$0742 (Fig.~\ref{fig:J1205$-$0742}) is concave, with evidence of variability at 1.4\,GHz. Using its spectral index between 1.4 and 350\,GHz, morphology, brightness temperature and linear size, MCP2005(V) showed that the radio emission from J1205$-$0742 is from a nuclear starburst, and that the source does not have a radio-loud AGN. This explains why J1205$-$0742 has a concave spectrum. At $\nu_{\mathrm o}<100$\,GHz, the negative spectral index is caused by starburst-driven radio synchrotron emission, while at $\nu_{\mathrm o}\gtrsim100\,{\mathrm {GHz}} \simeq \nu_{\mathrm r}\gtrsim570$\,GHz, the increase in flux density is the result of thermal dust emission \citep[e.g.][]{1994MNRAS.267L...9M,2000ApJ...528..171Y,2005AJ....129.1809M,2011A&A...536A...7P}.

%%%%%%%%%%%%%%%%%%%%%%%%%%%%%%
\subsubsection{J1311+2227}
\label{subsubsec: J1311+2227}
\begin{figure}
  \includegraphics[width=\columnwidth]{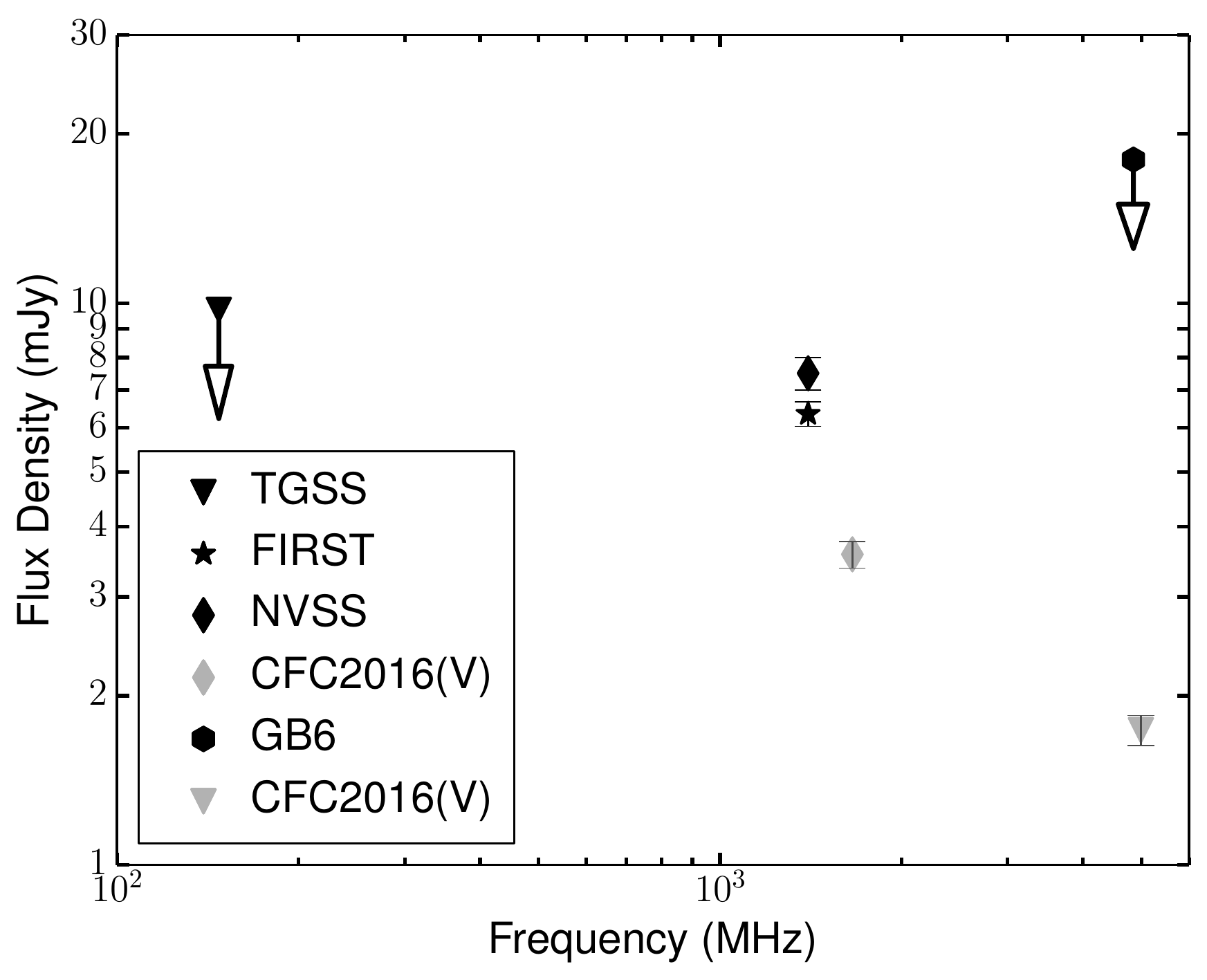}
  \caption{The radio spectrum of J1311+2227.}
  \label{fig:J1311+2227}
\end{figure}

Assuming that the spectrum of J1311+2227 (Fig.~\ref{fig:J1311+2227}) can be fitted with a single power law, and using the 1.4\,GHz FIRST flux density and the 148\,MHz TGSS and 4.9\,GHz GB6 upper limits, the spectral index is $-0.19<\alpha<0.84$. J1311+2227 can therefore either have a flat, inverted or peaked spectrum. 

%%%%%%%%%%%%%%%%%%%%%%%%%%%%%%
\subsubsection{J1454+1109}
\label{subsubsec: J1454+1109}
\begin{figure}
  \includegraphics[width=\columnwidth]{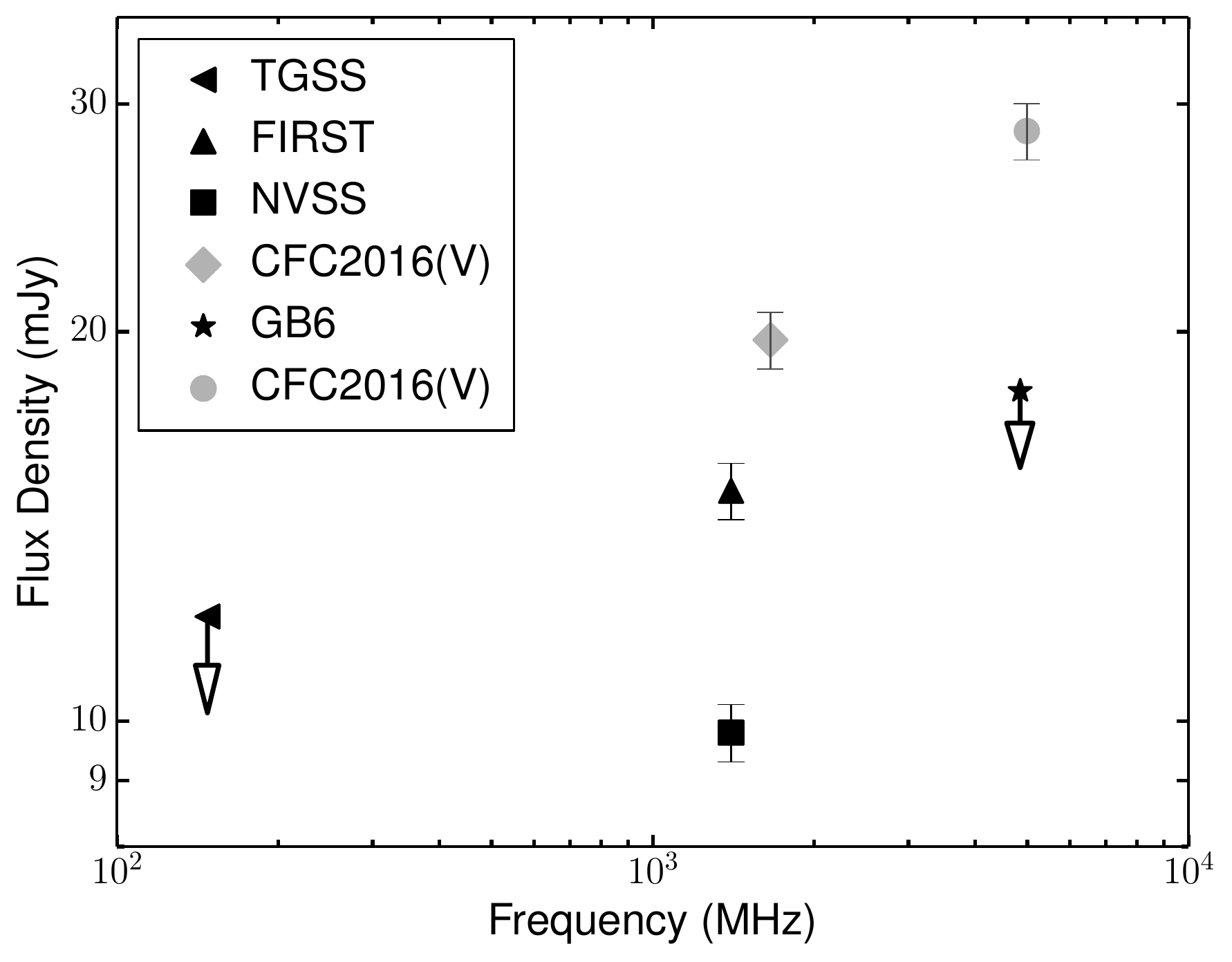}
  \caption{The radio spectrum of J1454+1109.}
  \label{fig:J1454+1109}
\end{figure}

Based on the VLBI flux densities being higher than the non-VLBI flux densities in the spectrum of J1454+1109 (Fig.~\ref{fig:J1454+1109}), and the 4.9\,GHz GB6 upper limit and the 1.4\,GHz FIRST flux density being higher than the 1.4\,GHz NVSS flux density, we conclude that J1454+1109 is variable. In addition, due to a lack of spectral coverage, we cannot constrain the spectrum. However, based on its variability, and the fact that the VLBI emission is Doppler-boosted \citep{2016MNRAS.tmp.1343C}, J1454+1109 is likely a blazar with a flat time-averaged spectrum.

%%%%%%%%%%%%%%%%%%%%%%%%%%%%%%
\subsubsection{J1611+0844}
\label{subsubsec: J1611+0844}
\begin{figure}
  \includegraphics[width=\columnwidth]{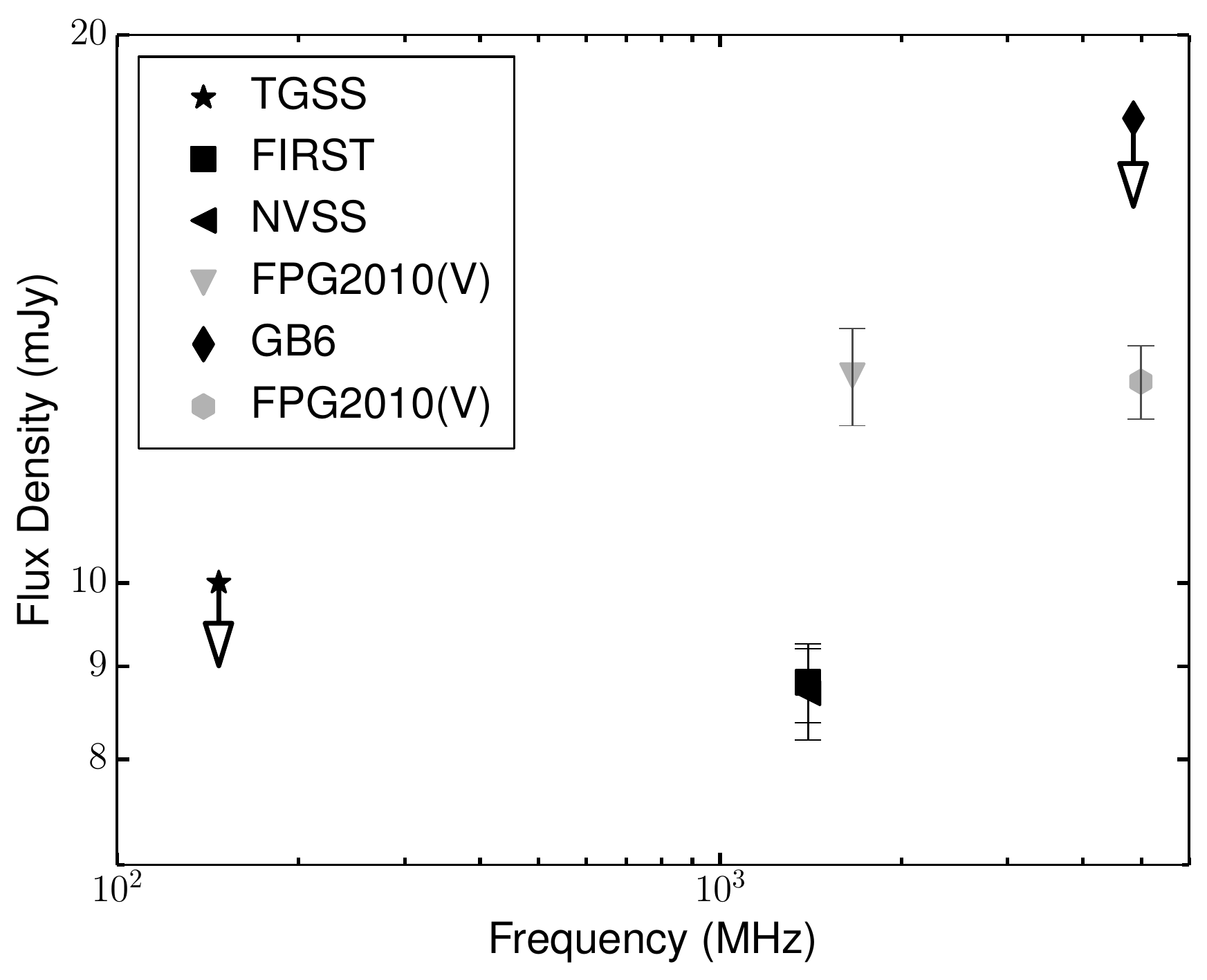}
  \caption{The radio spectrum of J1611+0844.}
  \label{fig:J1611+0844}
\end{figure}

Assuming that the spectrum of J1611+0844 (Fig.~\ref{fig:J1611+0844}) can be fitted with a single power law, and using the 1.4\,GHz FIRST flux density and the 148\,MHz TGSS and 4.9\,GHz GB6 upper limits, $-0.06<\alpha<0.57$. The time-averaged spectrum can therefore be either inverted, flat or peaked. Since the VLBI flux densities are higher than the non-VLBI flux densities, it is likely that J1611+0844 is variable. However, since the epochs when FIRST and NVSS (1.4\,,GHz) observed J1611+0844 differ by about 3.6\,years \citep{2011ApJ...737...45O,2015ApJ...801...26H}, if J1611+0844 is variable it means that the FIRST and NVSS observations were serendipitously done on two epochs when J1611+0844 happened to have the same flux density.

%%%%%%%%%%%%%%%%%%%%%%%%%%%%%%
\subsubsection{J1720+3104}
\label{subsubsec: J1720+3104}
\begin{figure}
  \includegraphics[width=\columnwidth]{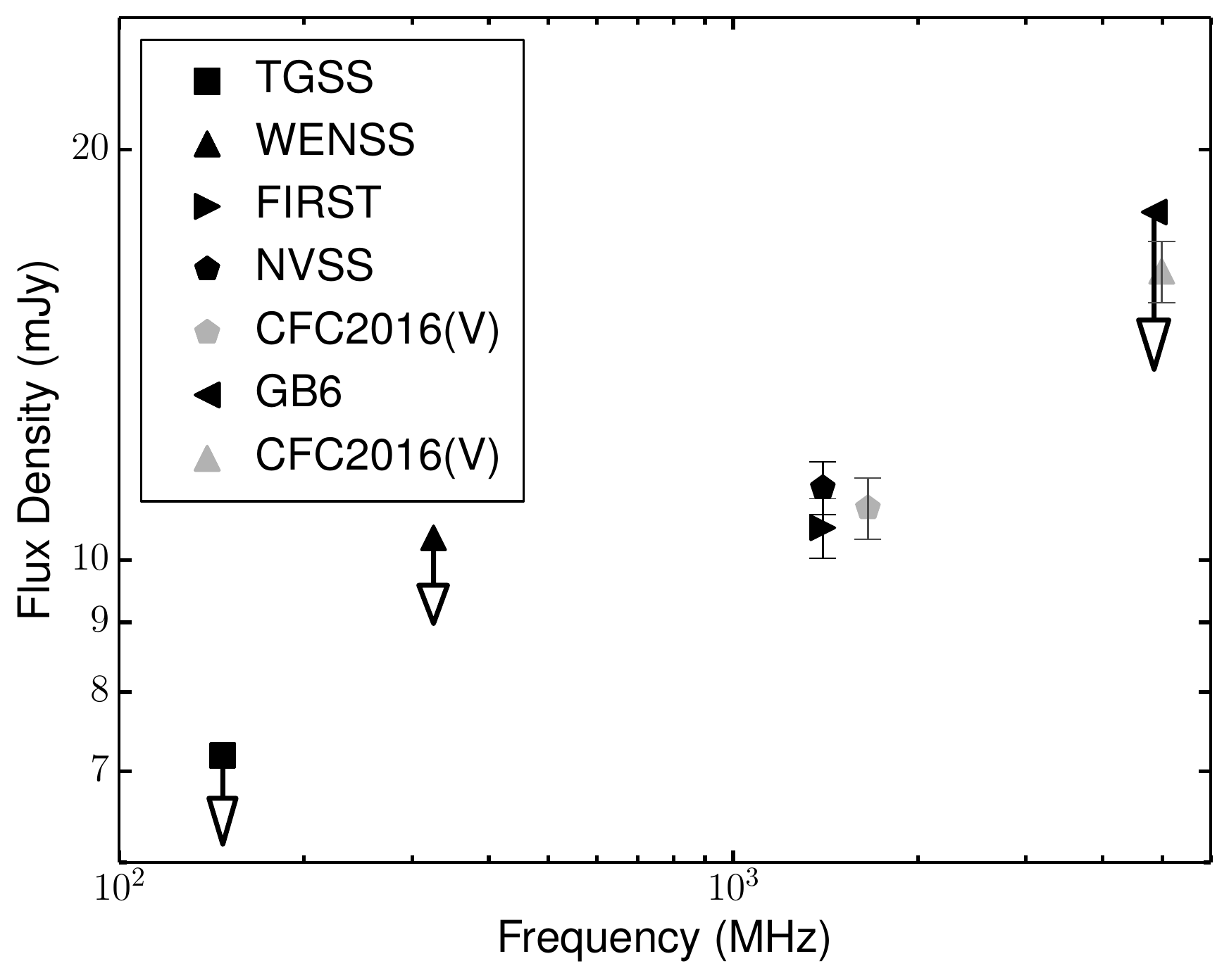}
  \caption{The radio spectrum of J1720+3104.}
  \label{fig:J1720+3104}
\end{figure}

Assuming that the spectrum of J1720+3104 (Fig.~\ref{fig:J1720+3104}) can be fitted with a single power law, and using the 1.4\,GHz FIRST flux density, and the 148\,MHz TGSS (which is more constraining than 325\,MHz WENSS value) and 4.9\,GHz GB6 upper limits, $0.17<\alpha<0.43$. This is consistent with the spectral index of $\alpha=0.36\pm0.07$ measured between the 1.7 and 5\,GHz CFC2016(V) VLBI flux densities. J1720+3104 can therefore have either a flat or a peaked spectrum.

%%%%%%%%%%%%%%%%%%%%%%%%%%%%%%%%%%%%%%%%%%%%%%%%%%%%%%%%%%%%%%%%%%%%%%%%%%%%%%%%%%%%%%%%%%%%%%%%%%%%%%%%%%%%%%%%%%%%%%%%%%%%%%%%%%%%%%%%%%%%%%%%%%%%%%%%%%%%%%%%%%%%%%%%%%%%%%%%%%%%%%%%%%%%%%%%%%%%%%%%%%%%%%%%%%%%%%%%%%
%%%%%%%%%%%%%%%%%%%%%%%%%%%%%%%%%%%%%%%%%%%%%%%%%%%%%%%%%%%%%%%%%%%%%%%%%%%%%%%%%%%%%%%%%%%%%%%%%%%%%%%%%%%%%%%%%%%%%%%%%%%%%%%%%%%%%%%%%%%%%%%%%%%%%%%%%%%%%%%%%%%%%%%%%%%%%%%%%%%%%%%%%%%%%%%%%%%%%%%%%%%%%%%%%%%%%%%%%%
\section{Discussion}
\label{sec:discus}
In Table~\ref{tbl:all clasification numbers}, the number and the percentage of sources in each spectral class are given for the full sample and unbiased sub-sample (which is described later in this section; see the table caption for a description of the nomenclature used). This table was compiled from the classifications in Table~\ref{tbl:source clasification} in the following way: (1) if a source is classified as e.g. `Flat' in Table~\ref{tbl:source clasification}, then the number of flat-spectrum sources is increased by one; (2) if a source is classified as `flat (steep)', then the number of flat-spectrum sources is increased by one, the lower uncertainty on the number of flat-spectrum sources is decreased by one, and the upper uncertainty on the number of steep-spectrum sources is increased by one; (3) if a source is classified as `flat or peaked', the upper uncertainty on the number of flat-spectrum and peaked-spectrum sources are both increased by one. Finally, the percentage of sources in each class of the full sample were calculated using a total number of 29 sources, since the spectrum of J1454+1109 is completely unconstrained (Section~\ref{subsubsec: J1454+1109}). We also point out again, that as discussed in Section \ref{sec:introduction}, in all of the sources except J1205$-$0742 (which has a concave spectrum), the radio emission is caused by AGN activity. In J1205$-$0742 the radio emission is caused by star formation.

\begin{table}
 \hspace{-1cm}
 \centering
 \begin{minipage}{\columnwidth}
  \caption{Spectral classification summary}
  \begin{tabular}{ccccc}
  \hline
                                 & \multicolumn{2}{c}{Full sample $^{\mathrm a}$} & \multicolumn{2}{c}{Unbiased sub-sample $^{\mathrm a}$} \\
  \cline{2-3}
  \cline{4-5}
  Spectral                       & Number of              & \% of                   & Number of              & \% of \\
  classification                 & sources  & sources  & sources  & sources      \\   
  \hline
  Inverted & $0^{+3}_{-0}$  & $0^{+10}_{-0}$    & $0^{+3}_{-0}$ & $0^{+14}_{-0}$\\
  Flat     & $6^{+5}_{-1}$  & $21^{+17}_{-3}$   & $5^{+5}_{-1}$ & $23^{+23}_{-5}$\\
  Steep    & $8^{+1}_{-3}$  & $28^{+3}_{-10}$   & $7^{+1}_{-2}$ & $32^{+5}_{-9}$\\
  USS      & $0^{+2}_{-0}$  & $0^{+7}_{-0}$     & $0^{+1}_{-0}$ & $0^{+5}_{-0}$\\
  Peaked   & $10^{+4}_{-1}$ & $34^{+14}_{-3}$   & $6^{+4}_{-1}$ & $27^{+18}_{-5}$\\
  Concave  & $1^{+0}_{-0}$  & $3^{+0}_{-0}$     & $0^{+0}_{-0}$ & $0^{+0}_{-0}$\\
  \hline
  \multicolumn{5}{p{9cm}}{\footnotesize{\textbf{Notes:} $^{\mathrm a}$ The format $b^{+c}_{-d}$ should be interpreted as follows: There are $b$ sources in the given spectral class, and an additional $c$ sources that are not in the class but could be. Of the $b$ sources, $d$ are in the class but could have a different spectral classification within the errors on their spectral indices.}}\\
  \end{tabular}
  \label{tbl:all clasification numbers}
 \end{minipage}
\end{table}

The primary selection effects in our sample of sources are that all of the sources have spectroscopic redshifts and were selected for follow-up high-resolution VLBI observations. In general the latter involves a flux density lower limit and the sources being compact on arcsec scales in previous (e.g. FIRST) observations. In addition, some authors selected sources for VLBI observations because of the shape of their radio spectra. Since this can bias the values of the full sample in Table~\ref{tbl:all clasification numbers}, we created a unbiased sub-sample of sources that were not selected for VLBI observation with a spectral bias. To do this we checked how each of the sources was selected for VLBI observation the first time that they were observed. If a source was selected for VLBI observations with a spectral bias it was not included in the unbiased sub-sample. This resulted in the following seven sources not being in the unbiased sub-sample: J0311+0507, J0324$-$2918, J0906+6930, J1026+2542, J1205$-$0742, J1606+3124 and J2102+6015. In columns 4 and 5 of Table~\ref{tbl:all clasification numbers} we re-calculated the values in columns 2 and 3 for our unbiased sub-sample. The percentage of sources in each class of the unbiased sub-sample was calculated using a total number of 22 sources, since the spectrum of J1454+1109 is completely unconstrained (Section~\ref{subsubsec: J1454+1109}).

In Table \ref{tbl:all clasification numbers}, the fact that we did not find a single USS sources is striking considering that the USS technique is specifically used to search for high-redshift sources. All of the VLBI observations of the sources were carried out above 1.4\,GHz (Table~\ref{tbl:flux ref}), where the flux densities of the USS sources are rapidly decreasing (Section~\ref{subsec:negative spectrum}). The lack of USS sources could therefore be the result of sources typically only being considered for VLBI observation if, in previous non-VLBI observations, they have flux densities above a certain minimum. 

To attempt to test if this is the case, we downloaded the 12th data release of the Sloan Digital Sky Survey quasar catalog \citep{2017A&A...597A..79P} and removed all sources with SDSS pipeline redshifts smaller than 4.5. Of the remaining 1054 sources, 16 are VLBI sources discussed in this paper. Using a search radius of 5\,arcsec, we matched all the sources in FIRST (1.4\,GHz) to the list of $z>4.5$ SDSS sources and the TGSS catalogue (148\,MHz). From this we found 22 sources which have both FIRST and TGSS flux densities, and of these, six are in this paper. Removing these six sources and calculating two-point spectral indices between FIRST and TGSS for the remaining sources, we found one USS source and one source that could be a USS source within its uncertainties. We do, however, note that since the FIRST and TGSS typical detection thresholds are 1 and 35\,mJy, respectively, only USS sources with FIRST flux densities above 4\,mJy will be detected in TGSS. Only $6^{+6}_{-0}$\,per\,cent of the FIRST--TGSS sources are USS sources. This is in agreement with the percentages of USS sources in Table \ref{tbl:all clasification numbers}. As the fraction of USS sources in these three samples are consistent, it is likely that the requirements for VLBI follow-up observations do not produce a bias against USS sources. 

The largest group of sources in the full sample, and the second largest group of sources in the unbiased sub-sample, are the peaked-spectrum sources. These sources are believed to be young AGN \citep[e.g.][]{o'dea1998,Conway2002,Murgia2002,2003PASA...20...19M,2016AN....337....9O} and make up more than a quarter of the sources in our unbiased sub-sample. Of the 10 peaked-spectrum sources in the sample, sufficient spectral coverage is available to determine the observed turnover frequency of seven of them to within $\sim1$\,GHz (Section~\ref{sec:spectra}). For two of these sources (J0913+5919 and J1659+2102), the observed spectral turnover lies below 1\,GHz, and for three more sources (J1242+5422, J2102+6015 and J2228+0110) the observed turnover could lie below 1\,GHz (but definitely lies below $\sim2$\,GHz). The final two sources (J0906+6930 and J1606+3124) both have observed spectral turnovers above $\sim3$\,GHz. Consequently, the peaked-spectrum sources show a wide range of observed turnover frequencies, and an even wider range of rest-frame turnover frequencies. Based on their observed turnover frequencies, the peaked-spectrum sources are MPS, GPS and HFP sources. This also shows that there are between two and four MPS sources in the unbiased sub-sample. Consequently, although there are more MPS sources than USS sources, neither of these methods would have selected more than $\sim18$\,per\,cent of the sources in the unbiased sub-sample. Interestingly, four of the sources (J0324$-$2918, J0906+6930, J1606+3124 and J2102+6015) that were excluded from the unbiased sub-sample were excluded because they were selected for VLBI observation based on having flat two-point spectral indices \citep{2002ApJS..141...13B,2004ApJ...610L...9R,2006AJ....131.1872P}. However, all four sources actually have peaked spectra and only appeared to have flat spectra because their spectral indices were determined close to the spectral peak.

It is worth noting that the spectra of the steep, and USS, sources have to turn over at some point due to synchrotron self-absorption. In addition, assuming $z=5$, any source with a rest-frame spectral turnover below $\sim3$\,GHz will appear as a steep-spectrum source in our sample since the observed frame turnover frequency will be below $\sim300$\,MHz. For six of the steep-spectrum and USS sources in the sample, the turnover has to be below an observed frequency 1\,GHz, while for two of the sources it has to be below 1.4\,GHz (Section~\ref{sec:spectra}). In total there are $13^{+5}_{-2}$ sources that are steep, USS or peaked in the unbiased sub-sample, which translates to $59^{+23}_{-9}$\,per\,cent of the sources in the unbiased sub-sample. It is therefore safe to say that, if the steep-spectrum sources are observed at lower frequencies ($\nu_{\mathrm o}<100$\,MHz), more of the sources in both the sample and unbiased sub-sample would be classified as peaked-spectrum sources, and there would likely be significantly more MPS sources. 

In CFC2016 we pointed out that the selection effects discussed previously likely bias the sample towards flat-spectrum sources in which the radio emission is Doppler-boosted (which increases the sources' flux density). It was therefore surprising that we found that less than half of the sources could be classified as flat-spectrum radio quasars \citep{2016MNRAS.tmp.1343C}. This conclusion is supported by our new finding that $23^{+23}_{-5}$\,per\,cent of the sources in the unbiased sub-sample have flat spectra.

%%%%%%%%%%%%%%%%%%%%%%%%%%%%%%%%%%%%%%%%%%%%%%%%%%%%%%%%%%%%%%%%%%%%%%%%%%%%%%%%%%%%%%%%%%%%%%%%%%%%%%%%%%%%%%%%%%%%%%%%%%%%%%%%%%%%%%%%%%%%%%%%%%%%%%%%%%%%%%%%%%%%%%%%%%%%%%%%%%%%%%%%%%%%%%%%%%%%%%%%%%%%%%%%%%%%%%%%%%
%%%%%%%%%%%%%%%%%%%%%%%%%%%%%%%%%%%%%%%%%%%%%%%%%%%%%%%%%%%%%%%%%%%%%%%%%%%%%%%%%%%%%%%%%%%%%%%%%%%%%%%%%%%%%%%%%%%%%%%%%%%%%%%%%%%%%%%%%%%%%%%%%%%%%%%%%%%%%%%%%%%%%%%%%%%%%%%%%%%%%%%%%%%%%%%%%%%%%%%%%%%%%%%%%%%%%%%%%%
\section{Summary and conclusions}
\label{sec:summary}
In this paper, we presented new multi-frequency GMRT observations at $\nu<1$\,GHz of eight $z>4.5$ VLBI sources. Matching these eight, and the remaining 22 $z>4.5$ VLBI sources, to the literature, we constructed broad-band radio spectra of all 30 $z>4.5$ VLBI sources. We then discussed and classified the spectra of each of the sources as flat, steep, peaked, unusual and unclassified. Next we looked at the properties of the sample -- particularly the fraction of sources in each spectral class. There are no USS sources in the sample, which we argued is not caused by the requirements for VLBI follow-up observations producing a bias against USS sources. We also show that the USS and MPS methods would each have selected less than $\sim5$ and $\sim18$\,per\,cent of the sources in the sample, respectively. This supports the argument by \citet{2003NewA....8..805P} that the USS sources are not representative of the entire high-redshift source population. We do note that because of the small number of MPS and USS sources in the sample, larger samples are required to draw a definitive conclusion.

The spectra of the steep-spectrum sources have to turn over at some point. If these sources are observed at lower frequencies ($\nu_{\mathrm o}<100$\,MHz), the percentage of peaked-spectrum and MPS sources in the sample would likely increase significantly. This would result in even more MPS than USS sources. We also note that due to a lack of spectral coverage, the classification of some of the sources is uncertain. This problem can be resolved with multi-frequency observations below 2\,GHz, since, for a source at $z=5$, its entire rest-frame spectrum below 12\,GHz will be shifted into observed frequencies below 2\,GHz.

The most striking feature of Table~\ref{tbl:all clasification numbers} is that there is no single spectral class that has the majority of sources. The sources are spread roughly evenly between the flat, steep and peaked classes. In addition, in one of the sources the radio emission is related to star-forming activity. Despite several selection effects, we have to conclude that the $z>4.5$ VLBI sources (and likely also the $z>4.5$ non-VLBI sources) have diverse radio spectra. Considering that we expect the Square Kilometre Array (SKA) to be able to detect sources out to beyond redshift 10 \citep[e.g.][]{falcke2004}, and knowing the general importance of these sources, it is critical that methods are found with which to reliably identify complete samples of high-redshift sources based on radio data.

%%%%%%%%%%%%%%%%%%%%%%%%%%%%%%%%%%%%%%%%%%%%%%%%%%%%%%%%%%%%%%%%%%%%%%%%%%%%%%%%%%%%%%%%%%%%%%%%%%%%%%%%%%%%%%%%%%%%%%%%%%%%%%%%%%%%%%%%%%%%%%%%%%%%%%%%%%%%%%%%%%%%%%%%%%%%%%%%%%%%%%%%%%%%%%%%%%%%%%%%%%%%%%%%%%%%%%%%%%
%%%%%%%%%%%%%%%%%%%%%%%%%%%%%%%%%%%%%%%%%%%%%%%%%%%%%%%%%%%%%%%%%%%%%%%%%%%%%%%%%%%%%%%%%%%%%%%%%%%%%%%%%%%%%%%%%%%%%%%%%%%%%%%%%%%%%%%%%%%%%%%%%%%%%%%%%%%%%%%%%%%%%%%%%%%%%%%%%%%%%%%%%%%%%%%%%%%%%%%%%%%%%%%%%%%%%%%%%%
\section*{Acknowledgements}
The authors wish to thank the referee, Heinz Andernach, and the editor for their suggestions and comments which helped to improve this paper.
We thank the staff of the GMRT who made these observations possible. The GMRT is run by the National Centre for Radio Astrophysics of the Tata Institute of Fundamental Research. Scientific results from data presented in this publication are derived from the following GMRT project codes: 21\_013 (PI: S. van Velzen) and 29\_007 (PI: R. Coppejans).
S.F., D.C., and K.\'E.G. thank the Hungarian National Research, Development and Innovation Office (OTKA NN110333) for their support. 
C.M. and H.F. are funded by the ERC Synergy Grant BlackHoleCam: Imaging the Event Horizon of Black Holes (Grant 610058).
WLW acknowledges support from the UK Science and Technology Facilities Council [ST/M001008/1].
This research has made use of the NASA/IPAC Extragalactic Database (NED) which is operated by the Jet Propulsion Laboratory, California Institute of Technology, under contract with the National Aeronautics and Space Administration, the VizieR catalogue access tool, CDS, Strasbourg, France and NASA's Astrophysics Data System.

%%%%%%%%%%%%%%%%%%%%%%%%%%%%%%%%%%%%%%%%%%%%%%%%%%%%%%%%%%%%%%%%%%%%%%%%%%%%%%%%%%%%%%%%%%%%%%%%%%%%%%%%%%%%%%%%%%%%%%%%%%%%%%%%%%%%%%%%%%%%%%%%%%%%%%%%%%%%%%%%%%%%%%%%%%%%%%%%%%%%%%%%%%%%%%%%%%%%%%%%%%%%%%%%%%%%%%%%%%
%%%%%%%%%%%%%%%%%%%%%%%%%%%%%%%%%%%%%%%%%%%%%%%%%%%%%%%%%%%%%%%%%%%%%%%%%%%%%%%%%%%%%%%%%%%%%%%%%%%%%%%%%%%%%%%%%%%%%%%%%%%%%%%%%%%%%%%%%%%%%%%%%%%%%%%%%%%%%%%%%%%%%%%%%%%%%%%%%%%%%%%%%%%%%%%%%%%%%%%%%%%%%%%%%%%%%%%%%%
%%%%%%%%%%%%%%%%%%%%%%%%%%%%%%%%%%%%%%%%%% BIBLIOGRAPHY %%%%%%%%%%%%%%%%%%%%%%%%%%%%%%%%%%%%%%%%%%

\bibliographystyle{mnras.bst}
\bibliography{references}

\label{lastpage}

%%%%%%%%%%%%%%%%%%%%%%%%%%%%%%%%%%%%%%%%%%%%%%%%%%%%%%%%%%%%%%%%%%%%%%%%%%%%%%%%%%%%%%%%%%%%%%%%%%%%%%%%%%%%%%%%%%%%%%%%%%%%%%%%%%%%%%%%%%%%%%%%%%%%%%%%%%%%%%%%%%%%%%%%%%%%%%%%%%%%%%%%%%%%%%%%%%%%%%%%%%%%%%%%%%%%%%%%%%
%%%%%%%%%%%%%%%%%%%%%%%%%%%%%%%%%%%%%%%%%%%%%%%%%%%%%%%%%%%%%%%%%%%%%%%%%%%%%%%%%%%%%%%%%%%%%%%%%%%%%%%%%%%%%%%%%%%%%%%%%%%%%%%%%%%%%%%%%%%%%%%%%%%%%%%%%%%%%%%%%%%%%%%%%%%%%%%%%%%%%%%%%%%%%%%%%%%%%%%%%%%%%%%%%%%%%%%%%%
\appendix
\section{Flux density references}

\label{appendix: Flux density references}
%\newpage

\begin{table*}
 \hspace{-4cm}
 \centering
 \begin{minipage}{\columnwidth}
  \caption{Flux density references}
  \begin{tabular}{cccc}
  \hline
  Observation name & $\nu$ [MHz] & Reference \\
  \hline  
  4C         & 178  & \citet{1967MmRAS..71...49G}\\
  7C         & 151  & \citet{1996MNRAS.282..779W}\\
  87GB       & 4850 & \citet {1991ApJS...75.1011G}\\
  AT20G      & 4800 \& 8640 \& 19904 & \citet{2010MNRAS.402.2403M}\\
  ATATS      & 1400 & \citet{2010ApJ...719...45C}\\
  B2.2       & 408  & \citet{1972AandAS....7....1C}\\
  B3         & 408  & \citet{1985AandAS...59..255F}\\
  BGP2002(V) & 2268 \& 8338 & \citet{2002ApJS..141...13B}\\
  CBR2001    & 1400 \& 250000 & \citet{2001ApJ...555..625C}\\
  CCW2015    & 325  & \citet{coppejans2015}\\
  CFC2016(V) & 1658 \& 4990 & \citet{2016MNRAS.tmp.1343C}\\
  CFG2014(V) & 1658 &  \citet{2014AandA...563A.111C}\\
  CKK2002    & 1390 \& 42828 & \citet{2002AJ....123.1838C}\\
  CLASS      & 8460 & \citet{2003MNRAS.341....1M}\\
  CMM1999    & 1400 & \citet{1999MNRAS.302..222C}\\
  CNSS       & 3000 & \citet{2016ApJ...818..105M}\\
  CRATES     & 8440 & \citet{2007ApJS..171...61H}\\
  CWH2007    & 233  & \citet{2007AJ....133.2841C}\\f
  FIRST      & 1400 & \citet{FIRST}\\
  FFP2013    & 43000 & \citet{2013MNRAS.431.1314F}\\
  FFP2013(V) & 4850 & \citet{2013MNRAS.431.1314F}\\
  FGP2008(V) & 1600 \& 5000 &  \citet{2008AandA...484L..39F}\\
  FMP2003(V) & 1600 & \citet{2003MNRAS.343L..20F}\\
  FPF2015(V) & 1658 \& 4990 & \citet{Frey2015}\\
  FPG2010(V) & 1658 \& 4990 &  \citet{2010AandA...524A..83F}\\
  FPG2011(V) & 1658 \& 4990 & \citet{2011AandA...531L...5F}\\
  FPM2005    & 1400 \& 5000 &  \citet{2005AandA...436L..13F}\\
  FPM2005(V) & 5000 &  \citet{2005AandA...436L..13F}\\
  GB6        & 4850 & \citet{1996ApJS..103..427G}\\
  GCF2015(V) & 1658 & \citet{2015MNRAS.450L..57G}\\
  GMRT610    & 608 or 612 & This publication\\ 
  GMRT325    & 323  & This publication\\
  GMRT235    & 235  & This publication\\
  GMRT150    & 147  & This publication\\
  GOC1999    & 222068 &   \citet{1999AandA...349..363G}\\
  HTT2007(V) & 4845 &\citet{2007ApJ...658..203H}\\
  LHC1990    & 4830 & \citet{1990ApJS...72..621L}\\
  LKR1997    & 4885 & \citet{1997AandAS..122..235L}\\
  M1972      & 408  & \citet{1972AuJPA..22....1M}\\
  MCM2008    & 8400 & \citet{2008AJ....136..344M}\\
  MCM2008(V) & 1400 & \citet{2008AJ....136..344M}\\
  MCP2005    & 1400 & \citet{2005AJ....129.1809M}\\
  MCP2005(V) & 1425 & \citet{2005AJ....129.1809M}\\
  MPC2004(V) & 1425 & \citet{2004AJ....127..587M}\\
  MOB1994    & 240000 & \citet{1994MNRAS.267L...9M}\\
  MOLONGLO   & 408  & \citet{1981MNRAS.194..693L}\\
  NVSS       & 1400 & \citet{nvss}\\
  OM1977     & 90000 & \citet{1977AJ.....82..776O}\\
  OP1987     & 4585 \& 15064 \& 90000 & \cite{1978AJ.....83..685O}\\
  OMC1996    & 239834 &  \citet{1996AandA...315....1O}\\
  OWB2013    & 250000 &  \citet{2013AandA...552A..43O}\\
  PBW1992    & 8400 & \citet{1992MNRAS.254..655P}\\
  PCB2003    & 1400 \& 5000 & \citet{2003AJ....126...15P}\\
  PFG1999(V) & 5000 &  \citet{1999AandA...344...51P}\\
  PK2012(V)  & 2300 \& 8600 &  \citet{2012AandA...544A..34P}\\
  PKF2006(V) & 2309 \& 8646 & \citet{2006AJ....131.1872P}\\
  PKF2008(V) & 2309 \& 8646 & \citet{2008AJ....136..580P}\\
  PKT2014(V) & 1658 \& 4994 & \citet{2014MNRAS.439.2314P}\\
  \hline
%  \multicolumn{4}{p{10cm}}{\footnotesize{\textbf{Columns:} Column 1 - ; Col. 2 - }}\\
  \end{tabular}
  \label{tbl:flux ref}
 \end{minipage}
\end{table*}

\begin{table*}
 \hspace{-4cm}
 \centering
 \begin{minipage}{\columnwidth}
  \contcaption{}
  \begin{tabular}{cccc}
  \hline
  Observation name & $\nu$ [MHz] & Reference \\
  \hline  
  PMN        & 4850 & \citet{1996ApJS..103..145W}\\
  PTK2014(V) & 1658 \& 4994 & \citet{2014MNRAS.439.2314P}\\
  R2006      & 1440 \& 4880 \& 8440 \& 14950 \& 22490 \& 43330 & \citet{2006AJ....132.1959R}\\
  RFR2000    & 2700 \& 5000 \& 10700 &  \citet{2000AandA...363..141R}\\
  RLM1994    & 1465  & \citet{1994AandAS..108...79R}\\
  RMP2011    & 15000 & \citet{2011ApJS..194...29R}\\
  RSG2004(V) & 15360 \& 43210 & \citet{2004ApJ...610L...9R}\\
  S1995      & 80 \& 160 & \citet{1995AuJPh..48..143S}\\
  STRIPE82   & 1425 & \citet{2011AJ....142....3H}\\
  TEXAS      & 365  & \citet{1996AJ....111.1945D}\\
  TGSS       & 148  & \citet{2016arXiv160304368I}\\
  VFP2010(V) & 15000 &  \citet{2010AandA...521A...6V}\\
  VLSSr      & 74   & \citet{2014MNRAS.440..327L}\\
  WENSS      & 325  & \citet{wenss}\\
  WFP2006    & 1425 \& 4860 \& 8460 \& 15200 \& 22460 \& 43340 & \citet{2006MNRAS.368..844W}\\
  WWC2011    & 32000 & \citet{2011ApJ...739L..34W} \\
  WWR2016    & 149  & \citet{2016MNRAS.460.2385W}\\
  WWT2014    & 62   & \citet{vanweeren2014}\\
  XSD2002(V) & 1660 & \citet{2002AandA...385..768X}\\
  YCK2000    & 1400 \& 4900 & \citet{2000ApJ...528..171Y}\\
  ZELENCHUK  & 3900 & \citet{1991SoSAO..68...14L}\\
  ZLJ2001(V) & 1657 & \citet{2001ChJAA...1..129Z}\\
  \hline
%  \multicolumn{4}{p{10cm}}{\footnotesize{\textbf{Columns:} Column 1 - ; Col. 2 - }}\\
  \end{tabular}
  \label{tbl:flux ref cont.}
 \end{minipage}
\end{table*}

%%%%%%%%%%%%%%%%%%%%%%%%%%%%%%%%%%%%%%%%%%%%%%%%%%%%%%%%%%%%%%%%%%%%%%%%%%%%%%%%%%%%%%%%%%%%%%%%%%%%%%%%%%%%%%%%%%%%%%%%%%%%%%%%%%%%%%%%%%%%%%%%%%%%%%%%%%%%%%%%%%%%%%%%%%%%%%%%%%%%%%%%%%%%%%%%%%%%%%%%%%%%%%%%%%%%%%%%%%
%%%%%%%%%%%%%%%%%%%%%%%%%%%%%%%%%%%%%%%%%%%%%%%%%%%%%%%%%%%%%%%%%%%%%%%%%%%%%%%%%%%%%%%%%%%%%%%%%%%%%%%%%%%%%%%%%%%%%%%%%%%%%%%%%%%%%%%%%%%%%%%%%%%%%%%%%%%%%%%%%%%%%%%%%%%%%%%%%%%%%%%%%%%%%%%%%%%%%%%%%%%%%%%%%%%%%%%%%%

\end{document}